\documentclass{article}
\usepackage{amsfonts,amssymb,eepic,quartic}
\begin{document}
\title{A family of SCFTs hosting all\\ ``very attractive'' relatives
of the $(2)^4$ Gepner model}
\author{Katrin Wendland\thanks{Dept.\ of Mathematics, UNC Chapel Hill,
CB\#3250 Phillips Hall, Chapel Hill, NC 27599-3250, USA.} \thanks{Mathematics Institute,
Univ. of Warwick, Coventry CV4 7AL, UK; wendland@maths.warwick.ac.uk.}}
\date{}
\maketitle
\abstract{This work gives a manual for constructing
superconformal field theories associated to a family of smooth
$K3$ surfaces. A direct method is not known, but a combination of orbifold
techniques with a non-classical duality turns out to yield
such models. A four parameter
family of superconformal field theories
associated to certain quartic $K3$ surfaces in $\CP^3$ is obtained,
four of whose complex structure
parameters give the parameters within superconformal field theory.
Standard orbifold techniques are used to construct these models, so
on the level of superconformal field theory they
are already well understood.

All ``very attractive" $K3$ surfaces
belong to the family of quartics underlying these theories, that is all
quartic hypersurfaces in $\CP^3$ with maximal Picard number
whose defining polynomial is given by the sum of two polynomials in
two variables.
A particular member of the family is the
$(2)^4$ Gepner model, such that these theories can be viewed
as complex structure deformations of $(2)^4$ in its
geometric interpretation on the Fermat quartic.}
\setlength{\parindent}{0pt}
\section*{Introduction}\label{intro}
Conformal field theory (CFT) should provide a natural link between mathematics
and physics: While mathematical definitions of CFT are available
\cite{frsh86,na87,mose88,mose89,se88b,gaga00,se04}, CFT has applications
both in statistical mechanics \cite{bpz84,gi89} and
in string theory \cite{fuve69,na69,gsw87,po98}. Superconformal field theory (SCFT)
can be viewed as a generalization of CFT which in many ways is better behaved,
not least since all consistent string theories
yield superconformally invariant field theories \cite{ra71,nesc71}. Nevertheless,
the mutual interactions between the mathematics and the physics of SCFT
remain limited. On the one hand, those approaches which are fully accepted
in mathematics  \cite{bo86,flm88,se88b,se04} do not succeed to
embody all examples that are of importance in string theory. On the other
hand, string theory has not yet matured to the status of a consistent theory
in the mathematical sense of the word.

Geometric methods yield a promising vehicle to bridge this gap: In physics
such methods have a good tradition, and they are built into string theory
by construction. In mathematics, an area which enjoys great
impact from SCFT is algebraic
geometry, where e.g.\ mirror symmetry tells its own well-known success story
\cite{lvw89,cogp91,grpl90}. To use geometric methods in SCFT a precise understanding
of the mechanisms by which SCFT enriches geometry is desirable. The present
work aims to make a contribution in that direction.

There are only very few examples where the encoding of geometry in SCFT is
understood to a satisfactory degree. When restricting to SCFTs associated to
Calabi-Yau $d$-folds, as seems natural from a string theorist's point of
view\footnote{Here and in the following I restrict my attention to unitary
SCFTs in two dimensions.},
then complex tori and their orbifolds almost exhaust the list of such
examples, to which the less conservative geometer will wish to add lattice and
WZW models along with their orbifolds and coset models. At complex dimension $d\geq3$
this still means that even the degree of our ignorance is hard to gauge.
On the other hand, at complex dimension $d=1$ with solely the elliptic curve to
account for
in the zoo of Calabi-Yau $d$-folds, the complete picture
is understood.

The complex two-dimensional case resides at the borderline when accounting
for ignorance: There are only two topological types of Calabi-Yau $2$-folds,
the complex two-torus and the $K3$ surface. Both the moduli spaces
of SCFTs associated to complex two-tori and to $K3$ surfaces are known to a high
degree of plausibility \cite{na86,se88,asmo94}. For complex two-tori all
associated SCFTs can be constructed explicitly, and their location within the
moduli space along with the translation from geometric to SCFT data is well
understood \cite{cent85,na86}. Within the $80$-dimensional moduli space of SCFTs
associated to $K3$ surfaces, only a finite number of
subvarieties of maximal dimension $16$
is known in the sense that the corresponding SCFTs can be constructed explicitly
(by orbifold techniques \cite{dhvw85,dhvw86,eoty89} or as Gepner models
\cite{ge87,ge88}), and their location within the moduli space along with a translation
between geometric and SCFT data is available \cite{nawe00,we01}. No
direct method is known for the construction of SCFTs associated to smooth $K3$ surfaces.
In this work I provide such a construction for a real four parameter family of
SCFTs associated to smooth quartic $K3$ surfaces. I combine orbifold techniques with
non-classical dualities thus not giving a new construction of SCFTs but rather singling
out a family of  theories which now is well under control from both a superconformal
field theorist's and an algebraic geometer's point of view: The relevant theories
are easy to construct as orbifolds and at the same time have a  parametrization
in terms of algebraic equations describing the underlying quartic $K3$ surfaces. In fact,
the latter geometric interpretation yields all four
real parameters as complex structure deformations, while the complexified K\"ahler
structure remains constant at a natural value.

The tools used here
combine a detailed understanding of the moduli space of SCFTs following Aspinwall
and Morrison \cite{asmo94} with orbifold techniques taken from the
physics literature and following Dixon, Harvey, Vafa, Witten \cite{dhvw85,dhvw86}
and Eguchi, Ooguri, Taormina, Yang \cite{eoty89} as well as their
mathematical predecessors, where particularly in the context of $K3$ surfaces
Nikulin's work \cite{ni75} is of importance. These joint forces have already led
to the appropriate description of all orbifold SCFTs obtained
from toroidal models within
the moduli space \cite{nawe00,we01}, which the present work is built on.
As a final ingredient Witten's results on the phase structure of the parameter
space of supersymmetric gauge theories \cite{wi93} comes to aid. The
family of SCFTs studied in this work allows a  description
within each of these settings.

SCFTs associated to Calabi-Yau 2-folds are comparatively tractable because
they enjoy extended $N=(4,4)$ supersymmetry beyond the usual $N=(2,2)$ supersymmetry
required for SCFTs associated to Calabi-Yau $d$-folds in general. Geometrically
this corresponds to the observation that all Calabi-Yau $2$-folds are hyperk\"ahler.
Hence many of my techniques will not generalize to higher dimensions.
However, the main result as stated addresses geometric interpretations of SCFTs on
$K3$ surfaces that are equipped with a complex structure, a K\"ahler class,
and a B-field. I call these data a ``refined geometric interpretation"
to distinguish them from the ordinary geometric interpretations of such theories which
amount to fixing a hyperk\"ahler structure, a volume, and a B-field.
Additionally specifying a complex structure within the data of such a
theory\footnote{Given a
hyperk\"ahler structure, for each compatible choice of complex structure
up to normalization there exists a unique compatible K\"ahler class.}
amounts to the choice of an $N=(2,2)$ subalgebra within the
given $N=(4,4)$ superconformal algebra (although vice versa not every choice of
an $N=(2,2)$ superconformal algebra induces the choice of a complex structure,
as we shall see and as was already pointed out in \cite{hu03}). Viewed as $N=(2,2)$ SCFTs
the main protagonist of this work, a four-parameter family of SCFTs associated
to a smooth family of quartic $K3$ surfaces, is understood to a degree which
should allow for applications
that may very well generalize to higher dimensions.

Particularly because this family of SCFTs simultaneously
allows a description in terms of representation theory through its orbifold construction
and in terms of algebraic geometry in a way which is compatible with linear sigma
model constructions, it yields a tailor made testing ground for
modern techniques in SCFT which so far have only been successfully applied
within one of these pictures or in simpler examples like toroidal SCFTs or
minimal models.
Indeed, my main protagonist family of SCFTs can be viewed as a complex structure
deformation of the $(2)^4$ Gepner model in its geometric interpretation on the Fermat
quartic. As such it should lend itself to a study of D-branes combining orbifold
techniques as in \cite{ber99} with modern techniques from matrix factorization
\cite{wa95,bhls03,kali03a,add04,hll04,brga05a,egj05,brga05b,err05},
not only for $(2)^4$ but for the
entire four-parameter family of SCFTs which deforms $(2)^4$. In terms of more
abstract approaches to SCFT it may also be interesting to study this
family from the viewpoint of the chiral de Rham complex \cite{msv98,boli00,goma03}:
All relevant vertex algebras should be accessible explicitly.

Concerning the title of this work let me briefly comment on ``very attractive"
$K3$ surfaces. Following Moore \cite{mo98b,mo98a}, I call a $K3$ surface attractive
iff it has maximal Picard number. If an attractive $K3$ surface can be given
as zero locus of a homogeneous polynomial of degree $4$ in $\CP^3$
which decomposes into a sum of two polynomials in two variables each, then I  call it
``very attractive". All ``very attractive" $K3$ surfaces belong to
the family of quartics  which I associate
SCFTs to in this work, forming a dense subset.
However, not all of the theories
associated to ``very attractive" $K3$ surfaces are rational.
This work originally arose from ideas concerning attractiveness in geometry
and rationality in SCFT. In particular, part of the results presented here were
already announced in \cite{we03}, where however I did not notice that the constructions
sketched there for ``very attractive" quartics
extend  to a smooth family of SCFTs. There, also only three  of
now four real parameters $\alpha,\,\beta,\,\beta^\prime,\,\gamma$
were explored. To reduce to the situation of \cite{we03}, set $\beta^\prime=\beta$
in the present work. Finally,
details and proofs were omitted in \cite{we03}
which I now provide in full generality. In fact, the present work aims to be essentially
self-contained.

It is organized as follows.

As a warmup, Section \ref{CY1} is devoted to SCFTs associated to Calabi-Yau 1-folds.
The material is well-established but is presented in a slightly more abstract form
than is common, to facilitate later reference in the higher dimensional case. I give
a representation theoretic definition of these theories and summarize
their properties as SCFTs and in relation to the geometry and the algebraic description
of elliptic curves.

Section \ref{CY2} also begins with the presentation of known
material concerning the moduli space of SCFTs associated to Calabi-Yau 2-folds.
Again I give a representation theoretic definition of such theories, and I
summarize the current state of knowledge concerning their moduli space and its
relation to geometric data. Particularly the notion of refined geometric interpretations
is discussed and compared to the generalized $K3$ structures of \cite{hi03,hu03}.
Moreover, two families of SCFTs are introduced, one associated to real four-tori
and one to $K3$ surfaces, yielding the main protagonists of this work. Both as a preparation
for the main result and as an example for the general techniques discussed before, two
distinct refined geometric interpretations are worked out for each of these families.

Section \ref{mainclaim} is devoted to the formulation of the main result
of this work and its discussion:
The family of SCFTs associated to $K3$ introduced previously allows a refined geometric
interpretation which associates it to a family of smooth quartic $K3$ surfaces,
given in terms of explicit algebraic equations. I provide a first step in the proof of
this claim and motivate it in terms of an extension of a construction by Inose
\cite{in76} to SCFTs: Inose's results concern complex structures of $K3$ surfaces only,
while on the level of SCFTs we deal with pairs of complex structures and
complexified K\"ahler structures.
Motivated by this interpretation of the main result
on a purely geometric level I deduce properties of
the natural K\"ahler class of our
quartic $K3$ surfaces in $\CP^3$, which descends from the class of
the Fubini-Study metric on $\CP^3$: The induced K\"ahler class on a $\Z_2$-orbifold
of such quartics is closely and explicitly related to a K\"ahler class which is induced
by a Kummer construction. This result makes the underlying K\"ahler-Einstein
metrics directly accessible to numerical approaches developed recently \cite{hewi05}
and may be interesting in its own right. I present a simple proof which does not use
results from SCFT.

The following Section \ref{proof} contains the remaining steps in the proof of the
main result. This largely amounts to understanding the particular model
$(2)^4$ within the family of SCFTs discussed here, along with its deformations.

I conclude with a discussion in Section \ref{discussion}, and four Appendices
contain details about the geometry of elliptic curves, about minimal models, and about
Gepner models, which are used in the main text.
\paragraph{Acknowledgments.}
It is a pleasure to thank Paul Aspinwall, Gavin Brown, Chuck Doran,
Viacheslav V.\ Nikulin, and James Smith
for helpful discussions. I particularly thank Matthew Headrick and Toby
Wiseman for communication about their work \cite{hewi05} and for asking
the crucial questions which led to the formulation of Proposition \ref{metric}.
The first foundations of this work were laid in \cite{nawe00}
in collaboration with Werner Nahm, which I gratefully acknowledge.
\section{Warmup: SCFTs associated to Calabi-Yau
1-folds}\label{CY1}
Before turning to the main topic of this work,
I discuss SCFTs associated to Calabi-Yau 1-folds. There is
only one topological type of Calabi-Yau 1-folds, namely the elliptic curve,
and SCFTs associated to elliptic curves are well understood:
These theories allow an abstract mathematical definition in terms
of representation theory, all such theories can be constructed
explicitly, and their moduli space is known, including the translation
from conformal field theoretic into geometric data. The
moduli space and the
notion of geometric interpretation for such theories bear some
resemblance to the corresponding notions for Calabi-Yau 2-folds,
which play center stage in this work. For this reason and also since
SCFTs associated to elliptic curves are
building blocks of the main protagonists of this work it
is worthwhile to describe these theories in some detail, even though the
material is  standard and can be found in various textbooks, see also
\cite{dvv87}.

Section \ref{CY1defprop} is devoted to the mathematical definition of
SCFTs on elliptic curves and the description of these theories. While my
definition is not completely standard, the expert will  notice that it
yields precisely those theories known as toroidal SCFTs with central charge
$c=3=\qu c$ in the physics literature. My definition has the advantage that
it can be completely paralleled when it comes to defining SCFTs associated
to Calabi-Yau 2-folds. Since some background knowledge in SCFT is assumed in
this section, the non-expert may choose to skip directly to
Section \ref{CY1modsp} and accept the claims made there as given facts.
That section is devoted to the discussion of the moduli space of SCFTs associated
to elliptic curves, including the notion or mirror symmetry and its cousins.
In Section \ref{algebraic} an
algebraic description for elliptic curves is introduced which is
needed later and which
differs from the standard Weierstra\ss\ form.
\subsection{Definition and properties}\label{CY1defprop}
I use the following definition for SCFTs associated
to elliptic curves:
\bdefi{toroidal}
An $N=(2,2)$ SCFT $\EEE$ with\footnote{Without further mention
SCFTs in this work always refer to unitary CFTs in two dimensions.} central
charges $c=\qu c=3$ is called \textsc{toroidal}
or \textsc{associated to an elliptic curve} iff the following holds:

The pre-Hilbert
space $\HHH$ of $\EEE$ decomposes into $\HHH=\NS\oplus\Ra$, where the
Neveu-Schwarz sector $\NS$ and the Ramond sector $\Ra$
are isomorphic under the spectral flow. Moreover, in $\NS$
all charges with respect
to the $\fu(1)$ currents $J,\,\qu J$ of the
superconformal algebras on the left and right are integral, where
the standard normalization
\beq{stno}
J(z) J(w) \sim {c/3\over(z-w)^2} + \OO(1),\quad\quad
\qu J(\qu z) \qu J(\qu w)
\sim {\qu c/3\over(\qu z-\qu w)^2} + \OO(1).
\eeq
is used.
\edefi
This definition implies that every
SCFT $\EEE$ associated to an elliptic curve contains the
operators of two-fold spectral flow as fermionic fields $\psi_\pm$
on the left and $\qu\psi_\mp$ on the right, and that
both the holomorphic and the anti-holomorphic
W-algebras contain a $\fu(1)^3$ current algebra: On each side
there is one $\fu(1)$ current
$J,\,\qu J$ belonging to the superconformal algebra, and two further
purely bosonic
currents, the superpartners $j_\pm,\,\qu\j_\pm$
of the $\psi_\pm,\,\qu\psi_\pm$. It is not hard to see that
$\psi_\pm,\,\qu\psi_\pm$ give an ordinary Dirac fermion, and\footnote{Here and
in the following, statements made for holomorphic (left-handed) fields hold
analogously for anti-holomorphic (right-handed) fields, though I will not
always mention this explicitly.}
$J=i\nop{\psi_-\psi_+}$. Moreover, $\EEE$ is a tensor product
of a bosonic toroidal CFT at $c=\qu c=2$
(most conveniently defined as CFT with central charges $c=\qu c=2$
such that both holomorphic and anti-holomorphic W-algebras contain
a $\fu(1)^2$ current algebra)
with the fermionic theory at
$c=\qu c=1$ given by the Dirac fermion.
We denote the real and imaginary parts of
$\sqrt{2} j_\pm,\,\sqrt{2} \qu\j_\pm$  by
$j_1,\,j_2,\,\qu\j_1,\qu\j_2$, normalized such that
\beq{bosnorm}
j_k(z) j_l(w) \sim {\delta_{kl}\over(z-w)^2} + \OO(1),\quad\quad
\qu\j_k(\qu z) \qu\j_l(\qu w)
\sim {\delta_{kl}\over(\qu z-\qu w)^2} + \OO(1).
\eeq
The left-handed Virasoro field of $\EEE$ hence is
\beq{virasoro}
T={1\over2}\nop{j_1 j_1}+{1\over2}\nop{j_2 j_2}
+{1\over2}\nop{\partial\psi_+ \psi_-}+{1\over2}\nop{\partial\psi_- \psi_+}.
\eeq
A toroidal theory $\EEE$ with central charges $c=\qu c=3$ is uniquely
determined by its charge
lattice $\Gamma\subset\R^{2,2}$ with respect to
$(j_1,\,j_2;\qu\j_1,\qu\j_2)$. Here $\R^2$ carries the standard
Euclidean scalar product, and
$$
\mb{for }\;(p;\qu p),\,(p^\prime;\qu p^\prime)\in\R^{2,2}\;\mb{ with }\;
p,\,\qu p,\,p^\prime,\,\qu p^\prime\in\R^2,\quad\quad
(p;\qu p)\cdot(p^\prime;\qu p^\prime)=p\cdot p^\prime-\qu p\cdot \qu p^\prime.
$$
Each $(p;\qu p)\in\Gamma$ labels a vertex operator of charge
$(p;\qu p)$ with respect to $(j_1,\,j_2;\qu\j_1,\qu\j_2)$, which by
\req{virasoro} has conformal weights $({p^2\over2},{\qu p^2\over2})$.
These vertex operators create the ground states with respect to the generic
W-algebras of toroidal SCFTs, which are generated by the superconformal
algebras together with the $\fu(1)$ currents $j_k,\,\qu\j_k$.
The total partition function of such a theory and its Neveu-Schwarz part
are  given by
\beqn{torpart}
Z(\tau^\prime,z) \>=\>
\tr[\HHH]\left[{1\over2}\left(1+\left(-1\right)^F\right) y^{J_0}\qu y^{\qu J_0}
q^{L_0-{1\over8}} \qu q^{\qu L_0-{1\over8}}\right]\\
\>=\>\sum_{\left(p;\qu p\right)\in\Gamma }
{q^{p^2\over2}\qu q^{\qu p^2\over2}\over\left|\eta(\tau^\prime)\right|^2}\cdot
\;{1\over2}
\sum_{i=1}^4 \left|{\theta_i(\tau^\prime,z)\over\eta(\tau^\prime)}\right|^2,
\quad\quad\quad\quad
(-1)^F=e^{i\pi(J_0-\qu J_0)},
\e
Z_{NS}(\tau^\prime,z) \>=\>
\tr[\NS] \left[ y^{J_0}\qu y^{\qu J_0}
q^{L_0-{1\over8}} \qu q^{\qu L_0-{1\over8}}\right]
=\sum_{\left(p;\qu p\right)\in\Gamma }
{q^{p^2\over2}\qu q^{\qu p^2\over2}\over\left|\eta(\tau^\prime)\right|^2}\cdot
\;\left|{\theta_3(\tau^\prime,z)\over\eta(\tau^\prime)}\right|^2,
\eeqn
where $\tau^\prime\in\H=\left\{\zeta\in\C\mid \Im(\zeta)>0\right\}$,
$z\in\C$,
$q=e^{2\pi i\tau^\prime}$, $y=e^{2\pi i z}$,
$\eta$ and $\theta_i$ denote the Dedekind
eta and the Jacobi theta functions, and $J_0,\,\qu J_0,\,L_0,\,\qu L_0$ the
respective zero modes of  $\fu(1)$ currents and the Virasoro fields
in the superconformal algebra.

For toroidal theories associated to elliptic curves the charge lattice $\Gamma$
can always be expressed in terms of
two moduli $\tau,\,\rho\in\H$ as follows:
\beq{chl}
\begin{array}{rclrcl}
\Gamma_{\tau,\rho}
\>\!\!\!\!\!\!=\!\!\!\!\!\!& \multicolumn{4}{l}{\left\{\ds
{1\over\sqrt{2}}\left(\lambda^*-B\lambda+\lambda;\lambda^*-B\lambda-\lambda\right)
\right|\left.\ds
\lambda^*=\sum_{k=1}^2 m_k\lambda^*_k,\;\lambda=\sum_{k=1}^2 n_k\lambda_k,\;
n_k,\,m_k\in\Z\right\}, }\\[14pt]
\ds\lambda_1\>\!\!\!\!\!\!:=\!\!\!\!\!\!\>\sqrt{\Im(\rho)\over\Im(\tau)}{1\choose0},
\>\lambda_2\>\!\!\!\!\!\!:=\!\!\!\!\!\!\>\sqrt{\Im(\rho)\over\Im(\tau)} {\Re(\tau)\choose \Im(\tau)},\,
\nonumber\e
\lambda^*_1\>\!\!\!\!\!\!:=\!\!\!\!\!\!\>{1\over\sqrt{\Im(\rho)\Im(\tau)}} {\Im(\tau)\choose -\Re(\tau)},
\>\lambda^*_2\>\!\!\!\!\!\!:=\!\!\!\!\!\!\>{1\over\sqrt{\Im(\rho)\Im(\tau)}} {0\choose 1},\quad
B:={\Re(\rho)\over \Im(\rho)}
\left(\begin{array}{cc}0\>-1\\1\>0\end{array}\right).
\end{array}
\eeq
\subsection{Moduli space and dualities}\label{CY1modsp}
Since by the above every SCFT $\EEE$ associated to an elliptic curve
is uniquely determined by its charge lattice $\Gamma_{\tau,\rho}$, which in
turn can be given in terms of a pair $\tau,\,\rho\in\H$, a parameter space
of all such theories is $\H\times\H$. In fact, inspection of
$\Gamma_{\tau,\rho}$ in \req{chl} shows (see \cite{na86})
\bprop{ellims}
The moduli space of SCFTs associated to elliptic curves according
to Definition \mbox{\rm\ref{toroidal}} is
$$
\MMM^\EEE
= \left( \vphantom{\sum}\PSL_2(\Z)\backslash\H\quad\times\quad\PSL_2(\Z)\backslash\H\right)
\big\slash\Z_2^2,
$$
where every pair $\tau,\,\rho\in\H$ determines the unique
such theory with charge lattice $\Gamma_{\tau,\rho}$ given in
\mbox{\rm\req{chl}}, $\PSL_2(\Z)$ acts by M\"obius transforms on $\H$,
and $\Z_2^2$ is generated by $U(\tau,\rho):=(\rho,\tau)$ and
$V(\tau,\rho):=(-\qu\tau,-\qu\rho)$.
\eprop
Though $U,\,V$ induce non-trivial actions on $\Gamma_{\tau,\rho}$,
these agree with the actions induced by reparametrizations
$(j_1,\,j_2;\qu\j_1,\qu\j_2)\mapsto (j_1,\,j_2;-\qu\j_1,\qu\j_2)$
and $(j_1,\,j_2;\qu\j_1,\qu\j_2)\mapsto (-j_1,\,j_2;-\qu\j_1,\qu\j_2)$,
yielding the associated SCFTs equivalent.

Traditionally, the parameter $\tau\in\H$ of a SCFT
$\EEE$ associated to an elliptic curve with charge lattice $\Gamma_{\tau,\rho}$
is interpreted as the \textsc{period} of an elliptic curve
$E_\tau$ fixing its complex structure, while $\rho\in\H$ determines a
\textsc{complexified K\"ahler} structure on $E_\tau$:
$\Im(\rho)>0$ gives the volume of $E_\tau$, thereby specifying
a K\"ahler structure
because $H^{1,1}(E_\tau,\C)\cap H^2(E_\tau,\R)=H^2(E_\tau,\R)\cong\R$
and in accord with $\det\left(\lambda_1,\lambda_2\right)=\Im(\rho)$
from \req{chl},
while $\Re(\rho)$ specifies
the so--called
B-field $B=\Re(\rho)\cdot\lambda_1^\ast\wedge\lambda_2^\ast\in H^2(E_\tau,\R)\cong\R$.
This justifies the
terminology in Definition \ref{toroidal}, and the pair
$(\tau,\rho)\in\H\times\H$ specifying a toroidal SCFT is referred
to as \textsc{geometric interpretation} of the theory. Note that
$U(\tau,\rho)=(\rho,\tau)$ exchanges  complex and complexified
K\"ahler structures of a given geometric interpretation
and thus yields the simplest form of \textsc{mirror symmetry},
while $V(\tau,\rho)=(-\qu\tau,-\qu\rho)$ is induced by an orientation
change of the ``target space" $E_\tau$.
\subsection{An algebraic description}\label{algebraic}
Instead of characterizing an elliptic curve $E_\tau$ by its period $\tau\in\H$
it is often more desirable to work with explicit equations. The
standard description gives an elliptic curve (with inflection point)
in $\CP^2$ in terms of its Weierstra\ss\ form
\beq{weierstrass}
\mb{with }a,\,b\in\C:\quad\quad
y^2 t= x^3 - 27a x t^2 -54 b t^3 \quad\mb{ for }\quad (x,y,t)\in\CP^2.
\eeq
The non-degenerate elliptic
curves, which I shall restrict to in the following,
are the ones which obey $a^3\neq b^2$. The period
$\tau\in\H$ of  \req{weierstrass} is
obtained by means of the $j$-function $j\colon\H\rightarrow\C$,
the unique modular
invariant biholomorphic function with $q$-expansion
$$
j(\tau) =  q^{-1} + 744 + 196884 q + \cdots,
\quad q:= e^{2\pi i \tau}.
$$
For the curve
\req{weierstrass} with period $\tau\in\H$ one has
$$
j(\tau) = {1728 a^3\over a^3-b^2}.
$$
Both for
the function $j$ and its inverse rapidly convergent
algorithms are available, see e.g.\ \cite[Section VI.9]{kn92}.

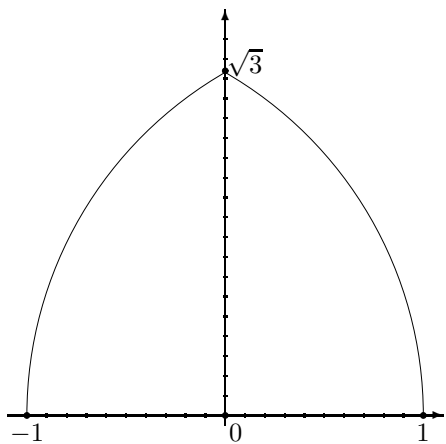
\begin{figure}[ht]\normalsize
\setlength{\unitlength}{0.75em}
\hspace*{\fill}
\begin{picture}(20,20)(-10,-0.5)
\put(-11,0){\vector(1,0){22}}
\put(0,-0.5){\vector(0,1){21}}
\multiput(-10,-0.1)(1,0){20}{\line(0,1){0.2}}
\multiput(-0.1,0)(0,1){20}{\line(1,0){0.2}}
\put(10,0){\arc{40}{9.42}{10.48}}
\put(-10,0){\arc{40}{5.23}{0}}
\put(0.1,18.5){\makebox(0,0)[tl]{$\sqrt3$}}
\put(-10,-0.4){\makebox(0,0)[tc]{$-1$}}
\put(10,-0.4){\makebox(0,0)[tc]{$1$}}
\put(0.2,-0.4){\makebox(0,0)[tl]{$0$}}
\put(-10,0){\circle*{0.3}}
\put(10,0){\circle*{0.3}}
\put(0,0){\circle*{0.3}}
\put(0,17.4){\circle*{0.3}}
\end{picture}
\hspace*{\fill}
\caption{\normalsize{}Fundamental domain for $\kappa\in\C$ in
$y_1^4 + 2\kappa y_1^2 y_2^2 + y_2^4$.}\label{domain}
\end{figure}

In the application below, the elliptic curves are given in weighted projective
space,
$$
E_f:\quad
y_0^2 = f(y_1,y_2) \quad\mb{ in }\quad \CP_{2,1,1}
$$
with $f$ a non-degenerate homogeneous polynomial of degree $4$,
i.e.\ such that no two roots of $f$ agree.
To arrive from the standard form \req{weierstrass}
at such a description one maps the four two-torsion points
to the four
solutions $(0,y_1,y_2)\in\CP_{2,1,1}$ of $f(y_1,y_2)=0$.
Without loss of generality $f$ has the form
\beq{quarticformulation}
f(y_1,y_2)
= y_1^4 + 2\kappa y_1^2 y_2^2 + y_2^4,\quad \kappa\in\C
\mb{ with } \Im(\kappa)\geq0, |\kappa\pm1|\leq2,
\eeq
where $\kappa\sim-\qu\kappa$ if $|\kappa\pm1|=2$, $\kappa\sim-\kappa$
if $\kappa\in\R$, and $\kappa=\pm1$ gives
a degenerate elliptic curve. See Figure \ref{domain} to picture the
fundamental domain for $\kappa$, and see
 \ref{quarelli} for  details.
Altogether the maps $\kappa=\kappa(\tau)$ and $\tau=\tau(\kappa)$
which relate the algebraic description \req{quarticformulation}
of elliptic curves in $\CP_{2,1,1}$ to their periods $\tau\in\H$
amount to combining the $j$-function or its inverse with solving
algebraic equations. Hence
$\kappa=\kappa(\tau)$ and $\tau=\tau(\kappa)$
can be  determined numerically. In some cases the
result is known explicitly, e.g.\
\beq{fermatelli}
\tau=i\in\H \quad\quad\longleftrightarrow\quad\quad E_{f_0}\colon\quad y_0^2=y_1^4+y_2^4\quad
\mbox{ in }\quad \CP_{2,1,1},
\eeq
see also  \ref{quarelli}.
\section{SCFTs associated to Calabi-Yau
2-folds}\label{CY2}
In this section I set the stage for the formulation of the
main Result \ref{main} of this work.
Namely, I discuss SCFTs associated to Calabi-Yau 2-folds. These
theories are defined
purely within representation theory,  paralleling  Definition \ref{toroidal}
of SCFTs associated to elliptic curves.
This definition is given in Section \ref{CY2defmod} along with the discussion
of the moduli space of SCFTs associated to Calabi-Yau 2-folds and
their (refined) geometric interpretations,  essentially summarizing the
results of \cite{asmo94}. Section \ref{protagonists}
is devoted to the introduction of the two families of SCFTs
associated to Calabi-Yau 2-folds which feature in
the main Result \ref{main}: One real four-parameter family of SCFTs associated to real
four-tori and one associated to $K3$ surfaces, where the latter is obtained
from the former by an orbifold construction. For later convenience a pair of
``dual" refined geometric interpretations for each member of both families
is  provided.
\subsection{Definition and moduli space}\label{CY2defmod}
To formally define SCFTs associated to Calabi-Yau 2-folds one can
straightforwardly parallel Definition \ref{toroidal}
\bdefi{CY2scft}
An $N=(4,4)$ SCFT$\,$\footnote{There are various extended $N=4$
superconformal algebras, see \cite{stp88}; the one needed here
is the one which contains a single $\fs\fu(2)$ Kac-Moody algebra at level $1$.}
$\CCC$ with central charges $c=\qu c=6$ is called \textsc{associated
to a Calabi-Yau 2-fold} iff the following holds:

The pre-Hilbert
space $\HHH$ of $\CCC$ decomposes into $\HHH=\NS\oplus\Ra$, where the
Neveu-Schwarz sector $\NS$ and the Ramond sector $\Ra$
are isomorphic under the spectral flow. Moreover, in $\NS$
all charges with respect
to the $\fu(1)$ currents of the
superconformal algebras on the left and right with standard normalizations
\mb{\req{stno}}
are integral.
\edefi
Every $N=(2,2)$ SCFT with central charges $c=\qu c=6$
which obeys the additional assumptions on the spectral flow and the
$\fu(1)$ charges of Definition \ref{CY2scft} automatically enjoys
$N=(4,4)$ supersymmetry of the type assumed in the
definition, see e.g.\ \cite{eoty89}: The assumptions ensure that the
operators of two-fold left and right handed spectral flows are operators
of these SCFTs; one checks that at central charges
$c=\qu c=6$ these operators furnish additional
currents which enhance the left and the right $\fu(1)$ subalgebras
of the $N=(2,2)$ superconformal algebra to an $\fs\fu(2)$ at level $1$
each, thus enhancing $N=(2,2)$ supersymmetry to $N=(4,4)$. I nevertheless
include $N=(4,4)$ supersymmetry in the assumptions of Definition \ref{CY2scft},
because these theories shall be viewed as $N=(4,4)$ SCFTs without an a priori
choice of an $N=(2,2)$ subalgebra of the $N=(4,4)$ superconformal algebra.
The condition on $\fu(1)$ charges makes sense without such a choice,
because all Cartan tori $\fu(1)\subset\fs\fu(2)$ are
conjugate such that the spectrum of $\fu(1)$ charges does not
depend on such a choice.

While definitions analogous to  \ref{toroidal} and \ref{CY2scft}
make sense at central charges $c=\qu c=3D$ for arbitrary $D\in\N$, the cases
$D\in\{1,2\}$ are special in that for them the moduli spaces of SCFTs associated to
Calabi-Yau $D$-folds are known and are expected to decompose into a finite number of
connected components. For $D=1$ we have seen this in Section \ref{CY1},
while for $D=2$ the statement is closely linked to
\blem{ellgen}
If $\CCC$ is a SCFT associated to a Calabi-Yau 2-fold according to
Definition \mb{\ref{CY2scft}}, consider its \textsc{conformal field
theoretic elliptic genus}
$$
\ZZZ(\tau^\prime,z) := \tr[\Ra]\left[\left(-1\right)^F y^{J_0}
q^{L_0-{1\over4}} \qu q^{\qu L_0-{1\over4}}\right],\quad\quad
(-1)^F=e^{i\pi(J_0-\qu J_0)},
$$
with $\tau^\prime\in\H,\, z\in\C,\,q=e^{2\pi i\tau^\prime},\,
y=e^{2\pi i z}$, $\Ra$ denoting the Ramond sector as in Definition
\mb{\ref{CY2scft}}, and $J_0,\,\qu J_0,\,L_0,\,\qu L_0$ the
respective zero modes of  $\fu(1)$ currents and the Virasoro fields
in the superconformal algebra.
Then either $\ZZZ(\tau^\prime,z)\equiv0$ or
\begin{eqnarray*}
\ZZZ(\tau^\prime,z)
&=& {2\over\eta^6(\tau^\prime)}
\left( \theta_2^2(\tau^\prime,z)\cdot\theta_3^2(\tau^\prime,0)\cdot\theta_4^2(\tau^\prime,0)
\right.\\
&&\quad\quad\left.
+\theta_3^2(\tau^\prime,z)\cdot\theta_4^2(\tau^\prime,0)\cdot\theta_2^2(\tau^\prime,0)
+\theta_4^2(\tau^\prime,z)\cdot\theta_2^2(\tau^\prime,0)\cdot\theta_3^2(\tau^\prime,0)
\right),
\end{eqnarray*}
where $\eta,\,\theta_l$ denote the Dedekind eta and the Jacobi theta functions.
In other words, the conformal field theoretic elliptic genus $\ZZZ$ agrees
with the geometric elliptic genus of a complex two-torus or a $K3$ surface, i.e.\
of one of the two topologically distinct Calabi-Yau 2-folds.
\elem
A proof of this Lemma follows from the modular properties of the conformal field theoretic
elliptic genus, which for SCFTs associated to Calabi-Yau 2-folds is a theta function
of degree $n=2$ and characteristic $(0,0;-4\pi in,-2\pi i\tau)$. The proof
can be found in \cite{ho91} and also in \cite{diss}. Lemma \ref{ellgen} allows us
to formally assign the label ``torus" or ``$K3$" to each SCFT associated to a Calabi-Yau
2-fold by means of the conformal field theoretic elliptic genus:
\bdefi{k34tori}
A SCFT associated to a Calabi-Yau 2-fold is said to be a
\textsc{SCFT on a real four-torus} iff its conformal field
theoretic elliptic genus vanishes. Otherwise, it is said to
be a \textsc{SCFT on a $K3$ surface}.
\edefi
While using \cite{cent85,na86} one can show that the
SCFTs associated to real four-tori form a connected component of
the moduli space of all SCFTs associated to Calabi-Yau 2-folds, a proof
of the analogous statement for SCFTs associated to $K3$ is not known.
In physics, one largely works under the assumption that indeed the space
of SCFTs associated to Calabi-Yau 2-folds has only two connected components,
and no counter example to this assumption is known. Below I will solely work
with smooth families of such SCFTs such that the possible existence
of further components of the moduli space is not relevant to the present work.

Essentially due to the extended
$N=(4,4)$ supersymmetry of SCFTs associated to Calabi-Yau 2-folds
it is possible to determine the
form of each connected component of their moduli space explicitly.
Here I restrict myself to stating the result; for more details see e.g.\
\cite{cent85,na86,se88,ce90,asmo94,diss,nawe00,we01}:
\btheo[\cite{cent85,na86,se88,ce90,asmo94,nawe00}]{modsp}
Every connected component of the moduli space of SCFTs associated to
Calabi-Yau 2-folds
is either of the form $\MMM^{tori}=\MMM^{0}$ or $\MMM^{K3}=\MMM^{16}$
with\footnote{For $\MMM^{16}$, a mathematical proof is not known
which
excludes the possibility that the actual moduli space is a quotient of
the one given here. However, as argued in \cite{asmo94}, any
such non-trivial
quotient carries a non-Hausdorff topology, in contradiction to expectations
from physics.}
$$
\MMM^\delta
\cong \OO^+(4,4+\delta;\Z) \backslash \OO^+(4,4+\delta;\R) / \SO(4)\times \OO(4+\delta),
\quad\quad
\delta\in\{0,16\}.
$$
There is only one connected component of the moduli space of type $\MMM^0$
and at least one such component of type $\MMM^{16}$. Here, $\MMM^{16}$
includes points where the SCFT description is expected to break down.
Namely, points $x$ in the Grassmannian $\wt\MMM^\delta$
of positive definite oriented four-planes in $\R^{4,4+\delta}$,
$$
\wt\MMM^\delta = \OO^+(4,4+\delta;\R) / \SO(4)\times \OO(4+\delta),
$$
are described
by their relative position with respect to the \mb{(}unique\mb{)}
even unimodular lattice $\Z^{4,4+\delta}\subset\R^{4,4+\delta}$. Then
$x\in\wt\MMM^{K3}$ is expected to correspond to an ill-defined SCFT
iff $x^\perp\subset\R^{4,4+\delta}$ contains roots, i.e.\
iff there exists an $e\in x^\perp\cap\Z^{4,4+\delta}$ with
$\langle e,e\rangle=-2$.
\etheo
Apart from the vocabulary -- using Calabi-Yau 2-folds, real four-tori, and
$K3$ surfaces -- the discussion, so far, has not made a connection to geometry.
However, Theorem \ref{modsp} is in accord with the expectation from string theory
that every SCFT associated to a Calabi-Yau 2-fold should allow for a non-linear
sigma model description on some Calabi-Yau 2-fold. Indeed, $\MMM^{tori}$
and $\MMM^{K3}$ agree with the moduli spaces of $N=(4,4)$ superconformal non-linear
sigma models on real four-tori and $K3$ surfaces, respectively, and
thanks to the high amount of supersymmetry the geometry of these moduli spaces
is not expected to receive
quantum corrections. The key to understanding this agreement can be found
in \cite{asmo94}, and it amounts to the observation that the Grassmannians in
Theorem \ref{modsp} can be modelled on the even cohomology of the respective
Calabi-Yau 2-folds:
$$
H^{even}(Y,\R)\cong\R^{4,4+\delta}\quad\mb{ for }\quad
Y=\left\{\begin{array}{ll}
A, \mb{ a real four-torus,}&\delta=0,\\[2pt]
X, \mb{ a $K3$ surface,}&\delta=16.\end{array}\right.
$$
Here and in the following $A,\,X,\,Y$ denote the diffeomorphism types
of the respective Calabi-Yau 2-folds as real four-manifolds, with all
additional structure to be introduced later. Moreover, on cohomology we
use the natural scalar product $\langle\cdot,\cdot\rangle$ induced by the
intersection form:
$$
\fa\alpha,\beta\in H^*(Y,\R):\quad\quad
\langle\alpha,\beta\rangle = \int_Y \alpha\wedge\beta.
$$
With this key in hands one can interpret the identification of the
spaces $\MMM^\delta$ of Theorem \ref{modsp} with spaces of superconformal
non-linear sigma model data on Calabi-Yau 2-folds as a generalization
of the following Torelli theorem for Calabi-Yau 2-folds:
\btheo[\cite{ku77,to80,lo81,si81,na83}]{Torelli}
Complex structures on a Calabi-Yau 2-fold $Y$ are in $1\colon1$
correspondence with positive definite oriented two-planes
$\Omega\subset H^2(Y,\R)\cong\R^{3,3+\delta}$ with $\delta=0$
for a real four-torus $Y=A$ and $\delta=16$ for a $K3$ surface
$Y=X$.
\etheo
Any orthonormal basis of a positive definite oriented
two-plane $\Omega\subset H^2(Y,\R)$ can be interpreted as giving
the real and the imaginary part
of a holomorphic volume form
$\wt\Omega\in H^{2,0}(Y,\C)$, thus explaining how $\Omega$ can
encode
a complex structure. In other words,
$\Omega\otimes\C\subset H^2(Y,\C)$ is the orthogonal complement
of the kernel of the period map. Any two-plane $\Omega\subset H^2(Y,\R)$
is specified by its relative position with respect to the even
unimodular lattice $H^2(Y,\Z)\subset H^2(Y,\R)$.
Appropriately decomposing the four-planes featuring
Theorem \ref{modsp} into pairs of perpendicular two-planes now yields
so-called refined geometric interpretations for them:
\bdefi{refinedgeom}
Given a Calabi-Yau 2-fold $Y$, let $x\subset H^{even}(Y,\R)$ denote
a positive definite oriented four-plane which according to Theorem
\mb{\ref{modsp}} specifies a SCFT on $Y$. A \textsc{refined
geometric interpretation} of this SCFT is a choice of null vectors
$\ups^0,\,\ups\in H^{even}(Y,\Z)$ along with a decomposition of $x$
into two perpendicular oriented two-planes, $x=\Omega\perp\mho$, such that
$$
(1)\quad
\langle\ups^0,\ups^0\rangle=\langle\ups,\ups\rangle=0,\quad
\langle\ups^0,\ups\rangle=1,\quad\quad\quad\mb{ and }\quad\quad\quad
(2)\quad\Omega\perp\ups^0,\ups.
$$
\edefi
Following \cite{asmo94}, a refined geometric
interpretation of a SCFT $x$ on $Y$ indeed assigns geometric data to $x$,
in fact precisely the data needed to specify a superconformal non-linear
sigma model on the complex Calabi-Yau 2-fold $Y$:
\blede{refgeom}
For a Calabi-Yau 2-fold $Y$, let $x\subset H^{even}(Y,\R)$
denote a positive definite oriented four-plane with refined
geometric interpretation $\ups^0,\,\ups\in H^{even}(Y,\Z)$,
$x=\Omega\perp\mho$ as in Definition \mb{\ref{refinedgeom}}.
Then $\ups^0,\,\ups$ are naturally interpreted as generators
of $H^0(Y,\Z)$ and $H^4(Y,\Z)$, respectively, and
one finds $\omega,\,B\in H^{even}(Y,\R)$ and $V\in\R$ such that
\begin{eqnarray*}
\mho&=&\spann_\R\left( \omega-\langle\omega,B\rangle\ups,\quad
\xi_4=\ups^0+B+\left(V-{1\over2}\langle B,B\rangle\right)\ups\right)\\
&&\mb{with }\;\omega,\,B\in H^2(Y,\R):=H^{even}(Y,\R)\cap(\ups^0)^\perp\cap(\ups)^\perp,\quad
V\in\R^+,\quad\langle\omega,\omega\rangle\in\R^+.
\end{eqnarray*}
While for every refined geometric interpretation $B$ and $V$ are uniquely defined,
$\omega$ is unique only up to scaling.
This allows to read from a refined geometric interpretation
the data $(\Omega,\omega,V,B)$ with natural interpretations in terms of
a complex structure $\Omega$ on $Y$, a K\"ahler class $\omega$ on $Y$ up to scaling,
a volume $V\in\R^+$, and a B-field $B\in H^2(Y,\R)$.

By abuse of language
I also call the data $(\Omega,\omega,V,B)$
a \textsc{refined geometric interpretation} of a given SCFT
$x\in\wt\MMM^\delta$.
$\mho$ or equivalently the data $(\omega,V,B)$ will be referred to as
\textsc{complexified K\"ahler structure}, and the class of $\omega$
will be called \textsc{normalized K\"ahler class}.
\elede
The statement of the Lemma is a consequence of the Torelli Theorem \ref{Torelli}
together with a bit of linear algebra using
$\langle\omega-\langle\omega,B\rangle\ups,\xi_4\rangle=0$ and
$\langle\xi_4,\xi_4\rangle=2V$.
For toroidal SCFTs one checks by
direct calculation
that the map from non-linear sigma model data $(\Omega,\omega,V,B)$
to $\MMM^{tori}$ encoded in Lemma/Definition \ref{refgeom}
preserves the respective natural metrics, given by the Zamolodchikov metric
on  $\wt\MMM^{tori}$. In \cite{asmo94} the same is claimed for
$\MMM^{K3}$.

Recall the comments made after Definition \ref{CY2scft} concerning
$N=(4,4)$ versus $N=(2,2)$ supersymmetry for SCFTs associated to Calabi-Yau
2-folds $Y$. According to Theorem \ref{modsp} every positive
definite oriented four-plane $x\subset H^{even}(Y,\R)$ specifies an $N=(4,4)$ SCFT
with central charges $c=\qu c=6$ without particular choice of an
$N=(2,2)$ subalgebra in the $N=(4,4)$ superconformal algebra. There is
an $\Sphere^2/\Z_2$ of such subalgebras both on the left and on the right,
specified by the choice of an unoriented Cartan torus $\fu(1)\subset\fs\fu(2)$
within the $N=4$ superconformal algebra on each side.
For later convenience, see \req{twoplanes}, my conventions
differ from \cite{asmo94,hu03} in that I do not impose an orientation on $\fu(1)$,
hence the division of
$\Sphere^2$ by $\Z_2$. The connected components of
the space of all $N=(2,2)$ superconformal field theories associated to
Calabi-Yau 2-folds then fiber over our spaces $\MMM^\delta$ with fibers
$\Sphere^2\times\Sphere^2/\Z_2^2$.

In \cite[Section 1]{nawe00} we have given an interpretation
of the four-plane $x$ in terms of the action of the $\fs\fu(2)\oplus\fs\fu(2)=\fs\fo(4)$
subalgebra of the $N=(4,4)$ superconformal algebra on the space of massless
fields in the respective SCFT.
Since
\begin{equation}\label{twoplanes}
\OO^+(2,2;\Z)\backslash \OO^+(2,2;\R) / \SO(2)\times \OO(2)
\quad\cong\quad
\left( \PSL_2(\Z) \backslash \H \;\times\; \PSL_2(\Z) \backslash\H\right) \;/ \;\Z_2^2,
\end{equation}
from this discussion one finds that the choice of an $N=(2,2)$
subalgebra in the $N=(4,4)$ superconformal algebra
amounts to a choice of decomposition
$x=\Omega\perp\mho$ of $x$ into two perpendicular two-planes, up to a choice of
their ordering and their individual orientations. In other words,
generators of $\Z_2^2$ in \req{twoplanes} act by interchanging
$\Omega$ and $\mho$ and by simultaneously reversing their orientations,
respectively. A decomposition $x=\Omega\perp\mho$ into oriented two-planes
with choice of ordering amounts to a choice of an $N=(2,2)$ subalgebra
of the $N=(4,4)$ superconformal algebra together with generators of its
$\fu(1)\oplus\fu(1)$ subalgebra. In \cite{hu03} the resulting ordered pairs
$(\Omega,\mho)$ are called \textsc{generalized $K3$ structures}, and the
picture from \cite{asmo94} drawn above  is confirmed and identified
with Hitchin's notion of generalized Calabi-Yau manifolds \cite{hi03}
in the case of Calabi-Yau 2-folds.

When working with a fixed grading
$H^{even}(Y,\R)=H^0(Y,\R)\oplus H^2(Y,\R)\oplus H^4(Y,\R)$
which amounts to the choice of two null vectors
$\ups^0,\,\ups\in\H^{even}(Y,\Z)$ as generators of
$H^0(Y,\Z)$ and $H^4(Y,\Z)$ as in Definition \ref{refinedgeom}
and subject to condition $(1)$ in that definition, then not
every ordered pair $(\Omega,\mho)$ of perpendicular oriented
positive definite two-planes in $H^{even}(Y,\R)$ gives a refined
geometric interpretation in terms of the data of a conformal
non-linear sigma model: Condition $(2)$ of Definition \ref{refinedgeom}
which ensures $\Omega\subset H^2(Y,\R)$ is crucial to that effect.
In particular, as observed in \cite{hu03}, one could say that not every
$N=(2,2)$ SCFT associated to a Calabi-Yau 2-fold arises from a
conformal non-linear sigma model construction on a (complex)
Calabi-Yau 2-fold like this.
However, if we temporarily assume that there is just one connected component
of type $\MMM^{K3}$ in the moduli
space of $N=(4,4)$ SCFTs associated to Calabi-Yau 2-folds,
then each $N=(4,4)$ SCFT arises from  a
non-linear sigma model construction: The relevant geometric data
only involve the choice of a hyperk\"ahler structure,
a volume, and a B-field, not the explicit choice of a complex structure.
This serves as justification for my Definition \ref{CY2scft} which
insists on extended $N=(4,4)$ supersymmetry. Huybrechts' observation
amounts to the fact that in such a non-linear sigma model construction,
not every choice of $N=(2,2)$ subalgebra can be interpreted as the choice
of a complex structure on the Calabi-Yau 2-fold. However, given
$x=\Omega\perp\mho$, pairs $\ups^0,\,\ups\in H^{even}(Y,\Z)$ with
(1) and (2) in Definition \ref{refinedgeom} can always be found.
In other words, as long as the grading of $H^{even}(Y,\R)$ is not fixed
a priori, indeed every pair $(\Omega,\mho)$ specifying an $N=(2,2)$ SCFT
can be interpreted in terms of non-linear sigma model data.
\eject
\subsection{The main protagonists}\label{protagonists}
Recall the definition of SCFTs associated to elliptic curves,
Definition \ref{toroidal}. One finds that the (fermionic)
tensor product of any two such theories is a SCFT associated
to a real four-torus according to Definition \ref{k34tori}.
Moreover, all known geometric orbifold constructions of $K3$
surfaces from real four-tori can be extended to constructions
in SCFT, producing SCFTs associated to $K3$ as orbifolds of
SCFTs associated to four-tori, see e.g.\ \cite{eoty89}.
As main protagonists of the present work I introduce
two families of SCFTs associated to Calabi-Yau 2-folds in this section.
The first, denoted
$\TTT_{\alpha,\beta,\beta^\prime,\gamma}$, is
associated to a family of real four-tori, and each theory is obtained
as a tensor product of theories associated to elliptic
curves. The second, denoted
$\CCC_{\alpha,\beta,\beta^\prime,\gamma}$, is
associated to a family of $K3$ surfaces, and each theory is obtained
as an orbifold of the corresponding $\TTT_{\alpha,\beta,\beta^\prime,\gamma}$.
\bdefi{protator}
Denote by $\TTT_{\alpha,\beta,\beta^\prime,\gamma}$ with
$$
\alpha,\,\beta,\,\beta^\prime,\,\gamma\in\R\quad\mb{ such that }\quad
\alpha,\gamma>0
\quad\mb{ and }\quad
\Delta:=\beta^2-4\alpha\gamma<0
$$
the \mb{(}fermionic\mb{)} tensor product
of the two SCFTs associated to elliptic curves with moduli $(\tau_k,\rho_k)$ given by
$$
\tau_1=\tau_2=i, \quad
\rho_1={-\beta+\sqrt\Delta\over2\alpha},\quad
\rho_2={-\beta^\prime+\sqrt\Delta\over2}\;.
$$
\edefi
By the discussion in Section \ref{CY1modsp}
the factor theories of $\TTT_{\alpha,\beta,\beta^\prime,\gamma}$ are
SCFTs on elliptic curves with square fundamental
cells ($\tau_1=\tau_2=i$), with radii $R_1,\,R_2$ such that
$R_1^2={\sqrt{-\Delta}\over2\alpha},\,
R_2^2 = {\sqrt{-\Delta}\over2}$, and with B-fields given by the
$-{\beta\over2\alpha}$ and the $-{\beta^\prime\over2}$-fold of a
generator of $H^2(E_{\tau_k},\Z)$,
respectively. Hence $\TTT_{\alpha,\beta,\beta^\prime,\gamma}$
is a toroidal SCFT with refined geometric interpretation
on a real four-torus $A_{\alpha,\beta,\gamma}$ with the  flat metric and
complex and K\"ahler structure induced by
\begin{eqnarray}\label{Aalbeg}
A_{\alpha,\beta,\gamma}&=&\R^4/\Lambda_{\alpha,\beta,\gamma},\quad
\Lambda_{\alpha,\beta,\gamma}=R_1\Z^2\oplus R_2\Z^2,\quad
R_1^2={\sqrt{-\Delta}\over2\alpha},\;
R_2^2 = {\sqrt{-\Delta}\over2},
\\
&&z_1=x_1+ix_2,\,z_2=x_3+ix_4,\quad
\omega_\alpha \sim dx_1\wedge dx_2+\alpha\, dx_3\wedge dx_4\nonumber
\end{eqnarray}
with respect to standard Cartesian coordinates
$x_1,\ldots,x_4$ on $\R^4$.
I use the standard basis $e_1,\ldots,e_4$ of $\R^4$
to introduce generators
$\lambda_1=R_1e_1,\,\lambda_2=R_1e_2,\,
\lambda_3=R_2e_3,\,\lambda_4=R_2e_4$ of $\Lambda_{\alpha,\beta,\gamma}$ and
view the vectors forming the dual basis $\lambda_1^\ast,\ldots,\lambda_4^\ast$
as generators of $H^1(A,\Z)$. Hence
$H^2(A,\Z)$ is generated by
\begin{equation}\label{rename}
\ups_1^0= \lambda^\ast_1\wedge \lambda^\ast_3,\;
\ups_1= \lambda^\ast_4\wedge \lambda^\ast_2,\;
\ups_2^0= \lambda^\ast_2\wedge \lambda^\ast_3,\;
\ups_2= \lambda^\ast_1\wedge \lambda^\ast_4,\;
\ups_3^0= \lambda^\ast_1\wedge \lambda^\ast_2,\;
\ups_3= \lambda^\ast_3\wedge \lambda^\ast_4,
\end{equation}
where the $\ups_k^0,\,\ups_l$ obey
$\langle\ups_k^0,\ups_l^0\rangle=0,\,
\langle\ups_k,\ups_l\rangle=0,\,
\langle\ups_k^0,\ups_l\rangle=\delta_{kl}$.
For later convenience let me give the location of each
$\TTT_{\alpha,\beta,\beta^\prime,\gamma}$ in the moduli space
$\MMM^{tori}$ of SCFTs associated to real four-tori, along with
two refined geometric interpretations:
\bprop{protatorgeom}
The SCFT $\TTT_{\alpha,\beta,\beta^\prime,\gamma}$ of Definition
\mb{\ref{protator}} has a refined geometric interpretation
$(\Omega_A^0, \omega_\alpha^A, V_{\alpha,\beta,\gamma}^A,$
$B_{\alpha,\beta,\beta^\prime}^A)$ given by
\begin{eqnarray*}
\Omega_A^0 &=& \spann_\R(\ups_1^0+\ups_1,\ups_2^0+\ups_2),\\
\omega_\alpha^A &=& \ups_3^0+\alpha\ups_3,\quad\quad
V^A_{\alpha,\beta,\gamma}=R_1^2 R_2^2 =\gamma - {\beta^2\over4\alpha},\quad\quad
B_{\alpha,\beta,\beta^\prime}^A
=-{\beta\over2\alpha}\ups_3^0-{\beta^\prime\over2}\ups_3.
\end{eqnarray*}
Within $\MMM^{tori}$ this theory is given by the four-plane
$x^A_{\alpha,\beta,\beta^\prime,\gamma}\subset H^{even}(A,\R)$ with
\begin{eqnarray*}
x^A_{\alpha,\beta,\beta^\prime,\gamma}
=\spann_\R\left(\xi_1,\xi_2,\xi_3,\wt\xi_4\right),
&\hspace*{-1em}&\xi_1=\ups_1^0+\ups_1,\;\; \xi_2\;=\;\ups_2^0+\ups_2,\\
&\hspace*{-1em}&\xi_3=\ups_3^0+\alpha\ups_3+{\beta+\beta^\prime\over2}\ups_4,\;\;
\wt\xi_4= \ups_4^0+{\beta-\beta^\prime\over2}\ups_3+\gamma \ups_4,
\end{eqnarray*}
where $\ups_4^0,\,\ups_4$ denote generators of $H^0(A,\Z)$ and
$H^4(A,\Z)$, respectively.

$\TTT_{\alpha,\beta,\beta^\prime,\gamma}$ has a ``mirror dual"
refined geometric interpretation $(\Omega^A_{\alpha,\beta,\beta^\prime,\gamma},
\omega_A,V_A,B_A)$ with complex structure given by the product of two
elliptic curves $E_{\rho_1}\times E_{\rho_2}$ at the moduli $\rho_1,\,\rho_2$
from Definition \mb{\ref{protator}},
with normalized K\"ahler form
$\omega_A= -{i\over2} (dz_1\wedge d\qu z_1+dz_2\wedge d\qu z_2)$ with respect
to complex coordinates $z_j$ of $E_{\rho_j}$
such that $-{i\over2} dz_j\wedge d\qu z_j$
generates $H^2(E_{\rho_j},\Z)$, and with
volume $V_A=1$ and B-field $B_A=0$.
\eprop
\bpr
That $\TTT_{\alpha,\beta,\beta^\prime,\gamma}$ has
refined geometric interpretation $(\Omega_A^0, \omega_\alpha^A, V_{\alpha,\beta,\gamma}^A,
B_{\alpha,\beta,\beta^\prime}^A)$
follows directly from the
construction of $\TTT_{\alpha,\beta,\beta^\prime,\gamma}$ and the given
geometric interpretation of its factor theories. From Definition
\ref{refgeom} we see that
$x^A_{\alpha,\beta,\beta^\prime,\gamma}$ is generated by
$\xi_1,\xi_2$ and
$$
\xi_3=\omega_\alpha^A-\langle\omega_\alpha^A,B_{\alpha,\beta,\beta^\prime}^A\rangle\ups_4,
\quad
\xi_4=\ups_4^0+B_{\alpha,\beta,\beta^\prime}^A
+\left(V_{\alpha,\beta,\gamma}
-{1\over2}\langle B_{\alpha,\beta,\beta^\prime}^A,B_{\alpha,\beta,\beta^\prime}^A\rangle
\right)\ups_4.
$$
One checks that $\xi_3$ has the claimed form and $\wt\xi_4=\xi_4+{\beta\over2\alpha}\xi_3$.

A ``mirror dual" geometric interpretation of $x^A_{\alpha,\beta,\beta^\prime,\gamma}$
is obtained by using $\ups^0=\ups_2^0,\,\ups=\ups_2$
as generators of $H^0(A,\Z)$ and $H^4(A,\Z)$ and letting
\beq{period}
\Omega^A_{\alpha,\beta,\beta^\prime,\gamma}
:=\spann_\R(\xi_3,\wt\xi_4)
=\spann_\R\left(\ups_3^0+\alpha\ups_3+{\beta+\beta^\prime\over2}\ups_4,\,
\ups_4^0+{\beta-\beta^\prime\over2}\ups_3+\gamma \ups_4\right)
\eeq
specify the complex structure. It immediately follows that $V_A=1,\,
B_A=0$ and $\omega_A=\ups_1^0+\ups_1$  in this geometric interpretation,
in accord with the claim. It remains to show that
$\Omega^A_{\alpha,\beta,\beta^\prime,\gamma}$ gives the claimed
complex structure. Although this should follow from consistency with the
explanations given in Section \ref{CY1modsp}, as a reality check I
include the detailed argument:

I show that the orthogonal complement of the
kernel of the period map in $H^2(A,\R)$ has generators which with respect
to standard generators $\ups_3^0,\,\ups_3,\,\ups^0_4,\,\ups_4$ of two copies
of a hyperbolic lattice $H$ have precisely the form given in \req{period}.
Because by \cite[Theorem 1.14.4]{ni80} the embedding
$H\oplus H\hookrightarrow H^2(A,\Z)$ is unique up to automorphisms of $H^2(A,\Z)$,
\req{period} actually determines the relative position of
$\Omega^A_{\alpha,\beta,\beta^\prime,\gamma}$ with respect to
$H^2(A,\Z)$, such that
the claim then follows from the  Torelli Theorem \ref{Torelli}.

Since $\rho_k\in\PSL_2(\Z)\backslash\H$ and
$(\tau_k,\rho_k)\sim(-\qu\tau_k,-\qu\rho_k)$
where $\tau_k=-\qu\tau_k$ for $\tau_k=i$ by Proposition \ref{ellims},
I can work with the complex two-torus $A$
obtained as product of two elliptic curves with
moduli
$$
\sigma_1={1\over\qu\rho_1}={-\beta+\sqrt\Delta\over2\gamma},\;\;
\sigma_2=-\qu\rho_2={\beta^\prime+\sqrt\Delta\over2}, \quad\quad
\Delta=\beta^2-4\alpha\gamma.
$$
Similarly to
\cite[p.\ 265 ff]{shmi74} I have $A=\C^2/L$, where
the lattice $L\subset\C^2$ is generated by
$$
l_1={1\choose0},\quad l_2={0\choose1},\quad
l_3={-\sigma_1\choose0},\quad l_4={0\choose\sigma_2}.
$$
With $m^1,\ldots,m^4$ the basis dual to the one given by the $l_k$,
$$
\ups_1^0:=m^{1}\wedge m^{3},\;\ups_1:=m^{4}\wedge m^{2},\;\ups_3^0:=m^{1}\wedge m^{2},\;
\ups_3:=m^{3}\wedge m^{4},\;
\ups_4^0:=m^{2}\wedge m^{3},\;\ups_4:=m^{1}\wedge m^{4}
$$
generate $H^2(A,\Z)$.
Then the period map is given by
$$
\sum_{i<j} \det\left( l_i l_j\right) m^{i}\wedge m^{ j}
= \ups_3^0 + \sigma_2 \ups_4 + \sigma_1 \ups_4^0 - \sigma_1\sigma_2 \ups_3.
$$
The kernel of the period map hence is generated by
$$
\ups_1^0,\quad\ups_1,\quad\ups_3^0-{\beta-\beta^\prime\over2}\ups_4-\alpha \ups_3,\quad
\ups_4^0-\gamma \ups_4 - {\beta+\beta^\prime\over2} \ups_3,
$$
and the orthogonal complement of the kernel of the period map indeed is
precisely the two-plane $\Omega^A_{\alpha,\beta,\beta^\prime,\gamma}$
of \req{period}.
\epr
By the above, $\TTT_{\alpha,\beta,\beta^\prime,\gamma}$
has a geometric interpretation on the torus
$A_{\alpha,\beta,\gamma}$ of \req{Aalbeg}
which in terms of standard Cartesian coordinates $x_1,\ldots,x_4$
of $\R^4$ enjoys the symmetry
\beq{z4}
\zeta_4\colon\quad (x_1,x_2,x_3,x_4)\longmapsto (-x_2,x_1,x_4,-x_3)
\eeq
of order four. According to \req{rename}, $\zeta_4$ leaves
$\ups_1^0+\ups_1,\,\ups_2^0+\ups_2$,
$\ups_3^0$, and $\ups_3$ invariant and thus induces a map on
$H^{even}(A,\R)$ which by the description in Proposition \ref{protatorgeom}
leaves invariant
the four-plane $x^A_{\alpha,\beta,\beta^\prime,\gamma}\subset H^{even}(A,\R)$
giving $\TTT_{\alpha,\beta,\beta^\prime,\gamma}$.
This means that $\zeta_4$ induces an automorphism of
$\TTT_{\alpha,\beta,\beta^\prime,\gamma}$, so that
a $\Z_4$-orbifold of this theory can be constructed:
\bdefi{protak3}
For $\alpha,\,\beta,\,\beta^\prime,\,\gamma\in\R$ as in Definition
\mb{\ref{protator}} let $\CCC_{\alpha,\beta,\beta^\prime,\gamma}$
denote the $\Z_4$-orbifold  of the SCFT
$\TTT_{\alpha,\beta,\beta^\prime,\gamma}$, where $\Z_4$ is
generated by the action induced by $\zeta_4$.
\edefi
The theories $\CCC_{\alpha,\beta,\beta^\prime,\gamma}$
of Definition \ref{protak3} are well understood and can be constructed
explicitly without difficulty. E.g.\ by the results of \cite{eoty89} each of these theories
is a SCFT associated to $K3$ according to Definition \ref{k34tori}.
The results of \cite{nawe00,we00} allow me to describe the four-plane
$x_{\alpha,\beta,\beta^\prime,\gamma}\subset H^{even}(X,\R)$
which specifies this theory in terms of the lattice $H^{even}(X,\Z)$,
using a refined geometric
interpretation on the $\Z_4$-orbifold limit $X_{\alpha,\beta,\gamma}$
of $K3$ obtained by minimally
resolving the singularities of $A_{\alpha,\beta,\gamma}/\Z_4$,
where again $\Z_4$ is the group generated by $\zeta_4$,
and $A_{\alpha,\beta,\gamma}$ carries the K\"ahler and complex
structure induced by \req{Aalbeg}.

Let me first
introduce some notation which I need in order to describe this
refined geometric interpretation:
Let $\pi\colon A_{\alpha,\beta,\gamma}\longrightarrow X_{\alpha,\beta,\gamma}$
denote the rational map obtained from the orbifold procedure, and
$\pi_\ast\colon H^2(A,\R)^{\Z_4}\longrightarrow H^2(X,\R)$ the induced
map on cohomology. Recall from \cite{be88,nawe00,we00} the description of
the lattice $H^{2}(X,\Z)$ in terms of $\pi_\ast H^2(A,\Z)^{\Z_4}$ and the
exceptional divisors coming from the resolution of
$A_{\alpha,\beta,\gamma}/\Z_4$:
Consider the action of
the subgroup $\Z_2$ of $\Z_4$ on $A_{\alpha,\beta,\gamma}$. It
has $16$ fixed points, labeled
by an affine $\F_2^4$ over the field $\F_2$ with two elements $0,\,1$,
where $i=(i_1,\ldots,i_4)\in\F_2^4$ with
$i_k\in\{0,1\}$ corresponds to the fixed point at
$(x_1,x_2,x_3,x_4)={1\over2}\sum_k i_k\lambda_k$.
For our $\Z_4$-orbifold we can use
the same notation, where four of the fixed points listed in $\F_2^4$, namely
those in $I^{(4)}:=\{(0000),\,(1100),\,(0011),\,(1111)\}$, are
fixed under $\Z_4$. The remaining twelve fixed points are
paired  to six fixed points under the $\Z_2$ subgroup of $\Z_4$, where
$(i_1,i_2,i_3,i_4)\sim(i_2,i_1,i_4,i_3)$.
We denote by $I^{(2)}$ the set of these six fixed points, i.e.\
$I^{(2)}=(\F_2^4-I^{(4)})/\sim$.

From the resolution of singularities of type $A_1$ at the
fixed points with labels $i\in I^{(2)}$ we obtain six lattice
vectors $\wh E_i\in H^2(X,\Z)$, $\langle\wh E_i,\wh E_i\rangle=-2$.
On the other hand, each
$i\in I^{(4)}$ gives a singularity of type $A_3$, yielding three
lattice vectors $\wh E_i^{(k)},\,k\in\{1,2,3\}$,
each. Their intersection matrix
is the negative of the Cartan matrix of the Lie algebra
$A_3$, while all pairwise scalar products between vectors associated
to different fixed points vanish. Moreover, all the $\wh E_i,\,\wh E_j^{(k)}$
are perpendicular to $\pi_\ast H^2(A,\R)^{\Z_4}$.
\bprop{protak3geom}
With $\alpha,\,\beta,\,\beta^\prime,\,\gamma\in\R$ and notations
as above and in particular as in Proposition \mb{\ref{protatorgeom}}
consider the $\Z_4$-orbifold SCFTs $\CCC_{\alpha,\beta,\beta^\prime,\gamma}$
of Definition \mb{\ref{protak3}}.

$\CCC_{\alpha,\beta,\beta^\prime,\gamma}$ has a refined geometric interpretation
on $X_{\alpha,\beta,\gamma}=\smash{\wt{A_{\alpha,\beta,\gamma}/\Z_4}}$
given by $(\Omega_X^0,\omega_\alpha,$
$V_{\alpha,\beta,\gamma}, B_{\alpha,\beta,\beta^\prime})$ as follows:
With $\wh\ups^0,\,\wh\ups$ generators of $H^0(X,\Z)$ and $H^4(X,\Z)$
such that $\langle\wh\ups^0,\wh\ups\rangle=1$, the lattice
$\pi_\ast H^2(A,\R)^{\Z_4}\cap H^2(X,\Z)$ has generators
$\wh\ups_3^0=\pi_\ast\ups_3^0,\;\wh\ups_3=\pi_\ast\ups_3,\;
\wt\Omega_1,$ $\wt\Omega_2$ such that
$$
\langle\wh\ups_3^0,\wh\ups_3^0\rangle=\langle\wh\ups_3,\wh\ups_3\rangle=0,\quad
\langle\wh\ups_3^0,\wh\ups_3\rangle=4,\quad
\langle\wt\Omega_1,\wt\Omega_1\rangle=\langle\wt\Omega_2,\wt\Omega_2\rangle=2,\quad
\langle\wt\Omega_1,\wt\Omega_2\rangle=0,
$$
and
\begin{eqnarray*}
\Omega_X^0 &=& \spann_\R\left(\wt\Omega_1,\wt\Omega_2\right),\quad
\omega_\alpha = \wh\ups_3^0+\alpha\wh\ups_3,\quad
V_{\alpha,\beta,\gamma}
= {1\over4}\left(\gamma-{\beta^2\over4\alpha}\right),\\
B_{\alpha,\beta,\beta^\prime}
&=&-{\beta\over8\alpha}\wh\ups_3^0-{\beta^\prime\over8}\wh\ups_3+{1\over4}\check B_4,\;\;
\check B_4 = -\sum_{i\in I^{(2)}}\wh E_i
- \sum_{i\in I^{(4)}}\left(
{3\over2}\left( \wh E_i^{(1)} + \wh E_i^{(3)}\right) + 2\wh E_i^{(2)}\right)
\end{eqnarray*}
with primitive $\check B_4\in H^2(X,\Z)$.
The four-plane describing $\CCC_{\alpha,\beta,\beta^\prime,\gamma}$
within $\MMM^{K3}$ is
$$
x_{\alpha,\beta,\beta^\prime,\gamma}
=\spann_\R\left( \wt\Omega_1,\,\wt\Omega_2,\,
\wh\ups_3^0+\alpha\wh\ups_3 +{\beta+\beta^\prime\over2}\wh\ups, \,
4\wh\ups^0+{\beta-\beta^\prime\over2}\wh\ups_3 +\check B_4 + \left({\gamma}+4\right)\wh\ups\right).
$$
$\CCC_{\alpha,\beta,\beta^\prime,\gamma}$ has a ``dual" refined
geometric interpretation $(\Omega^X_{\alpha,\beta,\beta^\prime,\gamma},
\omega_Q,V_Q,B_Q)$ with primitive $\omega_Q\in H^2(X,\Z)$,
\begin{eqnarray*}
\Omega^X_{\alpha,\beta,\beta^\prime,\gamma}
&=&\spann_\R\left(\wh\ups_3^0+\alpha\wh\ups_3 +{\beta+\beta^\prime\over2}\wh\ups, \,
4\wh\ups^0+{\beta-\beta^\prime\over2}\wh\ups_3 +\check B_4 + \left({\gamma}+4\right)\wh\ups
\right),\\
&&\langle\omega_Q,\omega_Q\rangle=4,\quad V_Q={1\over2},\quad
B_Q=-{1\over2}\omega_Q.
\end{eqnarray*}
The vectors $\wh\ups_3^0,\,\wh\ups_3,\,\wh\ups^0,\,\wh\ups$ generate a primitive sublattice
of $H^2(X,\Z)$ with quadratic form
$$
\left(\begin{array}{cc}4h&0\\0&h\end{array}\right),\quad
\mb{ where }\quad
h=\left(\begin{array}{cc}0&1\\1&0\end{array}\right).
$$
\eprop
\bpr
The claims about the lattice $\pi_\ast H^2(A,\R)^{\Z_4}\cap H^2(X,\Z)$
follow from \cite{shin77}, where $\Omega_X^0$ gives the complex structure
of $X_{\alpha,\beta,\gamma}$, see also \cite{nawe00,we00}.
Moreover, in \cite{nawe00}, \cite[Theorem 3.3]{we00} it is proved that
$\CCC_{\alpha,\beta,\beta^\prime,\gamma}$ has a geometric interpretation
$(\Omega_X^0,\pi_\ast\omega_\alpha^A,{1\over4} V_{\alpha,\beta,\gamma}^A$,
${1\over 4}\pi_\ast B^A_{\alpha,\beta,\beta^\prime}+{1\over4}\check B_4)$
with $\omega_\alpha^A$,
$V_{\alpha,\beta,\gamma}^A$,
and $B^A_{\alpha,\beta,\beta^\prime}$ as in Proposition \ref{protatorgeom},
from which the claims about the first refined geometric
interpretation are immediate. One checks $\langle\check B_4,\check B_4\rangle=-32$.
Using Definition \ref{refgeom}
for $x_{\alpha,\beta,\beta^\prime,\gamma}$ one thus finds generators
$$
\wt\Omega_1,\quad\wt\Omega_2,\quad
\wh\ups_3^0+\alpha\wh\ups_3 +{\beta+\beta^\prime\over2}\wh\ups,\quad
\wh\ups^0-{\beta\over8\alpha}\wh\ups_3^0
-{\beta^\prime\over8}\wh\ups_3
 + {1\over4}\check B_4
+ \left({\gamma\over4} - {\beta(\beta+\beta^\prime)\over16\alpha}+1\right)\wh\ups,
$$
which are seen to simplify to the form claimed.

To obtain the claimed ``dual" refined geometric interpretation,
for $i\in I^{(4)}$
let $\wh E_i:=\wh E_i^{(1)}+2\wh E_i^{(2)}+3\wh E_i^{(3)}$ and recall
from \cite[Proposition 2.1]{we00} that the lattice  $H^2(X,\Z)$
in particular contains the vectors
\begin{eqnarray*}
{1\over2}\wt\Omega_1 &-& {1\over2}\left( \wh E_{(0,0,0,0)}
+\wh E_{(1,0,0,0)}+\wh E_{(0,0,0,1)}+\wh E_{(0,1,0,1)}\right),\\
{1\over2}\wt\Omega_2 &-& {1\over2}\left( \wh E_{(0,0,0,0)}
+\wh E_{(1,0,0,0)}+\wh E_{(0,0,0,1)}+\wh E_{(1,0,0,1)}\right).
\end{eqnarray*}
Hence also
\begin{eqnarray*}
\ups_Q&:=& {1\over2}\left(\wt\Omega_1 -\wt\Omega_2\right)
 - {1\over2}\left(\wh E_{(0,1,0,1)}-\wh E_{(1,0,0,1)}\right),\\
\ups_Q^0&:=&{1\over2}\left(\wt\Omega_1 +\wt\Omega_2\right)
 + {1\over2}\left(\wh E_{(0,1,0,1)}-\wh E_{(1,0,0,1)}\right)
\end{eqnarray*}
are lattice vectors, and one checks that they are null vectors
obeying $\langle\ups_Q^0,\ups_Q\rangle=1$. To determine the
corresponding refined geometric interpretation, one first finds
\begin{eqnarray*}
\wt\Sigma_{\alpha,\beta,\beta^\prime,\gamma}
&=&x_{\alpha,\beta,\beta^\prime,\gamma}\cap(\ups_Q)^\perp\\
&=&\spann_\R\left(\wt\Omega_1 +\wt\Omega_2,\,
\wh\ups_3^0+\alpha\wh\ups_3 +{\beta+\beta^\prime\over2}\wh\ups, \,
4\wh\ups^0+{\beta-\beta^\prime\over2}\wh\ups_3 +\check B_4 + \left({\gamma}+4\right)\wh\ups
\right).
\end{eqnarray*}
Projection onto $H^2(X,\R)=H^{even}(X,\R)\cap(\ups_Q)^\perp\cap(\ups_Q^0)^\perp$
then shows that we can interpret
$$
\Omega_{\alpha,\beta,\beta^\prime,\gamma}^X
:=\spann\left(\wh\ups_3^0+\alpha\wh\ups_3 +{\beta+\beta^\prime\over2}\wh\ups, \,
4\wh\ups^0+{\beta-\beta^\prime\over2}\wh\ups_3 +\check B_4 + \left({\gamma}+4\right)\wh\ups
\right)\subset H^2(X,\R)
$$
as specifying the complex structure of this geometric interpretation,
while
$$
\omega_Q:= 2\wt\Omega_2 + \wh E_{(0,1,0,1)}-\wh E_{(1,0,0,1)}
$$
gives the normalized K\"ahler form. The latter is indeed a primitive
lattice vector with $\langle\omega_Q,\omega_Q\rangle=4$.
Moreover, $\wh\ups_3^0,\,\wh\ups_3,\,\wh\ups^0,\,\wh\ups\in H^2(X,\Z)$,
such that the claim about the lattice that these vectors generate is immediate
from the above. Next notice that
$\xi_4:={1\over2}\left(\wt\Omega_1 -\wt\Omega_2\right)$ obeys
$$
x_{\alpha,\beta,\beta^\prime,\gamma}
=\wt\Sigma_{\alpha,\beta,\beta^\prime,\gamma}\perp\langle\xi_4\rangle, \quad\quad
\langle\xi_4,\ups_Q\rangle=1,
$$
such that in this geometric interpretation our $K3$ surface has volume
$$
V_Q = {1\over2}\langle\xi_4,\xi_4\rangle = {1\over2}.
$$
Finally,
$$
B_Q:=\xi_4-\ups_Q^0
= -{1\over2}\left(2\wt\Omega_2 + \wh E_{(0,1,0,1)}-\wh E_{(1,0,0,1)}\right)
=-{1\over2}\omega_Q
$$
is perpendicular to both $\ups_Q^0,\,\ups_Q$ and hence gives
the B-field in this geometric interpretation.
\epr
\section{The main claim and its geometric background}\label{mainclaim}
I have now provided all the necessary background material to present
the main result of this paper. I do so in Section \ref{mainresult}:
The family $\CCC_{\alpha,\beta,\beta^\prime,\gamma}$ of SCFTs on
$K3$, which is obtained by means of an
orbifold construction, is given
a geometric interpretation on a smooth family of smooth algebraic
$K3$ surfaces. In this geometric interpretation, $\alpha,\,\beta,\,
\beta^\prime,\,\gamma$ give   complex structure
parameters. This family of SCFTs on $K3$ hence is well under control both from
a conformal field theorist's and from an algebraic geometer's point of
view. As such, it is a first example of its kind.

Section \ref{mainresult} also contains
a first part of the proof of this claim.
A geometric explanation arises by extending a construction due to Inose.
I therefore devote
Section \ref{Inose} to a summary of Inose's work \cite{in76}.
Section \ref{extension} explains how my main result extends Inose's
construction, using a specific (crude) version of mirror symmetry. As an implication,
which allows for a proof purely within geometry,
I show how the natural metric on the Fermat quartic, i.e.\ the
K\"ahler-Einstein metric in the class of the Fubini-Study metric
on $\CP^3$, is related to an orbifold limit of a metric on a Kummer surface. This
description makes the former metric accessible to numerical investigations
following \cite{hewi05}. I therefore find it interesting in its
own right and include the discussion in Section \ref{extension}.
\subsection{The main result}\label{mainresult}
As explained in Section \ref{CY2defmod}, the moduli space of SCFTs
associated to Calabi-Yau 2-folds is known,
at least to a high degree of plausibility (the open problems were pointed
out there).
Section \ref{protagonists}
was devoted to the discussion of two families of examples, one
in each connected component of the moduli space associated to real
four-tori and $K3$ surfaces, respectively. However,
further examples of such SCFTs where explicit constructions
are known are severely restricted:
While all SCFTs associated to real four-tori are known,
along with their locations within $\MMM^{tori}$ \cite{cent85,na86},
the only known constructions of SCFTs associated to $K3$ are orbifold
constructions and the Gepner construction \cite{ge87,ge88}.
For the former, the locations within the moduli space have been
worked out in \cite{nawe00,we00}. The latter give about $50$
discrete points in the moduli space known as Gepner or Gepner type
models, and for some examples the locations
have been determined in \cite{nawe00}. However, no direct construction
for SCFTs associated to smooth $K3$ surfaces is known, let alone for
a family of such surfaces. This is why I find the following
result surprising:
\bres{main}
For $\alpha,\,\beta,\,\beta^\prime,\,\gamma\in\R$ as in Definition
\mb{\ref{protator}}, the SCFT $\CCC_{\alpha,\beta,\beta^\prime,\gamma}$
of Definition \mb{\ref{protak3}} has a refined geometric interpretation
on the smooth quartic $K3$ surface
$$
X(f_1,f_2)\colon\quad f_1(x_0,x_1)+f_2(x_2,x_3) = 0 \quad\mb{ in }\quad \CP^3,
$$
where $f_1,\,f_2$ are homogeneous quartic polynomials such that the elliptic
curves
$$
E_{f_k}\colon\quad y_0^2=f_k(y_1,y_2)\quad \mb{ in }\quad \CP_{2,1,1}
$$
have periods $\rho_1,\,\rho_2\in\H$ with
$$
\rho_1={-\beta+\sqrt\Delta\over2\alpha},\quad
\rho_2={-\beta^\prime+\sqrt\Delta\over2},\quad\quad
\Delta=\beta^2-4\alpha\gamma
$$
as in Definition \mb{\ref{protator}},
thus defining an Abelian variety
$$
A(f_1,f_2):=E_{f_1}\times E_{f_2}.
$$
More precisely this refined geometric interpretation carries the natural
complex and K\"ahler structure induced by $X(f_1,f_2)\hookrightarrow\CP^3$,
i.e.\ the normalized K\"ahler class is the class $\omega_{FS}$ induced
by the Fubini-Study metric on $\CP^3$, and the volume and  B-field
are $V_{FS}={1\over2}$, $B_{FS}=-{1\over2}\omega_{FS}$.
\eres
Section \ref{proof} is devoted to the proof of this statement. However,
at this stage I can already prove the following weaker result
which also gives some insight into the geometric origin of the
main claim:
\blem{complexclaim}
For $\alpha,\,\beta,\,\beta^\prime,\,\gamma\in\R$ as in Definition
\mb{\ref{protator}} consider the SCFT $\CCC_{\alpha,\beta,\beta^\prime,\gamma}$
of Definition \mb{\ref{protak3}} and its refined geometric interpretation
$(\Omega^X_{\alpha,\beta,\beta^\prime,\gamma},
\omega_Q,V_Q,B_Q)$
of Proposition \mb{\ref{protak3geom}}. Then the complex structure
$\Omega_{\alpha,\beta,\beta^\prime,\gamma}^X$ agrees with the one
of $X(f_1,f_2)\subset\CP^3$ with $X(f_1,f_2)$ as in Result
\mb{\ref{main}}.
In fact this is true for any refined geometric interpretation
$(\Omega,\omega,V,B)$ of $\CCC_{\alpha,\beta,\beta^\prime,\gamma}$
with $\Omega=\Omega^X_{\alpha,\beta,\beta^\prime,\gamma}$.
\elem
\bpr
By Proposition \ref{protak3geom},
$$
\Omega^X_{\alpha,\beta,\beta^\prime,\gamma}
=\spann_\R\left(\wh\ups_3^0+\alpha\wh\ups_3 +{\beta+\beta^\prime\over2}\wh\ups, \,
4\wh\ups^0+{\beta-\beta^\prime\over2}\wh\ups_3 +\check B_4 + \left({\gamma}+4\right)\wh\ups
\right),
$$
where $\wh\ups_3^0,\,\wh\ups_3,\,\wh\ups^0,\,\wh\ups$ generate a primitive sublattice
of $H^2(X,\Z)$ with signature $(2,2)$, and $\check B_4\in\H^2(X,\Z)$ is primitive with
$\langle\check B_4,\check B_4\rangle=-32$. Setting
$\wh\ups_4^0:=4\wh\ups^0+\check B_4+4\wh\ups$ and $\wh\ups_4:=\wh\ups$
we obtain
\beq{omegaform}
\Omega^X_{\alpha,\beta,\beta^\prime,\gamma}
=\spann_\R\left(\wh\ups_3^0+\alpha\wh\ups_3 +{\beta+\beta^\prime\over2}\wh\ups_4, \,
\wh\ups^0_4+{\beta-\beta^\prime\over2}\wh\ups_3 + \gamma\wh\ups_4
\right),
\eeq
where $\wh\ups_3^0,\,\wh\ups_3,\,\wh\ups^0_4,\,\wh\ups_4$ generate a primitive sublattice
$\wh\Gamma^{2,2}$ of $H^2(X,\Z)$. By the results of Proposition \ref{protak3geom}
this lattice is $\wh\Gamma^{2,2}=\Gamma^{2,2}(4)$, the sum of two hyperbolic lattices
$\Gamma^{2,2}=H\oplus H$ with quadratic form rescaled by a factor of $4$.
By \cite[Theorem 1.14.4]{ni80} (see also \cite[Corollary 2.10]{mo84}),
the embedding $\wh\Gamma^{2,2}\hookrightarrow H^2(X,\Z)$ is
unique up to automorphisms of $H^2(X,\Z)$. Hence \req{omegaform}
fixes the location of $\Omega^X_{\alpha,\beta,\beta^\prime,\gamma}$
within $H^2(X,\R)$ with respect to $H^2(X,\Z)$. By the Torelli Theorem
\ref{Torelli} this uniquely identifies the complex structure.

Similarly, I showed in Proposition \ref{protatorgeom} that the complex
structure of the Abelian variety $A(f_1,f_2)=E_{f_1}\times E_{f_2}$
is given by the two-plane $\Omega^A_{\alpha,\beta,\beta^\prime,\gamma}\subset H^2(A,\R)$
whose relative position with respect to $H^2(A,\Z)$ is specified by
$$
\Omega^A_{\alpha,\beta,\beta^\prime,\gamma}
=\spann_\R\left(\ups_3^0+\alpha\ups_3+{\beta+\beta^\prime\over2}\ups_4,\,
\ups_4^0+{\beta-\beta^\prime\over2}\ups_3+\gamma \ups_4\right),
$$
where $\ups_3^0,\,\ups_3,\,\ups^0_4,\,\ups_4$ generate a primitive sublattice
$\Gamma^{2,2}\subset H^2(X,\Z)$ with $\Gamma^{2,2}=H\oplus H$ as above.
In \cite{in76} Inose shows that $A(f_1,f_2)$ and $X(f_1,f_2)$ are isogeneous,
namely the Kummer surface constructed from $A(f_1,f_2)$ is biholomorphic
to a $\Z_2$-orbifold of $X(f_1,f_2)$. Using
\cite[Lemma 5.7]{in76} in conjunction with \cite[Appendix \S5]{pss71}
the complex structure of $X(f_1,f_2)$ is thus described by a two-plane
in $H^2(X,\R)$ which has precisely the same form as
$\Omega^A_{\alpha,\beta,\beta^\prime,\gamma}\subset H^2(A,\R)$
but with the quadratic form of $\Gamma^{2,2}$ rescaled by a factor
of $4$. Since $\wh\Gamma^{2,2}=\Gamma^{2,2}(4)$, a comparison with
\req{omegaform} completes the proof.
\epr
The above proof is in line with the main idea of \cite{we00}, where for every
geometric $G$-orbifold construction of $K3$ from a complex two-torus $A$ the
rational map $\pi\colon A\longrightarrow X=\wt{A/G}$ induced from the
orbifold construction was studied. More precisely, the induced map
$\pi_\ast\colon H^2(A,\R)^G\longrightarrow H^2(X,\R)$ was extended to the
total even cohomology $H^{even}(A,\R)^G$.
The result \cite[(4.1)]{we00} shows that the vectors
$\wh\ups^0_4,\,\wh\ups_4$ used in the proof of the above Lemma \ref{complexclaim},
see \req{omegaform},
are the images of the vectors $\ups^0,\,\ups\in H^{even}(A,\Z)$ under $\pi_\ast$ which
in the geometric interpretation
$(\Omega_A^0,\omega^A_\alpha, V^A_{\alpha,\beta,\gamma}, B^A_{\alpha,\beta,\beta^\prime})$
of $\TTT_{\alpha,\beta,\beta^\prime,\gamma}$ in Proposition \ref{protatorgeom}
on $A_{\alpha,\beta,\gamma}$ generate
$H^0(A,\Z)$ and $H^4(A,\Z)$.

The result of Lemma \ref{complexclaim} is part of the claimed Result \ref{main}.
It seems to imply that the ``dual" refined geometric interpretation
of $\CCC_{\alpha,\beta,\beta^\prime,\gamma}$ in Proposition \ref{protak3geom}
is the desired one. Indeed, Lemma \ref{complexclaim}
says that the complex structure of that refined geometric interpretation is as
wanted, and Proposition \ref{protak3geom} confirms that its K\"ahler structure,
volume, and B-field are in accord with the claim in Result \ref{main}.
One would hence like to show that
$\omega_Q$ in Proposition \ref{protak3geom} is the K\"ahler class induced by
the Fubini-Study metric of $\CP^3$.
However, lattice calculations alone cannot yield such a proof, and I
cannot claim $\omega_Q=\omega_{FS}$.
The necessary
additional ingredients are explained in Section \ref{gep24}, and the proof
is completed in Section \ref{endofproof}.

The use of Inose's work \cite{in76} gives a lead to understand the geometry
underlying Result \ref{main},
which I shall follow on in Section \ref{Inose}.
Before doing so, let me put the statement of the result into context. Namely, a main
ingredient in the proof was the fact that all relevant complex structures are
given by two-planes $\Omega^Y_{\alpha,\beta,\beta^\prime,\gamma}$
($Y=X$ or $Y=A$) which can
be specified in terms of lattices of signature $(p,q)=(2,2)$, and that such lattices
have unique embeddings into $H^2(Y,\Z)$ as primitive sublattices
by Nikulin's results \cite{ni80}.
Here, $p,q<3$ is crucial; in fact, the two-planes
$\Omega^Y_{\alpha,\beta,\beta^\prime,\gamma}$ are generated by lattice vectors
iff $\alpha,\,\beta,\,\beta^\prime,\,\gamma\in\Q$, i.e.\ for a dense subset of
the parameter space.  In other words, iff $\alpha,\,\beta,\,\beta^\prime,\,\gamma\in\Q$,
then $A(f_1,f_2)$ and $X(f_1,f_2)$ are attractive according to
\bdefi{attractive}
A Calabi-Yau 2-fold $Y$ with complex structure given by a two-plane
$\Omega_Y\subset H^2(Y,\R)$ which is generated by lattice vectors
in $H^2(Y,\Z)$ is called \textsc{attractive}.

If $X$ is an attractive $K3$ surface with complex structure given by
$\Omega_X\subset H^2(X,\R)$, and if the quadratic form of the lattice
$\Omega_X\cap H^2(X,\Z)$ is $4Q_A$ with $Q_A$ an  even integral quadratic
form on that lattice, then $X$ with this complex structure
is called \textsc{very attractive}.
\edefi
For the family $A(f_1,f_2)=E_{f_1}\times E_{f_2}$ of Abelian varieties
with $E_{f_k}$ as in Result \ref{main} we see from \req{period} that for
$\alpha,\,\beta,\,\gamma\in\Z$ and $\beta^\prime=\beta$ the
quadratic form of
$\Omega^A_{\alpha,\beta,\beta^\prime,\gamma}\cap H^2(A,\Z)$ simply is
$\left(\begin{array}{cc}2\alpha&\beta\\\beta&2\gamma\end{array}\right)$.
This form however changes dramatically as
$\alpha,\,\beta,\,\beta^\prime,\,\gamma$ vary in $\Q$.

In the original mathematics literature, attractive Calabi-Yau 2-folds
are called singular. Since this word can be misleading,
I follow Moore's suggested terminology.
In \cite{mo98a,mo98b} Moore identifies such complex structures as
attractor points for the dynamical systems associated to extremal
static spherically symmetric supersymmetric black holes, which
explains his terminology, see also \cite{mo04}. The ``very attractive"
terminology of Definition \ref{attractive} is justified because
Inose shows in \cite[Theorem 1]{in76} that
``very attractive" $K3$ surfaces are the special attractive $K3$ surfaces of
the form $X(f_1,f_2)$. For the latter,  the quadratic
form of $\Omega_{X(f_1,f_2)}\cap H^2(X,\Z)$ is $4Q_A$ where $Q_A$ is the
quadratic
form of $\Omega_{A(f_1,f_2)}\cap H^2(A,\Z)$.
By the above the ``very attractive" $K3$ surfaces
are dense in the family $X(f_1,f_2)$. This statement
makes sense even though it is known \cite{at58} that the moduli space
of complex structures on $K3$ does not carry a Hausdorff topology:
We are varying surfaces $X(f_1,f_2)$ in $\CP^3$, giving complex structures
with a fixed polarization. In other words, in effect
we are varying ``marked pairs" of complex and K\"ahler structures
(c.f.\ \cite[p.\ 335]{bhpv84}), and their moduli space is indeed
Hausdorff \cite[Theorem VIII.12.3]{bhpv84}.

A note on rationality: A SCFT $\CCC_{\alpha,\beta,\beta^\prime,\gamma}$
of Result \ref{main} is rational iff the underlying toroidal SCFT
$\TTT_{\alpha,\beta,\beta^\prime,\gamma}$ is rational. For the latter,
equivalently the two tensor factors giving SCFTs associated to elliptic
curves are rational. A SCFT associated to an elliptic curve with geometric
interpretation given by $\tau,\,\rho\in\H$ is rational iff there exists
$D\in\Q$ such that $\tau,\,\rho\in\Q(\sqrt{-D})$ (see
\cite{mo98b}). In our example the parameters $\tau_1=i=\tau_2$
for the two tensor factors are fixed, so $\TTT_{\alpha,\beta,\beta^\prime,\gamma}$
and thereby $\CCC_{\alpha,\beta,\beta^\prime,\gamma}$ is rational iff
$\rho_1,\,\rho_2\in\Q(i)$, or equivalently $\alpha,\,\beta,\,\beta^\prime,
\sqrt{-\Delta}\in\Q$. It is easy to find examples of $\alpha,\,\beta,\,\beta^\prime,
\gamma\in\Q$ such that $\Delta=\beta^2-4\alpha\gamma$ has
$\sqrt{-\Delta}\not\in\Q$. In other words, by Proposition \ref{protak3geom}
one finds examples of \emph{non-rational} SCFTs in $\MMM^{K3}$ that are described
by a four-plane $x_{\alpha,\beta,\beta^\prime,\gamma}\subset H^{even}(X,\R)$
which is generated by lattice vectors in $H^{even}(X,\Z)$.
This contradicts one of the many beliefs about the relation between rationality
of SCFTs and the r\^ole of the lattice $H^{even}(X,\Z)\subset H^{even}(X,\R)$.
\subsection{Inose's construction}\label{Inose}
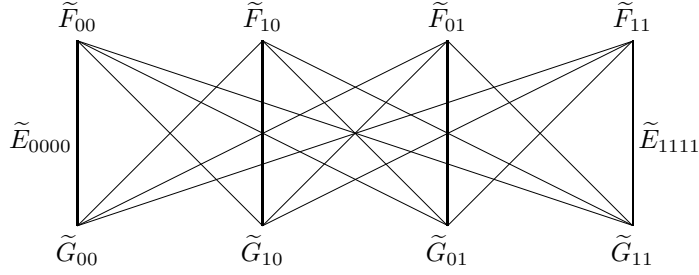
\begin{figure}[ht]
\setlength{\unitlength}{7em}
\hspace*{\fill}
\begin{picture}(4,1.5)(-0.5,-0.2)
\multiput(0,0)(1,0){4}{\line(0,1){1}}
\multiput(0,0)(1,0){3}{\line(1,1){1}}
\multiput(0,0)(1,0){2}{\line(2,1){2}}
\put(0,0){\line(3,1){3}}
\multiput(0,1)(1,0){3}{\line(1,-1){1}}
\multiput(0,1)(1,0){2}{\line(2,-1){2}}
\put(0,1){\line(3,-1){3}}
\put(0,-0.05){\makebox(0,0)[tc]{$\wt G_{00}$}}
\put(1,-0.05){\makebox(0,0)[tc]{$\wt G_{10}$}}
\put(2,-0.05){\makebox(0,0)[tc]{$\wt G_{01}$}}
\put(3,-0.05){\makebox(0,0)[tc]{$\wt G_{11}$}}
\put(0,1.05){\makebox(0,0)[bc]{$\wt F_{00}$}}
\put(1,1.05){\makebox(0,0)[bc]{$\wt F_{10}$}}
\put(2,1.05){\makebox(0,0)[bc]{$\wt F_{01}$}}
\put(3,1.05){\makebox(0,0)[bc]{$\wt F_{11}$}}
\put(-0.2,0.4){\makebox(0,0)[bc]{$\wt E_{0000}$}}
\put(3.2,0.4){\makebox(0,0)[bc]{$\wt E_{1111}$}}
\end{picture}
\hspace*{\fill}
\caption{The sixteen lines $\wt E_{j k}=\wt F_j\wt G_k$ in $X(f_1,f_2)$.}\label{lines}
\end{figure}

\begin{figure}[ht]
\setlength{\unitlength}{2em}\hspace*{\fill}
\begin{picture}(10,10)
\multiput(1,0.3)(2,0){4}{\line(0,1){8.5}}
\multiput(0,2)(0,2){4}{\line(1,0){0.6}}
\multiput(7.4,2)(0,2){4}{\line(1,0){1.2}}
\multiput(1.4,2)(0,2){4}{\line(1,0){1.2}}
\multiput(3.4,2)(0,2){4}{\line(1,0){1.2}}
\multiput(5.4,2)(0,2){4}{\line(1,0){1.2}}
\multiput(0.8,0.9)(0,2){4}{\line(1,1){1.4}}
\multiput(2.8,0.9)(0,2){4}{\line(1,1){1.4}}
\multiput(4.8,0.9)(0,2){4}{\line(1,1){1.4}}
\multiput(6.8,0.9)(0,2){4}{\line(1,1){1.4}}
\put(1,0){\makebox(0,0)[c]{$F_{00}$}}
\put(3,0){\makebox(0,0)[c]{$F_{11}$}}
\put(5,0){\makebox(0,0)[c]{$F_{10}$}}
\put(7,0){\makebox(0,0)[c]{$F_{01}$}}
\put(9.2,2){\makebox(0,0)[c]{$G_{00}$}}
\put(9.2,4){\makebox(0,0)[c]{$G_{11}$}}
\put(9.2,6){\makebox(0,0)[c]{$G_{10}$}}
\put(9.2,8){\makebox(0,0)[c]{$G_{01}$}}
\put(2,1.5){\makebox(0,0)[c]{$E_{0000}$}}
\put(4,1.5){\makebox(0,0)[c]{$E_{1100}$}}
\put(6,1.5){\makebox(0,0)[c]{$E_{1000}$}}
\put(8,1.5){\makebox(0,0)[c]{$E_{0100}$}}
\put(2,3.5){\makebox(0,0)[c]{$E_{0011}$}}
\put(4,3.5){\makebox(0,0)[c]{$E_{1111}$}}
\put(6,3.5){\makebox(0,0)[c]{$E_{1011}$}}
\put(8,3.5){\makebox(0,0)[c]{$E_{0111}$}}
\put(2,5.5){\makebox(0,0)[c]{$E_{0010}$}}
\put(4,5.5){\makebox(0,0)[c]{$E_{1110}$}}
\put(6,5.5){\makebox(0,0)[c]{$E_{1010}$}}
\put(8,5.5){\makebox(0,0)[c]{$E_{0110}$}}
\put(2,7.5){\makebox(0,0)[c]{$E_{0001}$}}
\put(4,7.5){\makebox(0,0)[c]{$E_{1101}$}}
\put(6,7.5){\makebox(0,0)[c]{$E_{1001}$}}
\put(8,7.5){\makebox(0,0)[c]{$E_{0101}$}}
\end{picture}
\hspace*{\fill}
\caption{The double Kummer pencil in  $Y(f_1,f_2)$.}\label{kummerpencil}
\end{figure}
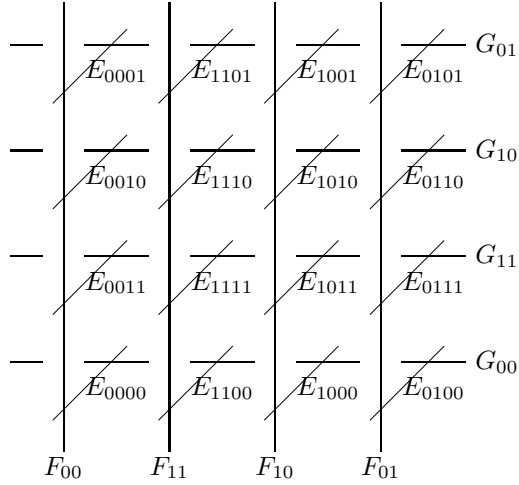

To allow insight into the geometry underlying Result \ref{main},
let me briefly summarize Inose's work \cite{in76}.
Choose two homogeneous polynomials $f_1,\,f_2$ of degree $4$ in
two variables each. I will assume that $f_1,\,f_2$ are non-degenerate,
i.e.\ that they do not have multiple roots.
As in Result \ref{main}
these polynomials  define elliptic curves
$$
E_{f_k}:\quad
y_0^2 = f_k(y_1,y_2) \quad\mb{ in }\quad \CP_{2,1,1}
$$
with moduli $\rho_1,\,\rho_2\in\H$, see
Section \ref{algebraic} and \ref{quarelli}. $\rho_1,\,\rho_2$
can always be brought into the form used in Result \ref{main}. The
polynomials $f_1,\,f_2$ also define
a smooth quartic $K3$ surface
$$
X(f_1,f_2):\quad\quad
f_1(x_0,x_1)+f_2(x_2,x_3)=0 \quad\mb{ in }\quad \CP^3.
$$
Note that all the surfaces $X(f_1,f_2)$ share the symplectic
automorphism $\sigma$ given by
\beq{sigma}
\sigma\colon(x_0,x_1,x_2,x_3)
\longmapsto (-x_0,-x_1,x_2,x_3).
\eeq
This automorphism generates a group $\langle\sigma\rangle$ of order $2$.
Now let $Y(f_1,f_2)$ denote the $K3$ surface obtained by blowing up
the eight nodal singularities of $X(f_1,f_2)/\langle\sigma\rangle$. On the other hand
let $\Km(E_{f_1}\times E_{f_2})$ denote the $K3$ surface obtained from the
Abelian variety $A(f_1,f_2)=E_{f_1}\times E_{f_2}$ by the  Kummer construction. In
other words, we represent
$E_{f_1}\times E_{f_2}$ as $\C^2/\!\sim$ with standard coordinates $(z_1,z_2)$
and $z_k\sim z_k+1\sim z_k+\rho_k$, to obtain a natural
$\Z_2$ action by multiplication by
$-1$ on $\C^2$. Now $\Km(E_{f_1}\times E_{f_2})$ is obtained
by blowing up the sixteen nodal singularities of
$E_{f_1}\times E_{f_2}/\Z_2$.
Hiroshi Inose has discovered
\btheo[\mbox{\cite[Theorem 2]{in76}}]{inoseresult}
The $K3$ surface $Y(f_1,f_2)$ obtained from $X(f_1,f_2)/\langle\sigma\rangle$ by
minimally resolving all singularities is canonically biholomorphic
to the Kummer surface $\Km(A(f_1,f_2))$ of
the Abelian variety $A(f_1,f_2)=E_{f_1}\times E_{f_2}$.
\etheo
The geometric situation
found in \cite{in76} is as follows:
Denote the roots of $f_l(1,\zeta)=0$
by $\zeta_j^l\in\C$ where for later convenience I use
indices $j\in\F^2_2=\{00,10,01,11\}$.
The quartic $X(f_1,f_2)$ contains sixteen
lines
$$
\wt E_{j k}: \quad
\left\{ (x_0,x_1,x_2,x_3) \mid x_1=\zeta_j^1x_0,\; x_3=\zeta_k^2x_2\right\}.
$$
The four lines $\wt E_{j00},\ldots,\wt E_{j11}$
intersect in the fixed point
$\wt F_j=(1,\zeta_j^1,0,0)$ of $\sigma$, while the four lines
$\wt E_{00k},\ldots,\wt E_{11k}$ intersect in the fixed point
$\wt G_k=(0,0,1,\zeta_k^2)$ of $\sigma$, forming a constellation as
depicted in Figure \ref{lines}.
Though the figure is note entirely suggestive, the
$\wt F_j,\wt G_k$ are the only intersection points of any two
lines $\wt E_{l m}$. These lines are mapped onto themselves under
the linear automorphism $\sigma$. In the resolved orbifold $Y(f_1,f_2)$ they
therefore give rational curves $E_{j k}$, while the fixed
points $\wt F_j,\wt G_k$ of $\sigma$ are blown up giving eight exceptional
rational curves $F_j,\,G_k$. Altogether Inose finds a double Kummer pencil
in $Y(f_1,f_2)$ as shown in Figure \ref{kummerpencil}.
Moreover,  the sixteen rational
curves $E_{j k}$ can be identified with the sixteen
irreducible components of the exceptional divisor
in  $\Km(E_{f_1}\times E_{f_2})$, while
the $F_j,\,G_k$ can be interpreted
in terms of two-cycles of $E_{f_1}\times E_{f_2}=\C^2/\!\sim$. Namely,
\beqn{mclass}
\mb{in } H_2(X,\Z):\quad\fa j,k\in\F^2_2:\quad\quad
2F_{j}+E_{j00}+E_{j10}+E_{j01}+E_{j11}&=&M_{12},\\[3pt]
2G_{k}+E_{00k}+E_{10k}+E_{01k}+E_{11k}&=&M_{34},
\eeqn
where $M_{12}$ is the class of the image of the cycle $z_1=\const$ in
$E_{f_1}\times E_{f_2}=\C^2/\!\sim$, while the class $M_{34}$
gives the image of the cycle $z_2=\const$
\subsection{Inose's construction extended?}\label{extension}
In view of Inose's geometric insights I can reformulate the Result \ref{main}
as follows: Since by Proposition \ref{protatorgeom} each
$\TTT_{\alpha,\beta,\beta^\prime,\gamma}$ has a refined geometric interpretation
on the Abelian variety $A(f_1,f_2)$, and $\CCC_{\alpha,\beta,\beta^\prime,\gamma}$
is claimed to have a refined geometric interpretation on $X(f_1,f_2)$, the assertion
amounts to the SCFTs $\CCC_{\alpha,\beta,\beta^\prime,\gamma}$,
$\TTT_{\alpha,\beta,\beta^\prime,\gamma}$, and the ordinary $\Z_2$-orbifold
$\TTT_{\alpha,\beta,\beta^\prime,\gamma}/\Z_2$ to provide an extension of
Inose's construction to the realm of Calabi-Yau 2-folds with complexified
K\"ahler structures, in other words to the realm of SCFTs. That
such extensions should exist is in itself not surprising. However, surprisingly
both on $X(f_1,f_2)$ and on $A(f_1,f_2)$ the most natural
K\"ahler structures turn out to occur, the ones arising from
$X(f_1,f_2)\hookrightarrow\CP^3$ and $A(f_1,f_2)\cong\C^2/\sim$, respectively.
In contrast, note that every ``very attractive" quartic (see Definition \ref{attractive})
is biholomorphic to some Kummer surface;
for instance $\CCC_{1,0,0,1}$ has a geometric
interpretation on the (very attractive) Fermat quartic
\beq{fermat}
X_{Fermat}=X(f_0,f_0)\colon\quad
x_0^4+x_1^4+x_2^4+x_3^4=0\quad\mb{ in }\quad\CP^3
\eeq
by Result \ref{main} and (\ref{fermatelli}) but nevertheless
cannot be constructed from any toroidal model by
a $\Z_2$-orbifold procedure \cite[p.\ 123]{nawe00}.
Furthermore, it is not obvious that in any given extension of
Inose's picture all associated SCFTs can be constructed explicitly, let
alone by geometric orbifolds like the one yielding
$\CCC_{\alpha,\beta,\beta^\prime,\gamma}$ from $\TTT_{\alpha,\beta,\beta^\prime,\gamma}$.
To understand why this is  possible in the present case, note that
assuming  Result \ref{main} it  follows that
$\TTT_{\alpha,\beta,\beta^\prime,\gamma}/\Z_2$
is also a $\Z_2$-orbifold of $\CCC_{\alpha,\beta,\beta^\prime,\gamma}$.
For every CFT $\CCC$, an orbifold $\CCC/G$ by a solvable group $G$
enjoys an action of $G$ such that orbifolding $\CCC/G$ by $G$
reproduces the original CFT $\CCC$ \cite[p.\ 126]{gi88b}.
It follows that any extension of Inose's construction to the level of
SCFTs must yield theories associated to $X(f_1,f_2)$ which are obtained
from theories associated to $A(f_1,f_2)$ by an
orbifold by a group of order $4$, i.e.\ by a $\Z_2\times\Z_2$ or by a
$\Z_4$-action on a family of toroidal SCFTs. Choosing the complexified
K\"ahler structure
$\omega_A=-{i\over2}\left(dz_1\wedge d\qu z_1+dz_2\wedge d\qu z_2\right),
V_A=1, B_A=0$ of Proposition \ref{protatorgeom} on $A(f_1,f_2)$ ensures that
the associated SCFTs $\TTT_{\alpha,\beta,\beta^\prime,\gamma}$ all enjoy
an automorphism of order $4$, namely the $\Z_4$-symmetry which is induced
by the geometric $\Z_4$-action $\zeta_4$ of \req{z4} in the
``mirror dual" refined geometric interpretation of
$\TTT_{\alpha,\beta,\beta^\prime,\gamma}$ on $A_{\alpha,\beta,\gamma}$
as in \req{Aalbeg} and with B-field $B_{\alpha,\beta,\beta^\prime}$
as in Proposition \ref{protator}.

In geometry, the reversal of an orbifold construction by another orbifold
is of course impossible. Indeed, if $\CCC$ is a CFT associated to some
Calabi-Yau variety $Y$ and $G$ is a solvable symmetry group of $\CCC$ which is induced
by symplectic automorphisms of $Y$, then $\CCC/G$ is a SCFT associated to
$\wt{Y/G}$ \cite{dhvw85,dhvw86}, and the symmetry of type $G$ which
$\CCC/G$ enjoys and which yields $\CCC$ back under orbifolding is not
a geometric symmetry of $\wt{Y/G}$. Such symmetries are known as
\textsc{quantum symmetries}.

That the theories $\CCC_{\alpha,\beta,\beta^\prime,\gamma}$
in Result \ref{main} are nevertheless obtained by a geometric orbifold
construction from the theories $\TTT_{\alpha,\beta,\beta^\prime,\gamma}$ is a
consequence of the geometric re-interpretation of $\TTT_{\alpha,\beta,\beta^\prime,\gamma}$
described in Proposition \ref{protatorgeom}. Indeed, one of the crucial ideas
from the early days of mirror symmetry is the observation that mirror symmetry
is a non-classical equivalence between SCFTs, which interchanges the
r\^oles of geometric and quantum symmetries \cite{grpl90}.
With an appropriate version of mirror symmetry it should therefore be possible to
find a geometric interpretation of $\TTT_{\alpha,\beta,\beta^\prime,\gamma}/\Z_2$
which has a geometric $\Z_2$-symmetry that upon orbifolding yields
$\CCC_{\alpha,\beta,\beta^\prime,\gamma}$. Because the ordinary $\Z_2$-orbifold
of a toroidal SCFT $\TTT$ descends to the Kummer construction in every refined
geometric interpretation of $\TTT$, we can equivalently expect to find a
refined geometric interpretation of $\TTT_{\alpha,\beta,\beta^\prime,\gamma}$
such that the total symmetry group of order $4$ which upon orbifolding
yields $\CCC_{\alpha,\beta,\beta^\prime,\gamma}$ acts geometrically.
Result \ref{main} consequentially claims that this desired refined geometric interpretation
is the ``mirror dual" of the one on $A(f_1,f_2)$ with complexified
K\"ahler structure $(\omega_A,V_A,B_A)$, as given in Proposition \ref{protatorgeom},
and that this geometric action is the standard action \req{z4} of $\Z_4$.

It is indeed natural to view the two geometric interpretations
$(\Omega_A^0, \omega_\alpha^A, V_{\alpha,\beta,\gamma}^A,
B_{\alpha,\beta,\beta^\prime}^A)$ and $(\Omega^A_{\alpha,\beta,\beta^\prime,\gamma},
\omega_A,V_A,B_A)$ of Proposition \ref{protatorgeom} as mirror duals:
On the one hand, by the proof of Proposition \ref{protatorgeom} exchanging these
two geometric interpretations amounts to interchanging the modular parameters
$\tau_k,\,\rho_k$ that specify the two tensor factor theories of
$\TTT_{\alpha,\beta,\beta^\prime,\gamma}$ which are SCFTs associated to
elliptic curves, where according to Section \ref{CY1modsp} a version of mirror
symmetry is given by $U(\tau_k,\rho_k)=(\rho_k,\tau_k)$. On the other hand, according to
Proposition \ref{protatorgeom} the exchange of the two geometric interpretations
of $\TTT_{\alpha,\beta,\beta^\prime,\gamma}$ amounts to interchanging the r\^ole
of the two-planes
$\Omega_A^0$ and $\Omega^A_{\alpha,\beta,\beta^\prime,\gamma}$ which
$x^A_{\alpha,\beta,\beta^\prime,\gamma}$ decomposes into, i.e.\ indeed to
interchanging complex and complexified K\"ahler structures. Note furthermore
that the $\Z_4$-orbifold procedure yielding the $K3$ surface
$X_{\alpha,\beta,\gamma}=\wt{A_{\alpha,\beta,\gamma}/\Z_4}$ from
$A_{\alpha,\beta,\gamma}=\R^4/\Lambda_{\alpha,\beta,\gamma}$ of \req{Aalbeg}
can indeed be performed in terms of two consecutive $\Z_2$-orbifolds, the
first one of which is the Kummer construction.

I hope to have convinced the reader that Result \ref{main}
does have a natural interpretation as extension of Inose's construction
to the realm of SCFTs. However, in the above explanation I have used a
very crude version of mirror symmetry, which amounts to interchanging
the two-planes of a refined geometric interpretation of a SCFT but
does not address the choice of null vectors as needed within any
refined geometric interpretation according to Definition \ref{refinedgeom}.
Likewise, the notion of ``quantum symmetries" was used in a slightly
obscure fashion without proper definition, and in particular without
giving a procedure to distinguish between ``geometric" and
``quantum" symmetries. Hence the above can only serve as a motivation,
not as a proof of Result \ref{main}.

On purely geometric grounds the above discussion naturally
raises the question whether Inose's construction
can be extended to the level of K\"ahler-Einstein metrics. More precisely,
the class of the most natural K\"ahler structure
on $X(f_1,f_2)$ is the class $\omega_{FS}\in H^2(X,\Z)$
of the Fubini-Study metric on $\CP^3$,
$$
\omega_{FS}(x)
={i\over2\pi} \partial\qu\partial
\log\left(\sum_{j=0}^3 \left|x_j\right|^2\right)
\quad\mb{ for } x=(x_0,x_1,x_2,x_3)\in\CP^3.
$$
By the Calabi-Yau theorem \cite{ya78} there is a unique
K\"ahler-Einstein metric on $X(f_1,f_2)$ with K\"ahler class $\omega_{FS}$.
Since $\omega_{FS}$ is invariant
under $\sigma$ it
descends to a class $\wh\omega_{FS}$
on  $Y(f_1,f_2)$. The class $\wh\omega_{FS}$ in turn represents
the orbifold limit of
an Einstein metric on $Y(f_1,f_2)$, which assigns vanishing volume to all
components of the exceptional divisor in the resolution of
$X(f_1,f_2)/\langle\sigma\rangle$:
\beq{vanish}
\fa j,k\in\F^2:\quad\quad
\langle \wh\omega_{FS}, \wh F_j\rangle =
\langle \wh\omega_{FS}, \wh G_k\rangle = 0
\eeq
with $\wh F_j,\,\wh G_k$ denoting the Poincar\'e duals of $F_j,\,G_k$,
respectively (see Figure \ref{kummerpencil}). One also checks
\beq{scpds}
\langle \wh F_j,\wh M_{12}\rangle
=\langle \wh G_k,\wh M_{34}\rangle= 0,\quad
\langle \wh F_j,\wh M_{34}\rangle
= \langle \wh G_k,\wh M_{12}\rangle=1, \quad
\langle \wh F_j,\sum_{l\in\F_2^4}\wh E_l \rangle
=\langle \wh G_i,\sum_{l\in\F_2^4} \wh E_l\rangle=4
\eeq
with $\wh E_l,\,l\in\F_2^4$ denoting the Poincar\'e duals of the rational
curves $E_l$, and $\wh M_j$ obtained as
Poincar\'e duals of the
classes $M_j$ introduced in \req{mclass},
$$
\wh M_j\in H^2(X,\Z),\quad\quad\langle\wh M_j,\wh M_j\rangle=0,\quad\quad
\langle\wh M_{12},\wh M_{34}\rangle=2,
$$
see e.g.\ \cite{ni75}. From what was said above one expects that it
should be possible to express $\wh\omega_{FS}$ in terms
of the simpler geometry of the Kummer surface $\Km(E_{f_1}\times E_{f_2})$.
Indeed, the result is remarkably simple\footnote{In \cite{hewi05}
a family of Ricci-flat K\"ahler-Einstein metrics  is determined
numerically which by Proposition \ref{metric} turns out to
approach the one represented
by  $\wh\omega_{FS}$. In fact, the explicit form of Proposition \ref{metric} arose
as a conjecture from a discussion
with the authors Matthew Headrick
and Toby Wiseman of \cite{hewi05}, and
I am grateful to them for raising
the relevant questions that led to this observation.}:
\bprop{metric}
Let $\wh\omega_{FS},\,\wh\omega_{\Km}$ represent the orbifold limits
of K\"ahler-Einstein metrics on $\wt{X(f_1,f_2)/\langle\sigma\rangle}=\Km(A(f_1,f_2))$
induced by the K\"ahler-Einstein metric with class
$\omega_{FS}$ of the Fubini-Study metric on $X(f_1,f_2)$, and $\omega_A$,
the class of the Euclidean metric on $\C^2$ with
$A(f_1,f_2)=\C^2/\sim$, respectively.
Then
$$
\wh\omega_{FS}=2\wh\omega_{\Km}-{1\over2}\sum_{i\in\F_2^4}\wh E_i.
$$
\eprop
\bpr
The key to the proof is the use of the explicit identifications
of cycles \cite{in76} given
in Section \ref{Inose}, along with a study of symplectic
automorphisms of the Fermat quartic hypersurface
$X_{Fermat}=X(f_0,f_0)$ of \req{fermat}. Indeed, one checks
$$
\wh\omega_{\Km}=\wh M_{12}+\wh M_{34},
$$
and since $\wh\omega_{\Km},\,\wh\omega_{FS}\in H^2(Y,\Z)$ do not change while
$f_1,\,f_2$ vary, a proof of the claim for the $\langle\sigma\rangle$-orbifold of
the Fermat quartic $X_{Fermat}$ is sufficient.
The group $G_{Fermat}$
of symplectic automorphisms of $X_{Fermat}$ is well known. It
is generated by phase symmetries
$$
[n_0,\ldots,n_3]\colon
(x_0,\ldots,x_3) \longmapsto (i^{n_0}x_0,\ldots,i^{n_3}x_3)\quad
\mb{ with } n_k\in\Z/4\Z\mb{ and }\sum_k n_k\equiv 0\mod 4
$$
along with permutations $\gamma\in S_4$ of the coordinates accompanied
by phase symmetries $[n_0,\ldots,n_3]$ as above such that
$\sum_k n_k\equiv (1-\det\gamma)\mod 4$. Since
$[1,\ldots,1]$ acts trivially on $\CP^3$ we find
$G_{Fermat}\cong\Z_4^3\rtimes S_4/\Z_4$.

The commutant of $\sigma\in G_{Fermat}$ in $G_{Fermat}$ gives the group
$\Z_4^3\rtimes D_4/\Z_4$ with $\Z_4^3$ as before and
generators $r,\,s$ of $D_4$, which acts as
the dihedral group of order $8$:
\begin{eqnarray*}
r\colon\quad (x_0,x_1,x_2,x_3) &&\longmapsto\quad (x_2,x_3,x_1,-x_0),\\
s\colon\quad  (x_0,x_1,x_2,x_3) &&\longmapsto\quad (x_1,-x_0,x_2,x_3).
\end{eqnarray*}
Each element of this group $\Z_4^3\rtimes D_4/\Z_4$ induces a
symplectic automorphism on the orbifold $Y_{Fermat}$ of $X_{Fermat}$
by $\langle\sigma\rangle$.
However, $\id$ and $\sigma\in \Z_4^3\rtimes D_4/\Z_4$ induce the
trivial automorphism, leaving us with the group
$G_{\Km} = (\Z_2\times \Z_4)\rtimes D_4$ generated by
$t_{1100}:=[1,3,0,0],\,t_{1000}:=s,\,r_{12}:=r\circ s,\,
r_{13}:=[1,0,0,3]\circ r^2$,
with notations
as above. For later convenience note
\begin{eqnarray*}
\mb{in }\CP^3/\langle\sigma\rangle:\quad\quad
t_{1100}\colon\quad (x_0,x_1,x_2,x_3) &&\longmapsto\quad (i x_0,-i x_1,x_2,x_3),\\
t_{1000}\colon \quad (x_0,x_1,x_2,x_3) &&\longmapsto\quad (x_1,-x_0,x_2,x_3),\\
r_{12}\colon \quad (x_0,x_1,x_2,x_3) &&\longmapsto\quad (x_2,x_3,x_0,x_1),\\
r_{13}\colon\quad  (x_0,x_1,x_2,x_3) &&\longmapsto\quad (-i x_1,x_0,x_3,i x_2).
\end{eqnarray*}
Let us now investigate these automorphisms in the light of the interpretation
of $Y_{Fermat}$ as Kummer surface as in Theorem \ref{inoseresult}.
More precisely, I will determine the
action on the cycles $F_j,\,G_k,\,E_l,\,j,k\in\F_2^2,\,l\in\F_2^4$,
introduced above. I will in
particular be interested in those cycles which are invariant under the
entire group $G_{\Km}$, since the class $\omega_{FS}$ is invariant
under $G_{Fermat}$ and hence the class
$\wh\omega_{FS}$ which we wish to express in terms of the Kummer geometry
is Poincar\'e dual to a cycle which is invariant under $G_{\Km}$.
All the rational curves $F_j,\,G_k,\,E_l$ are uniquely determined by the
positions of the fixed points $\wt F_j,\,\wt G_k$
of $\sigma$. Because $G_{\Km}$ acts projectively linearly, it suffices
to determine the action of $G_{\Km}$ on these fixed points. To this end
denote by $\eps$ a primitive eighth root of unity such that $\eps^2=i$.
Then for the Fermat quartic we denote
the roots $\zeta_j^l$ of the quartic polynomial
$f_{0}(1,\zeta)=1+\zeta^4=0$ by
$$
\zeta_{00}^l=\eps, \quad \zeta_{11}^l=-\eps,\quad
\zeta_{10}^l=i\eps, \quad \zeta_{01}^l=-i\eps.
$$
Consider the action of $t_{1100}$.
The fixed points $\wt G_k=(0,0,1,\zeta_k^2)$ are invariant under
this automorphism, while it interchanges $\wt F_{00}$ with
$\wt F_{11}$, and $\wt F_{10}$ with $\wt F_{01}$, respectively. In other words,
$t_{1100}$ acts by a shift by $(1100)$ on the index set $\F_2^4$ of the
$E_l$.
Similarly, $t_{1000}$ leaves
the fixed points $\wt G_k=(0,0,1,\zeta_k^2)$ invariant. It
interchanges $\wt F_{00}$ with
$\wt F_{10}$, and $\wt F_{01}$ with $\wt F_{11}$, respectively. In other words,
$t_{1000}$ acts by a shift by $(1000)$ on the index set $\F_2^4$ of the
$E_l$.
The action of $r_{12}$ is most easily determined - it acts on the indices
$l\in\F_2^4$ of the $E_l$ by the permutation
$(l_1,l_2,l_3,l_4)\mapsto(l_3,l_4,l_1,l_2)$.
Finally, one checks that $r_{13}$ leaves $\wt F_{00},\, \wt F_{11},\,
\wt G_{00},\,\wt G_{11}$ invariant while interchanging $\wt F_{10}$ with
$\wt F_{01}$, and $\wt G_{10}$ with $\wt G_{01}$. In other words,
$r_{13}$ acts on the indices
$l\in\F_2^4$ of the $E_l$ by the permutation
$(l_1,l_2,l_3,l_4)\mapsto(l_2,l_1,l_4,l_3)$.

Translating into cohomology by means of the Poincar\'e duality
we find a two-dimensional subspace of
$H^{1,1}(X,\C)\cap H^2(X,\Z)$ which is invariant under all of $G_{\Km}$,
with generators
$$
\wh\omega_{\Km}=\wh M_{12}+\wh M_{34}\quad
\mb{ and }\quad \wh E:=\sum_{l\in\F_2^4} \wh E_l, \quad\quad
\langle\wh\omega_{\Km},\wh E\rangle=0.
$$
On the other hand, recall that for an algebraic $K3$ surface $X$
with group $G$ of symplectic automorphisms the dimension of the
$G$-invariant subspace $\left(H^{1,1}(X,\C)\cap H^2(X,\Z)\right)^G$
of $H^{1,1}(X,\C)\cap H^2(X,\Z)$ can be determined by purely combinatorial
methods \cite{mu88}.
Namely, the group $G$ induces an action on the
total rational cohomology $H^\ast(X,\Q)$ given by a so-called
Mathieu representation
\cite[Theorem 1.4]{mu88}, which implies
\begin{eqnarray*}
\dim_\Q H^\ast(X,\Q)^G =
\mu(G)&:=& {1\over |G|} \sum_{g\in G} \mu(\ord(g)), \\
&&\mbox{ where for }
n\in \N: \;
\mu(n) := {24\over n\prod\limits_{
\stackrel{p\, \mbox{\tiny prime,}}{\scriptscriptstyle p|n} } (1+{1\over p})}
\end{eqnarray*}
\cite[Proposition 3.4]{mu88}. Since $G$ acts symplectically,
we have
$$
\dim_\Q H^\ast(X,\Q)^G =
\dim_\R H^\ast(X,\R)^G= \dim_\C H^\ast(X,\C)^G.
$$
By the definition of symplectic automorphisms $H^\ast(X,\C)^G\supset$
$H^0(X,\C)$ $\oplus$ $H^{2,0}(X,\C)$ $\oplus$ $H^{0,2}(X,\C)
\oplus H^{2,2}(X,\C)$, so
\beq{invdim}
\dim_\R \left(H^{1,1}(X,\C)\cap H^2(X,\R)\right)^G = \mu(G)-4 .
\eeq
For our group $G_{\Km}$ one checks
$$
\mu(G_{\Km}) = {1\over64}\left(\mu(1)+27\mu(2)+36\mu(4)\right)
= {24\over64} \left( 1 + {27\over3} + {36\over6}\right) =6.
$$
By \req{invdim} this implies that $H^{1,1}(X,\C)\cap H^2(X,\R)$ has only
a two-dimensional $G_{\Km}$ invariant subspace which hence is generated by
$\wh\omega_{\Km}$ and $\wh E$. Since the form
$\wh\omega_{FS}\in H^{1,1}(X,\C)\cap H^2(X,\R)$
is also invariant
under $G_{\Km}$ we can make an ansatz
$$
\wh\omega_{FS} = \lambda\left( \wh\omega_{\Km}+\alpha\wh E\right)
= \lambda\left( \wh M_{12}+ \wh M_{34}+\alpha\wh E\right).
$$
Now \req{vanish} and \req{scpds} imply
$\alpha=-{1\over4}$. Moreover, since $\wh\omega_{FS}$
is the image of a primitive lattice vector $\omega_{FS}\in H^2(X,\Z)$
in integral
cohomology, $\wh\omega_{FS}$ is a primitive lattice vector. Since
${1\over2}\wh E\in H^2(X,\Z)$ is primitive and
$\langle\wh M_j,\wh E\rangle=0$ with indecomposable $\wh M_j$
spanning a primitive sublattice of $H^2(X,\Z)$ \cite{ni75},
we find $\lambda=\pm2$. Finally,
$\lambda=2$ follows since $\wh\omega_{FS}$ is a K\"ahler class
and all $E_l$ are
effective, thus
$\langle\wh\omega_{FS},\wh E_l\rangle>0$ for all $l\in\F_2^4$.
\epr
The above proof shows that the action of $G_{\Km}$ on
$F_j,\,G_k,\,E_l$ could be induced by the group
$G_{Kummer}^+\cong\F_2^4\rtimes \Z_2^2\cong G_{\Km}$ of automorphisms that
fix the orbifold singular metric of a Kummer surface constructed from
$E_{f_0}\times E_{f_0}$, and which descend from
automorphisms of this variety \cite[Theorem 2.7]{nawe00}.
Namely, see  \ref{quarelli} for a proof of the well-known identification
$E_{f_0}\cong\C/\Z\oplus i\Z$ which implies
$E_{f_0}\times E_{f_0}\cong\C^2/\left(\Z^2\oplus i\Z^2\right)$.
Then elements of $\F_2^4\subset G_{Kummer}^+$ act by shifts by half periods on
$\C^2/\!\sim$, while
$\Z_2^2\subset G_{Kummer}^+$ is generated by $(z_1,z_2)\mapsto (z_2,-z_1)$ and
$(z_1,z_2)\mapsto (i z_1,-i z_2)$.
Although it seems likely that indeed $G_{\Km}=G_{Kummer}^+$, note that
we have only compared the actions on a sublattice of
$H^{1,1}(X,\C)\cap H^2(X,\Z)$ of rank $18$, while for the Fermat quartic
$H^{1,1}(X,\C)\cap H^2(X,\Z)$ has maximal
rank $20$. Hence we cannot conclude that
the actions of the two groups agree. Luckily the observation that
the actions agree on  $F_j,\,G_k,\,E_l$  has turned out to
suffice to prove Proposition \ref{metric}.
\section{Proof of the main result}\label{proof}
In this section I complete the proof of the main Result \ref{main}
of this work. The proof consists of three steps, the first one
of which I have already taken in Lemma \ref{complexclaim}
where I proved that for $\alpha,\beta,\beta^\prime,\gamma\in\R$
as in Definition \ref{protator} the SCFT
$\CCC_{\alpha,\beta,\beta^\prime,\gamma}$ of Definition \ref{protak3}
allows a geometric interpretation
$(\Omega_{\alpha,\beta,\beta^\prime,\gamma}^X,\omega_Q,V_Q,B_Q)$ with
$\Omega_{\alpha,\beta,\beta^\prime,\gamma}^X$ the complex structure of the
quartic $K3$ surface $X(f_1,f_2)\subset\CP^3$ specified by
$\alpha,\,\beta,\,\beta^\prime,\,\gamma$ as described in Result \ref{main}.
As a second step I show in Section \ref{gep24} that Result \ref{main}
holds for one special member of the family $\CCC_{\alpha,\beta,\beta^\prime,\gamma}$,
namely for $\CCC_{1,0,0,1}$: This model agrees with the Gepner model
$(2)^4$, for which indeed by a combination of results by Witten \cite{wi93} and
Aspinwall and Morrison \cite{asmo94} the claim follows. The third and final
step of the proof, which I explain in Section \ref{endofproof}, uses the observation
that the geometric interpretation of $\CCC_{\alpha,\beta,\beta^\prime,\gamma}$
in which  I have already found the desired complex structure by Lemma \ref{complexclaim},
has complexified K\"ahler structure which is independent of
$\alpha,\,\beta,\,\beta^\prime,\,\gamma$. I identify
the relevant deformations of $\CCC_{\alpha,\beta,\beta^\prime,\gamma}$
induced by varying $\alpha,\,\beta,\,\beta^\prime,\,\gamma$
and show that  they are compatible with keeping the
complexified
K\"ahler structure $(\omega_{FS},V_{FS},B_{FS})$
which was found for $(2)^4$ in Section \ref{gep24}
constant for the entire family.
\subsection{The Gepner model $(2)^4$}\label{gep24}
This section is devoted to a detailed study of one special member of the family
$\CCC_{\alpha,\beta,\beta^\prime,\gamma}$ of SCFTs introduced in Definition \ref{protak3},
namely the model $\CCC_{1,0,0,1}$ obtained from the toroidal SCFT
$\TTT_{1,0,0,1}$ on the standard torus $A_{1,0,1}=\R^4/\Z^4$ with vanishing
B-field by the $\Z_4$-orbifold procedure. As a first step, I rewrite this model
in a form which is more familiar to a certain class of string theorists:
\bprop[\mb{\cite[Theorem 3.5]{nawe00}}]{z4isgepner}
The $\Z_4$-orbifold CFT $\CCC_{1,0,0,1}$ of Definition \mb{\ref{protak3}}
agrees with the $(2)^4$ Gepner model.
\eprop
For a brief primer on Gepner models and its building blocks, the minimal
models, see \ref{minimal} and \ref{Gepner}. Specifically the models that are
relevant for Proposition \ref{z4isgepner} are discussed in \ref{gep2222}.
Proposition \ref{z4isgepner} was conjectured in \cite{eoty89} and a proof was
given in \cite{nawe00}. It
is based on an explicit field theory calculation which in fact simplifies
when one uses the identifications discussed in \ref{gep2222}:
The Gepner model $(2)^2$ agrees with the SCFT associated to the
elliptic curve $\R^2/\Z^2$ with complex structure given by introducing
a complex coordinate $z=x_1+ix_2$, where $x_1,\,x_2$ are the standard Cartesian
coordinates on $\R^2$, and with vanishing B-field. Though this identification
is well-known \cite{cln92}, \ref{gep22} recalls the explicit field identifications.
I show in \ref{geproof2222} how these identifications
imply that the ``Gepner orbifold"
\req{geporb}, which gives $(2)^2\otimes(2)^2/\Z_4=(2)^4$, is induced by the
geometric $\Z_4$-action \req{z4} on the four-torus $A_{1,0,1}=\R^4/\Z^4$
which underlies the toroidal model $(2)^2\otimes(2)^2=\TTT_{1,0,0,1}$,
showing $\TTT_{1,0,0,1}/\Z_4=(2)^4$.

This construction also allows to explicitly identify some of the deformations
of the model $(2)^4$, which will become useful below.
Namely, in SCFT, integrable deformations which preserve superconformal invariance
are given in terms of fields of conformal dimensions $h=\qu h={1\over2}$
and with $\fu(1)$-charges $(Q,\qu Q)$ such that $|Q|=|\qu Q|=1$, since the
superpartners of these fields are the integrable $(h,\qu h)=(1,1)$
marginal operators \cite{di87}.
For example, $(2)^2$ possesses four such linearly independent fields, as can be
seen from \req{weights}:
$$
\psi_\pm\qu\psi_\pm = \Phi^0_{\mp2,2;\pm2,2}\otimes\Phi^0_{\mp2,2;\pm2,2},\quad\quad
\psi_\pm\qu\psi_\mp = \Phi^0_{\mp2,2;\mp2,2}\otimes\Phi^0_{\mp2,2;\mp2,2},
$$
where $\psi_\pm,\,\qu\psi_\pm$ denote the left- and the right handed Dirac
fermions as in \ref{gep22}, and where I have used \req{Diracid}.
This is in
accord with the dimension $4$ of the moduli space of SCFTs associated to elliptic
curves as stated in Proposition \ref{ellims}.
In $\TTT_{1,0,0,1}=(2)^2\otimes(2)^2$, these fields also give deformations, where only
\beqn{nicedefs}
V_\pm^{(1)} \>:=\>\Phi^0_{\pm2,2;\pm2,2}\otimes\Phi^0_{\pm2,2;\pm2,2}
\otimes\Phi^0_{0,0;0,0}\otimes\Phi^0_{0,0;0,0},\\[2pt]
V_\pm^{(2)} \>:=\>\Phi^0_{0,0;0,0}\otimes\Phi^0_{0,0;0,0}\otimes
\Phi^0_{\pm2,2;\pm2,2}\otimes\Phi^0_{\pm2,2;\pm2,2}
\eeqn
are invariant under the action \req{geporb} which yields the $\Z_4$-orbifold
$(2)^2\otimes(2)^2/\Z_4=(2)^4$. In other words, $V_\pm^{(1)}$ and $V_\pm^{(2)}$
are the deformations that $(2)^4$ has in common with $(2)^2\otimes(2)^2$,
and we have
\bprop{def2222}
The fields $V_\pm^{(1)},\,V_\pm^{(2)}$ of \mb{\req{nicedefs}} give
deformations of the Gepner model $(2)^4=\CCC_{1,0,0,1}$ which in the
geometric interpretation $(\Omega_X^0,\omega_1,V_{1,0,1},B_{1,0,0})$
of Proposition \mb{\ref{protak3geom}} on the
$\Z_4$-orbifold $X_{1,0,1}=\wt{A_{1,0,1}/\Z_4}$
\mb{(}with $A_{1,0,1}=\R^4/\Lambda_{1,0,1}$, $\Lambda_{1,0,1}=R_1\Z^2\oplus R_2\Z^2$
as in \mb{\req{Aalbeg}}, $R_1=R_2=1$\mb{)} amount to deformations of the
radii $R_1,\, R_2$ of the torus $A_{\alpha,\beta,\gamma}$ and of the B-field to
$B_{\alpha,\beta,\beta^\prime}$ as in Proposition \mb{\ref{protak3geom}}.

For any refined geometric interpretation $(\Omega,\omega,V,B)$ of
the four-plane
$$
x_{1,0,0,1}=\spann_\R\left(
u_1=\wt\Omega_1,\, u_2=\wt\Omega_2,\,
u_3=\wh\ups_3^0+\wh\ups_3,\, u_4=4\wh\ups^0+\check B_4+5\wh\ups\right)
\in\wt\MMM^{K3}
$$
specifying $\CCC_{1,0,0,1}$ in $\MMM^{K3}$
as in Proposition \mb{\ref{protak3geom}}, the
following holds:
If $\Omega$ gives the complex structure of the Fermat quartic
$X_{Fermat}=X(f_0,f_0)$ as in \mb{\req{fermat}}, then
\begin{eqnarray*}
\Omega&=&
\Omega_{1,0,0,1}^X = \spann_\R\left(u_3=\wh\ups_3^0+\wh\ups_3,\,
u_4=4\wh\ups^0+\check B_4+5\wh\ups\right),\\
\mho_X^0&:=&\spann_\R\left(\omega-\langle\omega,B\rangle\ups,\,
\ups^0+B+\left(V-{1\over2}\langle B,B\rangle\right)\ups\right)
=\spann_\R\left(
\wt\Omega_1,\, \wt\Omega_2\right)=\Omega_X^0,
\end{eqnarray*}
with $\ups^0,\,\ups$ generators of $H^0(X,\Z)$ and $H^4(X,\Z)$
in this geometric interpretation. Furthermore, the fields
$V_\pm^{(1)},\,V_\pm^{(2)}$ of \mb{\req{nicedefs}} give
deformations which leave invariant the two-plane $\mho_X^0$ that
encodes the complexified K\"ahler structure of this geometric
interpretation.
\eprop
\bpr
The statements about the interpretation of deformations in
terms of the $\Z_4$-orbifold construction follow from the above
discussion, because solely the deformations listed are compatible
with the $\Z_4$-action.
Note that the induced deformation of the four-plane $x_{1,0,0,1}$
leaves the two-plane
$\Omega_X^0=\spann_\R\left(\wt\Omega_1,\wt\Omega_2\right)$
invariant, as this plane is shared by all $x_{\alpha,\beta,\beta^\prime,\gamma}$
according to Proposition \ref{protak3geom}. Moreover, by the same proposition
the lattice $\QQQ:=x_{1,0,0,1}\cap H^{even}(X,\Z)$ has rank four and is generated
by the pairwise perpendicular lattice vectors $u_1,\,u_2,\,u_3,\,u_4$
with
$$
\langle u_1,u_1\rangle = 2 = \langle u_2,u_2\rangle,
\quad\quad\langle u_3,u_3\rangle = 8 = \langle u_4,u_4\rangle.
$$
For the Fermat quartic $X_{Fermat}=X(f_0,f_0)$ with complex structure
$\Omega_{Fermat}\subset H^2(X,\R)$ by \cite{in76} the
quadratic form associated to $\Omega_{Fermat}\cap H^2(X,\Z)$
(see Definition \ref{attractive}) is
$\diag(8,8)$. However, the only primitive sublattice of $\QQQ$
with this quadratic form is the one generated by $u_3,\,u_4$.
It follows that for every refined geometric interpretation
$(\Omega,\omega,V,B)$ of $\CCC_{1,0,0,1}$
with $\Omega=\Omega_{Fermat}$, the complex structure of the
Fermat quartic, we must have $\Omega_{Fermat}=\Omega_{1,0,0,1}^X$
as claimed and thus also $\mho_X^0=\Omega_X^0$ as claimed.
Above I have already argued that the deformations given by
$V_\pm^{(1)},\,V_\pm^{(2)}$ of \mb{\req{nicedefs}}
leave the plane $\mho_X^0$ invariant, completing the proof.
\epr
Note that the result of Proposition \ref{def2222} does not imply that all
refined geometric interpretations of $\CCC_{1,0,0,1}$ with complex structure
of the Fermat quartic agree: According to Definition \ref{refinedgeom},
such a refined geometric interpretation is given by a decomposition
$x_{1,0,0,1}=\Omega\perp\mho$ into perpendicular oriented two-planes together
with an appropriate choice of null vectors $\ups^0,\,\ups$ which by
Lemma \ref{refgeom} allows to read off the data $(\omega,V,B)$ from $\mho$.
However, given the decomposition $x_{1,0,0,1}=\Omega^X_{1,0,0,1}\perp\mho_X^0$,
an infinity of pairs of null vectors obeying conditions $(1)$ and $(2)$
of Definition \ref{refinedgeom} exists. Therefore I will not be able
to show that the refined geometric interpretation of
$\CCC_{\alpha,\beta,\beta^\prime,\gamma}$ on the quartic $X(f_1,f_2)$
given in Lemma \ref{complexclaim} and Proposition \ref{protak3geom}
agrees with the one claimed to exist in Result \ref{main}. The proof
of Result \ref{main} does not require such an identification.

Proposition \ref{z4isgepner} allows to study the model
$\CCC_{1,0,0,1}=(2)^4$ from a different perspective, namely as a model
arising as orbifold of a certain Landau-Ginzburg model at criticality
\cite{ma89,vawa89}.
This viewpoint, taken from \cite{wi93}, implies (see also \cite[(74)]{as96})
\bfact[\cite{ge87,wi93}]{Witten}
The parameter space $\wt\MMM^{K3}$ of SCFTs on $K3$ contains a subspace of the
form
$$
\OO^+(2,19;\R)/\SO(2)\times\OO(19)\quad\times\quad
\OO^+(2,1;\R)/\SO(2)\times\OO(1)
$$
of SCFTs associated to quartic $K3$ surfaces in $\CP^3$ with normalized
K\"ahler class $\omega=\omega_{FS}$, the class of the Fubini-Study metric,
and B-field $B=b\omega_{FS}$ for some $b\in\R$. It is the space of models
which arise as infrared fixed points of the renormalization group flow
from linear sigma models in $\CP^3$ according to \mb{\cite{wi93}}.
The first factor of this space
accounts for the
choice of the complex structure of the quartic $K3$ surface, while the
second factor captures the parameters $V\in\R^+$ of the volume and $b\in\R$ of the
B-field.

Fixing the complex structure to that of the Fermat quartic
$X_{Fermat}=X(f_0,f_0)$ of \mb{\req{fermat}} and identifying any two
equivalent SCFTs in the resulting space, one obtains a space
$$
\MMM^{Fermat} \quad\cong\quad\OO^+(2,1;\Z)\backslash\OO^+(2,1;\R)/\SO(2)\times\OO(1)
\quad\cong\quad \Sphere^2-\{\infty\}
$$
of SCFTs associated to the Fermat quartic with normalized K\"ahler
class $\omega=\omega_{FS}$ and B-field $B_{FS}=b\omega_{FS},\, b\in\R$.
It has two special points
with non-trivial monodromy: There is one point with monodromy
of order $2$ where the SCFT description is expected to break down, while
the second special point has monodromy of order $4$ and gives the
Gepner model $(2)^4$. The deformations of this model given by fields
$$
\Phi^{n_0}_{\pm n_0,0;\pm n_0,0}\otimes\Phi^{n_1}_{\pm n_1,0;\pm n_1,0}\otimes
\Phi^{n_2}_{\pm n_2,0;\pm n_2,0}\otimes\Phi^{n_3}_{\pm n_3,0;\pm n_3,0}
\;\;\mb{\slshape with }\;\; n_i\in\{0,1,2\},\; \;\sum_{i=0}^3n_i=4
$$
amount to deformations of the defining polynomial of $X(f_0,f_0)$ by monomials
of the form
$\delta\, x_0^{n_0}x_1^{n_1}x_2^{n_2}x_3^{n_3}$,
$\delta\in\C$.
\efact
It is not hard to translate Fact \ref{Witten} into the language of
our moduli space, i.e.\ to combine the results of \cite{wi93} with
those of \cite{asmo94}:
\bcor{onepoint}
The Gepner model $(2)^4$ has a refined geometric interpretation
$(\Omega_{Fermat},\omega_{FS},V_{FS},B_{FS})$ with $\Omega_{Fermat}$
the complex structure of the Fermat quartic $X(f_0,f_0)\subset\CP^3$
of \mb{\req{fermat}}, $\omega_{FS}$ the K\"ahler class induced by the
Fubini-Study metric in $\CP^3$, $V_{FS}={1\over2}$, and
$B_{FS}=-{1\over2}\omega_{FS}$. Furthermore, the deformations
$V_\pm^{(1)},\,V_\pm^{(2)}$ of \mb{\req{nicedefs}}
give pure complex structure deformations
within the family
$$
X(f_1,f_2)\colon\; f_1(x_0,x_1)+f_2(x_2,x_3)=0 \;\mb{ in }\;\CP^3,
\quad f_k(y_1,y_2)=y_1^4+y_2^4+\delta_k\, y_1^2y_2^2,\quad\delta_k\in\C
$$
in this geometric interpretation.
\ecor
\bpr
Using Definition \ref{refinedgeom} and Lemma \ref{refgeom} one finds
that the family of SCFTs $\MMM^{Fermat}$ is given by four-planes
$x=\Omega_{Fermat}\perp\mho_{V,b}\subset H^{even}(X,\R)$
with fixed complex structure $\Omega_{Fermat}$
of $X_{Fermat}$ and complexified K\"ahler structure
$(\omega_{FS},V,B=b\omega_{FS})$, $V\in\R^+,\,b\in\R$,
i.e.\ with
$$
\mho_{V,b}=\spann_\R\left( \omega_{FS}-4b\ups_{FS},\,
\ups^0_{FS}+b\omega_{FS}+\left(V-2b^2\right)\ups_{FS}\right),
$$
where $\ups^0_{FS},\,\ups_{FS}$ generate $H^0(X,\Z)$ and $H^4(X,\Z)$,
respectively. It is convenient to introduce the complex parameter
$$
\tau:=b+i\sqrt{V\over2}\in\H,
$$
and one finds that $\tau,\,\tau^\prime\in\H$ corresponding to four-planes
$x=\Omega_{Fermat}\perp\mho_{V,b}$
and $x^\prime=\Omega_{Fermat}\perp\mho_{V^\prime\!,b^\prime}$ specify
the same SCFT iff $\tau=\gamma\tau^\prime$ for some $\gamma\in\Gamma_0(2)_+$,
the normalizer of $\Gamma_0(2)$ in $\PSL_2(\R)$ \cite[\S3]{we03}.
In other words, $\MMM^{Fermat}$ of Fact \ref{Witten} is given by
$\Gamma_0(2)_+\backslash\H\cong\Sphere^2-\{\infty\}$. This space indeed has
two special points with non-trivial stabilizer in $\Gamma_0(2)_+$, i.e.\
with non-trivial monodromy in $\MMM^{Fermat}$. One of these points
has monodromy
of order $4$, namely $\tau=-{1\over2}+{i\over2}$, which according to
Fact \ref{Witten} gives the Gepner point. Hence $(2)^4$ has refined
geometric interpretation on the Fermat quartic with complexified
K\"ahler structure encoded in $\mho_{{1\over2},-{1\over2}}$,
amounting to $V={1\over2},\,B=-{1\over2}\omega_{FS}$ as claimed.

The claim about the deformations corresponding to
$V_\pm^{(1)},\,V_\pm^{(2)}$ of \mb{\req{nicedefs}} follows directly
from Fact \ref{Witten} together with the result of \ref{quarelli}
that in $X(f_1,f_2)$ by appropriate coordinate transformations the
polynomials $f_1,\,f_2$ can be brought into the
form $f_k(y_1,y_2)=y_1^4+y_2^4+\delta_k\, y_1^2y_2^2$ with
$\delta_k\in\C$.
\epr
As a compatibility check also note the following:
By Proposition \ref{z4isgepner} we can construct the Gepner model
$(2)^4$ in terms of an orbifold of the toroidal model $\TTT_{1,0,0,1}$
which can be carried out in two steps: With $\iota$ the
``Gepner orbifold" of \req{geporb},
$\TTT_{1,0,0,1}/\langle\iota^2\rangle=(\wh 2)^4$ by the discussion in
\ref{geproof2222}, and then $(\wh 2)^4/\langle\iota\rangle=(2)^4$.
This second orbifold has an ``inverse" obtained by the action
\beq{sigmahat}
\wh\sigma\colon\quad
\bigotimes_{j=1}^4\Phi^{l_j}_{m_j,s_j;\qu m_j,\qu s_j}
\quad\longmapsto\quad
e^{{2\pi i\over4}[(\qu m_1+m_1)+(\qu m_2+m_2)]}
\bigotimes_{j=1}^4\Phi^{l_j}_{m_j,s_j;\qu m_j,\qu s_j}
\eeq
on the fields of  $(2)^4$. Together with the identifications
made in Fact \ref{Witten} this action descends to the automorphism $\sigma$
of \req{sigma} which was used in Inose's construction. In other words,
$(2)^4/\langle\wh\sigma\rangle =(\wh 2)^4= (2)^2\otimes(2)^2/\langle\iota^2\rangle$
is indeed a lift of Inose's construction to the SCFT level.
\subsection{The proof}\label{endofproof}
To complete the proof of Result \ref{main} let me first take stock of what we
have achieved so far. By Corollary \ref{onepoint} the claim is true for the special
SCFT $\CCC_{1,0,0,1}$ which agrees with the Gepner model $(2)^4$ by Proposition
\ref{z4isgepner}. By Proposition \ref{def2222} this implies that for the
four-plane $x_{1,0,0,1}\in\wt\MMM^{K3}$ specifying this model within the moduli
space, $x_{1,0,0,1}=\Omega_{1,0,0,1}^X\perp\mho_X^0$ with
$\Omega_{1,0,0,1}^X=\Omega_{Fermat}$  (notations as in Propositions
\ref{protak3geom}, \ref{def2222}, and Corollary \ref{onepoint}). Moreover, the
variation of the parameters $\alpha,\,\beta,\,\beta^\prime,\,\gamma$ in
$\CCC_{\alpha,\beta,\beta^\prime,\gamma}$ away from
$(\alpha,\beta,\beta^\prime,\gamma)=(1,0,0,1)$ leave the two-plane
$\mho_X^0$ invariant,
$x_{\alpha,\beta,\beta^\prime,\gamma}
=\Omega_{\alpha,\beta,\beta^\prime,\gamma}^X\perp\mho_X^0$.
Finally, by Lemma \ref{complexclaim} any refined geometric interpretation of
$x_{\alpha,\beta,\beta^\prime,\gamma}$ using
$\Omega_{\alpha,\beta,\beta^\prime,\gamma}^X$ to specify the complex structure
gives the complex structure of the quartic $X(f_1,f_2)$ obtained from
$\alpha,\,\beta,\,\beta^\prime,\,\gamma$ as in Result \ref{main}.

Recall from the discussion at the end of Section \ref{CY2defmod} that a decomposition
of a four-plane $x\in\wt\MMM^{K3}$ into two oriented two-planes
$x=\Omega\perp\mho$ with choice of ordering amounts to an interpretation of the
corresponding SCFT on $K3$ in terms of a generalized $K3$ structure. Hence
the above already shows that $\CCC_{\alpha,\beta,\beta^\prime,\gamma}$
can be interpreted in terms of generalized $K3$ structures in accord with the
claim of Result \ref{main}. However, the claim made there is stronger in that
it refers to a refined geometric interpretation rather than a generalized $K3$
structure. See also the end of Section \ref{CY2defmod} for a discussion
of this distinction:
It remains to show that the null vectors $\ups_{FS}^0,\,\ups_{FS}$ needed for
the refined geometric interpretation of $\CCC_{1,0,0,1}$ in terms of the Fermat
quartic with normalized K\"ahler class $\omega_{FS}$, volume $V_{FS}={1\over2}$,
and B-field $B_{FS}=-{1\over2}\omega_{FS}$ (Corollary \ref{onepoint}) are
compatible with interpreting $\Omega_{\alpha,\beta,\beta^\prime,\gamma}^X$
in $x_{\alpha,\beta,\beta^\prime,\gamma}
=\Omega_{\alpha,\beta,\beta^\prime,\gamma}^X\perp\mho_X^0$
as two-plane yielding a complex structure for all admissible
$\alpha,\,\beta,\,\beta^\prime,\,\gamma$. In other words, we need to
show that $\Omega_{\alpha,\beta,\beta^\prime,\gamma}^X\perp\ups_{FS}^0$ and
$\Omega_{\alpha,\beta,\beta^\prime,\gamma}^X\perp\ups_{FS}$ for all admissible
$\alpha,\,\beta,\,\beta^\prime,\,\gamma$.

This follows by means of the identifications of deformations
of $\CCC_{1,0,0,1}$ that I have given in Section \ref{gep24}:
By Proposition \ref{def2222}, the fields $V_\pm^{(1)}$ and
$V_\pm^{(2)}$ of \req{nicedefs} give the deformations of
$x_{1,0,0,1}$  into the four parameter family $x_{\alpha,\beta,\beta^\prime,\gamma}$.
On the other hand, by Corollary \ref{onepoint} within the refined
geometric interpretation
$(\Omega_{Fermat}=\Omega_{1,0,0,1}^X,\omega_{FS},V_{FS},B_{FS})$ of $\CCC_{1,0,0,1}$
these fields induce pure complex structure deformations of $X_{Fermat}=X(f_0,f_0)$
to $X(f_1,f_2)$ with $f_k(y_1,y_2)=y_1^4+y_2^4+\delta_k y_1^2y_2^2$, $\delta_k\in\C$.
This amounts to $\ups_{FS}^0,\ups_{FS}\perp\Omega_{\alpha,\beta,\beta^\prime,\gamma}^X$
as needed.
\vspace{\myskip}\qed\noindent
The use of the results
of \cite{wi93} in the above proof
ties this work to seminal insights from the physics literature.
However,
one would hope to be able to find a proof completely within the language of
algebraic geometry instead of having to mix two viewpoints.
A possible strategy for such a proof involves a more
detailed study of the model $(\wh 2)^4$ (see \ref{geproof2222}) and its
two refined
geometric interpretations induced from its two $\Z_2$-orbifold constructions:
One
arising from the Kummer construction for the standard torus $A_{1,0,1}=\R^4/\Z^4$
with vanishing B-field, $(\wh 2)^4=\TTT_{1,0,0,1}/\Z_2
=(2)^2\otimes(2)^2/\langle\iota^2\rangle$ with $\iota$ as in \req{geporb},
and the other
as $\Z_2$-orbifold CFT arising from the extension of the  orbifold
construction $\wt{X(f_1,f_2)/\langle\sigma\rangle}$ to SCFT level,
$(\wh 2)^4=(2)^4/\langle\wh\sigma\rangle$ with $\sigma,\,\wh\sigma$
as in \req{sigma}, \req{sigmahat}, respectively. One should find
the appropriate lattice automorphism of $H^{even}(X,\Z)$ which relates
these two refined geometric interpretations of the relevant
four-plane $x_{(\wh 2)^4}\in\wt\MMM^{K3}$ to one another. Starting from
the Kummer construction the resulting normalized K\"ahler class
in the geometric interpretation on $\wt{X(f_1,f_2)/\langle\sigma\rangle}$
needs to be characterized by its intersection numbers with all other two-cycles
in $\wt{X(f_1,f_2)/\langle\sigma\rangle}$ (or otherwise)
to show that it agrees with the
class of the orbifold limit of an Einstein metric descending from
$\omega_{FS}$ on $X(f_1,f_2)\subset\CP^3$, the class induced by the
Fubini-Study metric on $\CP^3$. The result of Proposition \ref{metric}
was obtained as a welcome side effect of my quest for such a proof.
\section{Discussion}\label{discussion}
This work aims to provide a self-contained description of how to construct
SCFTs $\CCC_{\alpha,\beta,\beta^\prime,\gamma}$ associated to the smooth
quartic $K3$ surfaces
\beq{anotherquartic}
X(f_1,f_2)\colon\quad f_1(x_0,x_1)+f_2(x_2,x_3)=0\quad\mb{ in }\quad\CP^3
\eeq
with normalized K\"ahler class $\omega_{FS}$
induced by the Fubini-Study metric on $\CP^3$,
volume $V={1\over2}$, and B-field $B=-{1\over2}\omega_{FS}$. The construction
itself is simple, since $\CCC_{\alpha,\beta,\beta^\prime,\gamma}$ turns out to be
a standard $\Z_4$-orbifold of a toroidal SCFT.
I regard this as a virtue rather than a disadvantage, since it implies that the family
$\CCC_{\alpha,\beta,\beta^\prime,\gamma}$ does not only lend itself to all
field theory techniques that are linked to the algebraic description through
\req{anotherquartic} but also that the underlying vertex operator algebras are
completely explicitly accessible. Furthermore $\CCC_{1,0,0,1}$ agrees with
the $(2)^4$ Gepner model, such that the family $\CCC_{\alpha,\beta,\beta^\prime,\gamma}$
can be viewed as a deformation of that model. Altogether the four-parameter
family $\CCC_{\alpha,\beta,\beta^\prime,\gamma}$ is well under control,
both from a SCFT and an algebraic point of view, and as such it is the
first known example of its kind.

My construction can be viewed as a generalization to SCFTs of a classical construction
by Inose \cite{in76} by employing a crude version of mirror symmetry.
As a by-product, motivated by discussions with M.\ Headrick and T.\ Wiseman,
a characterization  in terms of a Kummer construction is obtained for
the K\"ahler class induced by the class of the Fubini
Study metric on an orbifold of, say, the Fermat quartic. This makes the
K\"ahler-Einstein metric in the former class accessible to numerical approaches
developed in \cite{hewi05}. One may hope that such numerical approaches
can be generalized to the level of SCFT to begin an analysis of as yet
unexplored SCFTs which have no orbifold description.

Rational SCFTs seem not to play a central r\^ole within the family
$\CCC_{\alpha,\beta,\beta^\prime,\gamma}$.
While not all theories with $\alpha,\,\beta,\,\beta^\prime,\,\gamma
\in\Q$ are rational, these are the parameter values at which the corresponding
quartic hypersurfaces \req{anotherquartic} are ``very attractive", i.e.\ they
have maximal Picard number. It would be interesting to know whether
any particular intrinsic property of the underlying SCFTs distinguishes
rational from non-rational values of $\alpha,\,\beta,\,\beta^\prime,\,\gamma$.
After all, within $\wt\MMM^{K3}$ these theories are characterized by the
fact that the four-plane $x_{\alpha,\beta,\beta^\prime,\gamma}\subset H^{even}(X,\R)$
is generated by lattice vectors in $H^{even}(X,\Z)$.

The proof for the main result of this work links my construction to Witten's
results on gauged linear sigma models \cite{wi93}. An independent
proof would be desirable, but this link could be of considerable use in
applications: Although the relation between Landau-Ginzburg models
and SCFTs has been known for a long time \cite{ma89,vawa89}, this
connection has only rarely been put to use in SCFT. Recent
exceptions to this rule are novel techniques to construct D-branes
by using matrix factorizations, where by an unpublished result of Kontsevich
topological D-branes in
Landau-Ginzburg models are classified
in terms of matrix factorizations \cite{ei80} and therefore are
expected to translate to boundary states in SCFT
\cite{wa95,bhls03,kali03a,add04,hll04,brga05a,egj05,brga05b,err05}. While for supersymmetric
minimal models this correspondence is fully confirmed and understood, for
Gepner models a number of problems remain open. E.g.\ a special
class of matrix factorizations is expected to correspond to arbitrary
permutation branes \cite{re02,add04,brga05a,err05},
but the full correspondence is not yet established. The family
$\CCC_{\alpha,\beta,\beta^\prime,\gamma}$ studied in the present
work seems to provide a promising testing ground for these methods:
Its algebraic description is tailor made for a study in the language
of Landau-Ginzburg models, while its SCFT construction makes it
accessible to all techniques provided by representation theory.
Moreover, since $\CCC_{\alpha,\beta,\beta^\prime,\gamma}$
is a family of deformations of the Gepner model $(2)^4$, such a
study would surpass known results. Very recently a step
in this direction has been carried out in \cite{de05}. There the
model $(2)\otimes(2)$ is  investigated
which can be viewed as a $\Z_4$-orbifold
of the Gepner model $(2)^2$; note $(2)^2\otimes(2)^2/\Z_4=(2)^4$.

While a large part of the tool-set used for the proof of my
main result relies on the particularities of SCFTs associated to $K3$,
above all on the high amount of supersymmetry which these models enjoy,
insights into techniques like matrix factorization or the chiral de Rham
complex as briefly mentioned in the Introduction can be
hoped to generalize to higher dimensions. Indeed, all
these applications intrinsically use a description of the relevant
SCFTs in terms of $N=(2,2)$ supersymmetry. In the geometric interpretation
of $\CCC_{\alpha,\beta,\beta^\prime,\gamma}$ this corresponds to the
fact that I explicitly determine a complex structure for the underlying
$K3$ surfaces. From this viewpoint the family $\CCC_{\alpha,\beta,\beta^\prime,\gamma}$
is special solely because we have several useful descriptions
for it, not because its target space has complex dimension $2$,
and it should be possible to take profit from these descriptions
which can be hoped to generalize to higher dimensions.
\setcounter{section}{0}
\renewcommand\thesection{Appendix \Alph{section}}
\renewcommand\theequation{\Alph{section}.\arabic{equation}}
\section{\hspace*{-0.5em}Quartic representation of elliptic curves}\label{quarelli}
Consider an elliptic curve in Weierstra\ss\ form \req{weierstrass}.
To express this curve  within $\CP_{2,1,1}$, factorize the right hand side of
\req{weierstrass},
$$
y^2t=\prod_{i=1}^3 \left( x-\xi_i t\right)
\quad \Longleftrightarrow\quad
(t y)^2= t\prod_{i=1}^3 \left( x-\xi_i t\right)
\quad \mb{ if } t\neq 0.
$$
Set $y_0=t y$ and with suitable $\alpha,\,\beta,\,\gamma,\,\delta
\in\C$ let
$t=\alpha y_1 + \beta y_2 ,\, x=\gamma y_1+\delta y_2$
to obtain an equation
$$
E_f:\quad
y_0^2 = f(y_1,y_2) \quad\mb{ in }\quad \CP_{2,1,1}
$$
with $f$ a homogeneous polynomial of degree $4$.

As a helpful example consider the elliptic curve with period $\tau=i$.
Its $j$-invariant is well-known, $j(i)=1728$, so its Weierstra\ss\ form can be
taken as
$$
y^2t=x(x-t)(x+t).
$$
Let $\eps$ denote a primitive eighth root of unity, $\lambda\in\C$ such that
$\lambda^{-3}=2i$, and set $t y=y_0,\,t=\lambda (y_1 - \eps y_2),\,
x=-i\lambda( y_1 + \eps y_2)$ as above. This yields
\begin{eqnarray*}
y_0^2 &=& \lambda^3 \left(y_1 - \eps y_2\right) i \left(y_1 + \eps y_2\right)
(1+i) \left(y_1 + i\eps y_2\right) (1-i) \left(y_1 - i\eps y_2\right)\\
&=& y_1^4 + y_2^4\quad =:\quad f_0(y_1,y_2).
\end{eqnarray*}
In general for non-degenerate elliptic curves we can assume without
loss of generality  that
$f$ has the form
$$
f(y_1,y_2) = y_1^4 + 2\kappa y_1^2 y_2^2 + y_2^4,\quad \kappa\in\C.
$$
Indeed, one first finds $\alpha,\,\beta,\,\gamma,\,\delta$ above such that
$f(y_1,y_2) = \nu_1 y_1^4 + 2\kappa^\prime y_1^2 y_2^2 + \nu_2 y_2^4$:
Assuming $\alpha\beta\gamma\delta=1$ with $A:=\alpha\beta,\,
B:=\alpha\delta$ and inserting $t=\alpha y_1 + \beta y_2,\, x=\gamma y_1+\delta y_2$
directly into \req{weierstrass} one needs to solve
\begin{eqnarray*}
0&=&B^2 A^{-1} + 3A^{-1} -27a\left( B^{-2}A+3A \right) -216b A^2B^{-1},\\
0&=&B^{-2} A^{-1} + 3A^{-1} -27a\left( B^{2}A+3A \right) -216b A^2B.
\end{eqnarray*}
The matrix with coefficients $\alpha,\,\beta,\,\gamma,\,\delta$ needs
to be invertible, which implies $B^2\neq1$. Hence we can divide by
$(B-B^{-1})$, and setting $C:=B+B^{-1}$
the above system of equations is equivalent to
\begin{eqnarray*}
D&=&27a A ^2,\\
0&=&\left( 1+D\right) C - 216 b A^3,\\
0&=&\left( 1-D\right) C^2 - 216 b A^3 C+ 4\left( 1-D\right).
\end{eqnarray*}
This system can be  solved in terms of a quartic equation for $D$.
Having brought $f$ to the form
$\nu_1 y_1^4 + 2\kappa^\prime y_1^2 y_2^2 + \nu_2 y_2^4$, where non-degeneracy
implies $\nu_1,\,\nu_2\neq0$,
one merely
needs to rescale the $y_k$ to obtain the desired form
$y_1^4 + 2\kappa y_1^2 y_2^2 + y_2^4$.

To determine under which circumstances
two different values of $\kappa\in\C$ in
$f(y_1,y_2) = y_1^4 + 2\kappa y_1^2 y_2^2 + y_2^4$ yield the same elliptic
curve first restrict to $\Im(\kappa)\geq0$ by employing
$(y_1,y_2)\mapsto(i y _1,y_2)$. For real $\kappa$ one can furthermore
assume $\kappa\geq0$.
By a similar calculation to the above one finds that $\kappa$ and
$\kappa^\prime$ with non-negative imaginary parts
yield the same elliptic curve iff
$\kappa^\prime = {\kappa+3\over1-\kappa}$ or
$\kappa = {\kappa^\prime+3\over1-\kappa^\prime}$. In other words, a
fundamental domain for $\kappa$ is
$\Gamma\backslash\left\{z\in\C\mid z\sim-z\right\}$
where $\Gamma\subset\PSL_2(\R)$
acts by M\"obius transforms and
is generated by $\kappa\mapsto{\kappa+3\over1-\kappa}$. This
transformation has order $3$ and correctly identifies the three values
$\kappa\in\{\pm1,\infty\}$ for which the elliptic curve degenerates.
Its unique fixed point with non-negative imaginary part is $\kappa=i\sqrt3$.
The circle about $\kappa=1$ of radius $2$ contains both
$\kappa=-1$ and $\kappa=i\sqrt3$. Hence a fundamental domain for
those $\kappa$ which yield non-degenerate elliptic curves
is bounded by the interval $(-1,1)$ on the real axis together with the
two circle arcs $|\kappa\pm1|=2$ between the real
axis and $\kappa=i\sqrt3$. These latter two arcs are glued together, while
on $(-1,1)$ we impose $z\sim-z$.
Summarizing one obtains \req{quarticformulation}, as
depicted in Figure \ref{domain}.
\section{\hspace*{-0.5em}Minimal models}\label{minimal}
Let me recall the construction of the $N=(2,2)$ superconformal minimal models
\cite{bfk86,dpz86,zafa86,qi87}. In fact
we will only be concerned with the so-called $A$-series
of minimal models, so by abuse of notation I use
$(k)$ for $k\in\N$ to denote the
coset model
$$
{\SU(2)_k\otimes \U(1)_2 \over \U(1)_{k+2,diag}}
$$
at central charges
$$
c=\qu c={3k\over k+2}
$$
\cite{bfk86,dpz86,kasu89a,kasu89b}.
Following \cite{digr88,fks92} I use the most convenient  description
of the field content of $(k)$ in terms of a free boson $\phi$ and
the parafermion model at level $k$ found in \cite{zafa86,geqi87}:
Let $\psi_l,\,l\in\{1,\ldots,k-1\}$
denote the $\Z_k$ parafermion algebra, i.e.\
$$
\psi_l(z)\psi_{l^\prime}(w)
\sim \left\{
\begin{array}{ll}
c_{l,l^\prime} (z-w)^{-2l l^\prime/k} \left( \psi_{l+l^\prime}(w) +\cdots \right)
& \mb{ if } l+l^\prime <k,\\[2pt]
c_{l,l^\prime} (z-w)^{-2l l^\prime/k} \left( \psi_{l+l^\prime-k}(w)
+\cdots \right)
& \mb{ if } l+l^\prime >k,\\[2pt]
(z-w)^{-2l l^\prime/k} \left( \id + c_k (z-w)^2 T_{pf} (w)
+\cdots \right)
& \mb{ if } l+l^\prime =k,
\end{array}\right.
$$
where $T_{pf}$ is the Virasoro field of the parafermion model.
I denote by $\xi^l_{m,\qu m}$ the primary fields of the parafermion theory,
$l\in\{0,\ldots,k\}$, $m,\qu m\in\Z/2(k+2)\Z$. In particular,
$\psi_l=\xi^0_{2l,0}$. The primary fields
$\Phi^l_{m,s;\qu m,\qu s}(z,\qu z)
=\psi^l_{m,s}(z)\otimes \psi^l_{\qu m,\qu s}(\qu z)$
of the minimal
model $(k)$ can then be expressed as follows:
\begin{eqnarray*}
\mb{for }  l\in\{0,\ldots,k\},&&
\hspace*{-1.5em}m,\qu m\in\Z/2(k+2)\Z,\;\;s,\qu s\in\Z/4\Z,\;\;
l+m+s\equiv\qu l+\qu m+\qu s\equiv0\mod2:\\
\Phi^l_{m,s;\qu m,\qu s}(z)
&=& \xi^l_{m-s,\qu m-\qu s}
e^{-i\beta_k Q_{m,s}\phi(z) -i\beta_k Q_{\qu m,\qu s}\qu\phi(\qu z)},\\
&&\quad\mb{where }\;\;\beta_k := \sqrt{k+2\over k},\;
Q_{m,s}:={m\over k+2}-{s\over2}.
\end{eqnarray*}
I use the same symbols $\Phi^l_{m,s;\qu m,\qu s}$ to label the
conformal families of these primary fields.
More precisely, $\left[\Phi^l_{m,s;\qu m,\qu s}\right]$ denotes the representation
built on the primary $\Phi^l_{m,s;\qu m,\qu s}$ with respect
to the bosonic subalgebra of the superconformal algebra.
One finds the fusion rules
\beq{minifusion}
\mb{for }\Phi^l_{m,s;\qu m,\qu s}(z,\qu z)
=\psi^l_{m,s}(z)\otimes \psi^l_{\qu m,\qu s}(\qu z):
\quad
\left[ \vphantom{\psi^{l^\prime}_{m^\prime,s^\prime}} \psi^l_{m,s}\right]
\times \left[ \psi^{l^\prime}_{m^\prime,s^\prime}\right]
=
\sum_{ \stackrel{\wt{l} = |l-l^\prime| }{
\scriptscriptstyle\wt{l}\equiv l+l^\prime (2)} }^{
\min{(l+l^\prime, 2k-l-l^\prime) }}
\left[ \psi^{\wt{l}}_{m+m^\prime,s+s^\prime}\right],
\eeq
which enjoy the following $\Z_{k+2}$ symmetry:
\beq{phase}
\Phi^l_{m,s;\qu m,\qu s}(z)
\longmapsto
e^{{2\pi i\over 2(k+2)}(m+\qu m)}\cdot
\Phi^l_{m,s;\qu m,\qu s}(z) .
\eeq
The left-handed superconformal algebra is generated by
$$
j(z)={i\over\beta_k}\partial\phi(z), \quad
G^+(z) = {1\over\beta_k}\psi_1(z) e^{i\beta_k\phi(z)},\quad
G^-(z) = {1\over\beta_k}\psi_{k-1}(z) e^{-i\beta_k\phi(z)},
$$
and analogously on the right hand side, i.e.\ $G^+$
and $G^-$ belong to the same conformal family $\left[\Phi^0_{0,2;0,0}\right]$,
where it should be kept in mind that $\Phi^0_{0,2;0,0}$ is primary only
with respect to the bosonic subalgebra of the superconformal algebra,
as mentioned above.
Moreover, up to  shifts by even integers,
$Q_{m,s}$ is the charge of $\Phi^l_{m,s;\qu m,\qu s}$ with respect to
the $\fu(1)$ current $j$ of the superconformal
algebra. All charges $Q$ of primaries $\Phi^l_{m,s;\qu m,\qu s}$
obey $|Q|\leq1$. One has
$$
\Phi^l_{m,s;\qu m,\qu s} = \Phi^{k-l}_{m+k+2,s+2;\qu m+k+2,\qu s+2},
$$
and moreover,
$$
\mb{when } |m-s|\leq l:\quad
h^l_{m,s} := {l(l+2)-m^2\over4(k+2)}+{s^2\over8}
$$
gives the conformal
dimensions $h^l_{m,s},\,h^{l}_{\qu m,\qu s}$ of
$\Phi^l_{m,s;\qu m,\qu s}$. The above formula holds in general
up to shifts by integers. If neither $\Phi^l_{m,s;\qu m,\qu s}$
nor $\Phi^{k-l}_{m+k+2,s+2;\qu m+k+2,\qu s+2}$ lie in the regime where the
formula holds precisely, then one uses it for the representative with
$m-s=l-2$, or, if this does not exist, for the one with $m-s=l+2$, and
adds $1$ to the result \cite{te92}.

As mentioned above, $(k)$ denotes the  $A$-model at level $k$, i.e.\ this
theory has primaries $\Phi^l_{m,s;m,\qu s}$, where $s\equiv\qu s\mod2$.
Fields with even $s$ live in the Neveu-Schwarz sector, while fields with odd
$s$ live in the Ramond sector. Moreover, fields with $s-\qu s\equiv0\mod4$
are bosonic, while those with $s-\qu s\equiv2\mod4$ are fermionic.
Equivalently
and more conveniently, a field is bosonic iff its left and right handed
charges differ by an even integer.
In particular,
$$
\Phi^{l_1}_{m_1,s_1;{m}_1,\qu{s}_1}
\circ \Phi^{l_2}_{m_2,s_2;\qu{m}_2,\qu{s}_2}
= (-1)^{{1\over 4}(s_1-\qu{s}_1)(s_2-\qu{s}_2)}
\Phi^{l_2}_{m_2,s_2;\qu{m}_2,\qu{s}_2}
\circ\Phi^{l_1}_{m_1,s_1;{m}_1,\qu{s}_1}.
$$

For example, at level $k=2$ one gets the following values
for the conformal dimensions and charges $(h^l_{m,s},Q_{m,s})$
of the primary bosonic fields in the Neveu-Schwarz sector:
\beq{weights}
\begin{tabular}[t]{||c||c|c||c|c||c|c||}
\hline
$\;\,\backslash l$&$0,$&$0,$&$1,$&$1,$&$2,$&$2,$\\
$m\backslash
$&$\;\;s=0$&$\;\;s=2$&$\;\;s=0$&$\;\;s=2$&$\;\;s=0$&$\;\;s=2$\\
\hline\hline
$-3$&&&$\ds\left({5\over8},-{3\over4}\right)$&$\ds\left({1\over8},{1\over4}\right)$&&\vphantom{$\ds\sum_7^7$}\\\hline
$-2$&$\ds\left({3\over4},-{1\over2}\right)$&$\ds\left({1\over4},{1\over2}\right)$&&&$\ds\left({1\over4},-{1\over2}\right)$&$\ds\left({3\over4},{1\over2}\right)$\vphantom{$\ds\sum_7^7$}\\\hline
$-1$&&&$\ds\left({1\over8},-{1\over4}\right)$&$\ds\left({5\over8},{3\over4}\right)$&&\vphantom{$\ds\sum_7^7$}\\\hline
$0$&$\left(0,0\right)$&$\ds\left({3\over2},\pm1\right)$&&&$\ds\left({1\over2},0\right)$&$\left(1,\pm1\right)$\vphantom{$\ds\sum_7^7$}\\\hline
$1$&&&$\ds\left({1\over8},{1\over4}\right)$&$\ds\left({5\over8},-{3\over4}\right)$&&\vphantom{$\ds\sum_7^7$}\\\hline
$2$&$\ds\left({3\over4},{1\over2}\right)$&$\ds\left({1\over4},-{1\over2}\right)$&&&$\ds\left({1\over4},{1\over2}\right)$&$\ds\left({3\over4},-{1\over2}\right)$\vphantom{$\ds\sum_7^7$}\\\hline
$3$&&&$\ds\left({5\over8},{3\over4}\right)$&$\ds\left({1\over8},-{1\over4}\right)$&&\vphantom{$\ds\sum_7^7$}\\\hline
$4$&$\left(1,\pm1\right)$&$\ds\left({1\over2},0\right)$&&&$\ds\left({3\over2},\pm1\right)$&$\left(0,0\right)$\vphantom{$\ds\sum_7^7$}\\\hline
\end{tabular}
\eeq
The character of the conformal family
$\Phi^l_{m,s;\qu m,\qu s}$ is given by
\begin{eqnarray}\label{minichar}
X^l_{m,s;\qu{m},\qu{s}} (\tau^\prime,z)
& = & \chi^l_{m,s}(\tau^\prime,z) \cdot
\chi^l_{\qu{m},\qu{s}}(\qu\tau^\prime,\qu{z}) ,\nonumber\\
\chi^l_{m,s}(\tau^\prime,z)
& = &
\sum_{j=1}^k c^l_{4j+s-m}(\tau^\prime) \Theta_{2m-(k+2)(4j+s), 2k(k+2)}
\left(\tau^\prime, {z\over k+2}\right),
\end{eqnarray}
with $\tau^\prime\in\H,\,z\in\C$, and
where $c^l_j, l\in \{0,\dots,k\}, j\in\Z/{2k}\Z$ are
the level $k$ string functions of $\SU(2)_k$, and
$\Theta_{a,b}, a\in\Z/{2b}\Z$ denote classical
theta functions of level $b\in\N$
\cite{ge88,raya87,qi87}.

All minimal models are invariant under simultaneous left and right handed
spectral flow, where the simple current
$\Phi^0_{1,1;0,0}$ is the operator which generates the  spectral flow on
the left.
Hence the Neveu-Schwarz part of the partition function is
given by
$$
Z_{NS}(\tau^\prime,z)
= {1\over2} \hspace*{-1.5em}
\sum_{ \stackrel{
\stackrel{l=0,\dots, k}{m=-k-1,\dots,k+2} ,
}{\scriptscriptstyle l+m\equiv 0(2)}}
\left( \chi_m^{l,0}(\tau^\prime,z) + \chi_m^{l,2}(\tau^\prime,z) \right)
\left( \chi_{m}^{l,0}(\qu\tau^\prime,\qu{z})
+ \chi_{m}^{l,2}(\qu\tau^\prime,\qu{z}) \right),
$$
while the partition functions for the remaining sectors can be obtained
from $Z_{NS}$ by means of spectral flow.
\section{\hspace*{-0.5em}Gepner models}\label{Gepner}
In the main body of this paper I
study SCFTs which can be viewed as internal theories of
type IIA string theories.
Gepner models \cite{ge87,ge88,gvw89}
are heterotic string theories, which are obtained from
certain type IIA models by a trick called heterosis.
However,
by abuse of terminology I instead
call the internal parts of these type IIA theories
Gepner models. To construct these models one first
forms the fermionic tensor product of a number of $N=(2,2)$ minimal models
$(k_1),\ldots,(k_r)$ as discussed in \ref{minimal},
i.e.\ the $\NS$ and the $\Ra$ sectors are
tensorized separately to obtain
$(k_1)\otimes\cdots\otimes(k_r)$. For the construction to work one needs to ensure
that the total central charge of this model is a multiple of $3$,
$$
c=\qu c =
\sum_{i=1}^r {3k_i\over k_i+2} = 3D,\quad D\in\N.
$$
Our model $(k_1)\otimes\cdots\otimes(k_r)$ enjoys a cyclic symmetry
$\Z_M$ induced by \req{phase}
with $M=\lcm\{k_i+2, i=1,\ldots,r\}$. The symmetry is generated by
\beq{gso}
\zeta_M\colon\quad
\bigotimes_{j=1}^r\Phi^{l_j}_{m_j,s_j;m_j,\qu s_j}
\longmapsto e^{{2\pi i\over6}(cs_1)}
\left(\prod_{j=1}^r e^{{2\pi i\over 2(k_j+2)}(m_j+m_j)}\right)
\bigotimes_{j=1}^r\Phi^{l_j}_{m_j,s_j;m_j,\qu s_j}.
\eeq
The Gepner model $(k_1)\cdots(k_r)$ is the orbifold of
$(k_1)\otimes\cdots\otimes(k_r)$ by this symmetry. For
calculations it is useful
to note that the $\Z_M$ invariant part of $(k_1)\otimes\cdots\otimes(k_r)$
is given by those $\NS$ states with integral left and right handed
charges and those $\Ra$ states with
integral (half integral) left and right handed
charges if $D$ is even (odd). Moreover,
the operator of two-fold left handed spectral flow,
$$
U:=\bigotimes_{j=1}^r \Phi^{0}_{2,2;0,0},
$$
in this orbifold
maps the sector twisted by $\zeta_M^m$ to the one twisted by
$\zeta_M^{m+1}$.
In other words, $(k_1)\cdots(k_r)$ is obtained from
$(k_1)\otimes\cdots\otimes(k_r)$ by projecting onto those states
with the correct charges and then generating all remaining states by
repeated action of the two-fold left handed
spectral flow $U$. This process is also known
as GSO projection or as Gepner's $\beta$ method.
Note that $U$ has $\fu(1)$ charge $(-D)$, so that our
condition $D\in\N$ ensures that all $\fu(1)$ charges in a Gepner model
are integral in the $\NS$ sector and integral or half integral in the $\Ra$ sector.
Moreover, \req{minifusion} shows that $U$
is a simple current,
\beq{simplecur}
[U]\times \left[\bigotimes_{j=1}^r\Phi^{l_j}_{m_j,s_j;\qu m_j,\qu s_j}\right]
= \left[\bigotimes_{j=1}^r\Phi^{l_j}_{m_j+2,s_j+2;\qu m_j+2,\qu s_j+2}\right].
\eeq
Note that the bosonic fields in a Gepner model are precisely those fields
whose left and right handed charges differ by an even integer.

The Gepner model $(k_1)\cdots(k_r)$ enjoys many symmetries, in particular
phase symmetries inherited from \req{phase},
%
\begin{eqnarray}\label{gepnerphase}
\mb{for }a_j\in\Z,\;\;
[a_1,\ldots,a_r]\colon&&\hspace*{-1.5em}
\bigotimes_{j=1}^r\Phi^{l_j}_{m_j,s_j;\qu m_j,\qu s_j}\\
&&\longmapsto e^{{2\pi i\over6}(cs_1)}
\left(\prod_{j=1}^r e^{{2\pi i\over 2(k_j+2)}a_j(m_j+\qu m_j)}
\right)
\bigotimes_{j=1}^r\Phi^{l_j}_{m_j,s_j;\qu m_j,\qu s_j}.\nonumber
\end{eqnarray}
As an example, to calculate the partition function of the Gepner model
$(2)^2$ one uses
the characters as obtained from \req{minichar}, which
with $y=e^{2\pi i z}$ yield
\begin{eqnarray*}
(\chi^0_{0,0} + \chi^0_{0,2})(\tau^\prime,z)
\!\!\!\!\!&=&\!\!\!\!\! {1\over2\eta(\tau^\prime)} \left( \sqrt{\theta_3(\tau^\prime,0)\over\eta(\tau^\prime)} \theta_3(2\tau^\prime,z)
+ \sqrt{\theta_4(\tau^\prime,0)\over\eta(\tau^\prime)} \theta_4(2\tau^\prime,z) \right),\e
(\chi^0_{-2,0} + \chi^0_{-2,2})(\tau^\prime,z)
\!\!\!\!\!&=&\!\!\!\!\! {1\over2\eta(\tau^\prime)} \left( \sqrt{\theta_3(\tau^\prime,0)\over\eta(\tau^\prime)} \theta_2(2\tau^\prime,z)
+ \sqrt{\theta_4(\tau^\prime,0)\over\eta(\tau^\prime)} i\theta_1(2\tau^\prime,z) \right),\e
(\chi^0_{4,0} + \chi^0_{4,2})(\tau^\prime,z)
\!\!\!\!\!&=&\!\!\!\!\! {1\over2\eta(\tau^\prime)} \left( \sqrt{\theta_3(\tau^\prime,0)\over\eta(\tau^\prime)} \theta_3(2\tau^\prime,z)
- \sqrt{\theta_4(\tau^\prime,0)\over\eta(\tau^\prime)} \theta_4(2\tau^\prime,z) \right),\e
(\chi^0_{2,0} + \chi^0_{2,2})(\tau^\prime,z)
\!\!\!\!\!&=&\!\!\!\!\! {1\over2\eta(\tau^\prime)} \left( \sqrt{\theta_3(\tau^\prime,0)\over\eta(\tau^\prime)} \theta_2(2\tau^\prime,z)
- \sqrt{\theta_4(\tau^\prime,0)\over\eta(\tau^\prime)} i\theta_1(2\tau^\prime,z) \right),\e
(\chi^1_{1,0} + \chi^1_{1,2})(\tau^\prime,z)
\!\!\!\!\!&=&\!\!\!\!\! {q^{1\over16} y^{1\over4}\over2\eta(\tau^\prime)} \sqrt{\theta_2(\tau^\prime,0)\over\eta(\tau^\prime)}
\theta_3(2\tau^\prime,z+{\tau^\prime\over2}),\e
(\chi^1_{3,0} + \chi^1_{3,2})(\tau^\prime,z)
\!\!\!\!\!&=&\!\!\!\!\! {q^{1\over16} y^{1\over4}\over2\eta(\tau^\prime)} \sqrt{\theta_2(\tau^\prime,0)\over\eta(\tau^\prime)}
\theta_2(2\tau^\prime,z+{\tau^\prime\over2}).
\end{eqnarray*}
For the partition function of $(2)^2$ with some patience from this one obtains

\beq{pf22}
Z_{NS}^{(2)^2}(\tau^\prime,z)
= {1\over2}\left[
\left|{\theta_2(\tau^\prime,0)\over\eta(\tau^\prime)}\right|^4
+\left|{\theta_3(\tau^\prime,0)\over\eta(\tau^\prime)}\right|^4
+\left|{\theta_4(\tau^\prime,0)\over\eta(\tau^\prime)}\right|^4\right]
\left|{\theta_3(\tau^\prime,z)\over\eta(\tau^\prime)}\right|^2.
\eeq
\section{\hspace*{-0.5em}The Gepner models $(2)^2$, $(2)^4$, and $(\wh2)^4$}\label{gep2222}
\subsection{\hspace*{-0.5em}The Gepner model $(2)^2$}\label{gep22}
The partition function \req{pf22} of $(2)^2$
has the form \req{torpart} of the partition
function of a toroidal SCFT. Indeed, every $N=(2,2)$ SCFT at central charges
$c=\qu c=3D$ with $D\in\N$ which is invariant under spectral flow and
only has integral $\fu(1)$
charges in the $\NS$ sector with respect to the $\fu(1)$ currents of the left
and right handed superconformal algebras is expected to have a non-linear
sigma model description on a Calabi-Yau manifold of complex dimension $D$.
Moreover, if for the Gepner model $(k_1)\cdots(k_r)$ one has
$r\leq D+2$, then this model is expected to have a non-linear sigma model
realization on the Calabi-Yau hypersurface
$$
z_1^{2+k_1}+\cdots+ z_{D+2}^{2+k_{D+2}} = 0\quad
\mb{ in } \CP_{{M\over 2+k_1},\ldots,{M\over 2+k_{D+2}}},\
$$
where we set
$k_{r+1}=\cdots=k_{D+2}:=0$ and $M:=\lcm\{2+k_i, , i=1,\ldots,D+2\}$
\cite{ge87}. This claim has been considerably substantiated in \cite{wi93},
though here we will not go into details of its precise meaning in the presence
of quantum corrections. For small $D$, $D\in\{1,2\}$,
however, quantum corrections are not expected in the description of the
relevant moduli spaces, so that this claim can be made much more
precise. Indeed, the Definitions \ref{toroidal} and \ref{k34tori} together
with the properties of Gepner models discussed in \ref{Gepner} ensure
that these models are associated to elliptic curves if $D=1$ or a real
four-torus or $K3$ surface if $D=2$.
Specifically for the Gepner model $(2)^2$ we hence expect a geometric
interpretation on the elliptic curve
$$
y_0^2 = y_1^4+y_2^4  \quad\mb{ in }\quad \CP_{2,1,1},
$$
i.e.\ on an elliptic curve with modulus $\tau=i$ by \ref{quarelli}.
In fact, $(2)^2$ agrees with the toroidal SCFT at central charges $c=\qu c=3$
which is specified by the two moduli $\tau=\rho=i$. This claim is well
established in the literature \cite{cln92}. However, since I will need the explicit
identifications of fields in these two theories,
let me sketch the proof.\footnote{The proof I gave
together with W.\ Nahm in \cite[Theorem 3.2]{nawe00} unfortunately
contains typos and a gap, which I also wish to correct here. As we shall
see, these mistakes
do not influence any other results in that publication.}

We wish to identify two $N=(2,2)$ SCFTs at central charges $c=\qu c=3$,
both of which are
invariant under spectral flow and contain only fields with integral $\fu(1)$
charges in their $\NS$ sectors. Moreover, one checks that for $\tau=\rho=i$
the partition function
of the toroidal theory, which can be obtained from \req{torpart}, agrees
with the one constructed for $(2)^2$ in \req{pf22}. It remains to be shown
that $(2)^2$ decomposes into the tensor product of the fermionic theory at
$c=\qu c=1$ describing a Dirac fermion, and a bosonic theory
at $c=\qu c=2$ with two further $\fu(1)$ currents on each side, such that
the relevant charge lattice is $\Gamma_{i,i}$ as given in \req{chl}.
To this end, one starts by
using \req{weights} to determine all fermionic holomorphic
fields of $(2)^2$ with conformal weights $({1\over2},0)$.
There are only two such linearly independent fields,
realized by the operators of
two-fold left-handed spectral flows. Hence taking $\fu(1)$ charges into
account we readily identify
\beq{Diracid}
\psi_+=\Phi^0_{-2,2;0,0}\otimes \Phi^0_{-2,2;0,0}, \quad\quad
\psi_-=\Phi^0_{2,2;0,0}\otimes \Phi^0_{2,2;0,0}.
\eeq
Recall that these fields are simple currents, and by \req{simplecur}
they indeed realize the OPE of a Dirac fermion. Moreover, since
the analogous simple currents exist on the right hand side,
$(2)^2$ splits into a tensor product of a bosonic theory $\BBB$ at
central charges $c=\qu c=2$ with the fermionic theory which describes
the Dirac fermion. The superpartners of the $\psi_\pm$ give two further
Hermitean conjugate $\fu(1)$ currents on each side of the
theory $\BBB$,
\beq{bosid}
j_\pm = \Phi^0_{\mp2,0;0,0}\otimes \Phi^0_{\mp2,2;0,0}
-\Phi^0_{\mp2,2;0,0}\otimes \Phi^0_{\mp2,0;0,0},
\eeq
such that the real and imaginary parts of $\sqrt2 j_\pm$ obey \req{bosnorm},
and similarly on the right-hand side.
Hence this theory indeed is a bosonic toroidal CFT, and it remains
to determine its charge lattice $\Gamma_{\tau,\rho}$.
The primary fields of $\BBB$ are
obtained as follows: Each orbit under the action
of $\psi_+$ and its right handed analog on the $\NS$ sector of $(2)^2$
contributes one such primary field, namely the one with lowest
conformal weights in the orbit. Using \req{weights} we therefore find
that the following  fields contribute as left or
right hand components of primary fields in $\BBB$, with
notations as in \req{minifusion}:
\begin{eqnarray*}
&&\psi^0_{0,0}\otimes\psi^0_{0,0},\quad
\psi^0_{4,2}\otimes\psi^0_{0,0},\quad
\psi^0_{0,0}\otimes\psi^0_{4,2},\quad
\psi^0_{4,2}\otimes\psi^0_{4,2},\\
&&\psi^0_{2,0}\otimes\psi^0_{2,2}, \quad
\psi^0_{2,2}\otimes\psi^0_{2,0},\quad
\psi^0_{-2,0}\otimes\psi^0_{-2,2}, \quad
\psi^0_{-2,2}\otimes\psi^0_{-2,0},\\
&&\psi^1_{1,0}\otimes\psi^1_{-1,0}, \quad
\psi^1_{-1,0}\otimes\psi^1_{1,0}, \quad
\psi^1_{3,2}\otimes\psi^1_{-3,2},\quad
\psi^1_{-3,2}\otimes\psi^1_{3,2},
\\
&&\psi^1_{1,0}\otimes\psi^1_{3,2}, \quad
\psi^1_{-1,0}\otimes\psi^1_{-3,2}, \quad
\psi^1_{3,2}\otimes\psi^1_{1,0}, \quad
\psi^1_{-3,2}\otimes\psi^1_{-1,0}.
\end{eqnarray*}
Let us denote the $\fu(1)$ currents of the two minimal model factors
of $(2)^2$ by
$j={i\over\sqrt2}\partial\phi,\,j^\prime={i\over\sqrt2}\partial\phi^\prime$,
respectively, such that $j+j^\prime = J = i\nop{\psi_-\psi_+}$. Then
apart from the $\fu(1)^2\oplus\fu(1)^2$ current algebra generated by
$j_\pm,\,\qu\j_\pm$ the theory
$\BBB$ allows a further four-dimensional space of holomorphic primaries
at weight $(1,0)$, generated by
\begin{eqnarray*}
&&j-j^\prime,\;\; \Phi^0_{4,2;0,0}\otimes\Phi^0_{4,2;0,0},\\
&&\Phi^0_{2,0;0,0}\otimes \Phi^0_{2,2;0,0}
+\Phi^0_{2,2;0,0}\otimes\Phi^0_{2,0;0,0},\;\;
\Phi^0_{-2,0;0,0}\otimes \Phi^0_{-2,2;0,0}
+\Phi^0_{-2,2;0,0}\otimes\Phi^0_{-2,0;0,0},
\end{eqnarray*}
and similarly for antiholomorphic fields of weights $(0,1)$.
Hence without loss of generality we can assume that
the lattice $\Lambda$
generated by $\lambda_1,\,\lambda_2$ in  \req{chl} contains vectors
$e_1={1\choose0}$ and $e_2={a\choose b}$ with $a^2+b^2=1$.

On the other hand, using \req{minifusion} one finds that
the eight primaries
\begin{eqnarray*}
&&\hspace*{-0.5em}
\Phi^1_{1,0;1,0}\otimes \Phi^1_{-1,0;-1,0}, \;\;
\Phi^1_{-1,0;-1,0}\otimes \Phi^1_{1,0;1,0}, \;\;
\Phi^1_{1,0;-3,2}\otimes \Phi^1_{-1,0;3,2}, \;\;
\Phi^1_{-1,0;3,2}\otimes \Phi^1_{1,0;-3,2}, \\
&&\hspace*{-0.5em}
\Phi^1_{1,0;-1,0}\otimes \Phi^1_{-1,0;-3,2}, \;\;
\Phi^1_{-1,0;1,0}\otimes \Phi^1_{1,0;3,2}, \;\;
\Phi^1_{1,0;3,2}\otimes \Phi^1_{-1,0;1,0}, \;\;
\Phi^1_{-1,0;-3,2}\otimes \Phi^1_{1,0;-1,0}
\end{eqnarray*}
in $\BBB$ together with the Virasoro fields
generate all fields in $\BBB$ by means of the OPE. All the eight
primaries in the above list have the minimal non-zero
conformal weights which occur
in $\BBB$, $h=\qu h={1\over4}$. Denoting by $\Lambda^*$ the lattice
generated by $\lambda^\ast_1,\,\lambda^\ast_2$ in \req{chl}, this
means that the lattice $\Gamma_{\tau,\rho}$ is generated by
four vectors of the form
$$
\textstyle
{1\over\sqrt2}(\lambda-B\lambda,-\lambda-B\lambda),\,
{1\over\sqrt2}(\wt\lambda-B\wt\lambda,-\wt\lambda-B\wt\lambda),\,
{1\over\sqrt2}(\lambda^\ast,\lambda^\ast),\,{1\over\sqrt2}(\wt\lambda^\ast,\wt\lambda^\ast),\quad
\lambda,\wt\lambda\in\Lambda,\,\lambda^\ast,\,\wt\lambda^\ast\in\Lambda^*,
$$
where $\lambda^\ast$ and $\wt\lambda^\ast$ are generators of
the lattice $\Lambda^*$ with length $1$. Together with $e_1,\,e_2\in\Lambda$,
and since $\Lambda^*$ is indeed the dual lattice of $\Lambda$ when we identify
$\R^2\cong(\R^2)^\ast$ by means of the standard Euclidean scalar product,
this implies
that without loss of generality $\lambda^\ast={1\choose0}, \,\wt\lambda^\ast={0\choose1}$ and
$\lambda^\ast=e_1, \,\wt\lambda^\ast=e_2$.
Hence $\Lambda=\Lambda^*=\Z^2$, meaning $\tau=\rho=i$
as claimed.
\subsection{\hspace*{-0.5em}The Gepner models $(2)^4$ and $(\wh2)^4$}\label{geproof2222}
By the above the fermionic tensor product
$\TTT_{1,0,0,1}:=(2)^2\otimes(2)^2$ of two Gepner models $(2)^2$
is the toroidal SCFT on a complex two-torus $A_{1,0,1}=\C^2/\!\!\sim$ which is
the product of two elliptic curves with moduli $\tau=\rho=i$ each.
On $A_{1,0,1}$ and
with respect to standard coordinates $(z_1,z_2)$ of $\C^2$ we
have $z_k\sim z_k+1\sim z_k+i$, and the theory $\TTT_{1,0,0,1}$
has vanishing B-field.
In the following we
denote the left handed Dirac fermions of the two tensor factors $(2)^2$
of $\TTT_{1,0,0,1}$ by
$\psi_\pm^1,\,\psi_\pm^2$, respectively, and their superpartners by
$j_\pm^1,\,j_\pm^2$.

The theory $\TTT_{1,0,0,1}$
enjoys a natural symmetry of order $4$, which is induced
by the geometric symmetry $(z_1,z_2)\mapsto(i z_1,-i z_2)$ of $A_{1,0,1}$,
or more precisely
by
$$
(\psi_\pm^1,\psi_\pm^2)\longmapsto (\pm i\psi_\pm^1,\mp i\psi_\pm^2),
\quad
(j_\pm^1,j_\pm^2)\longmapsto (\pm i j_\pm^1,\mp i j_\pm^2).
$$
Given the identifications \req{Diracid} and \req{bosid} this means that
in Gepner language on $(2)^2\otimes(2)^2$ we are using the symmetry
\beq{geporb}
\iota\colon\quad
\bigotimes_{j=1}^4\Phi^{l_j}_{m_j,s_j;\qu m_j,\qu s_j}
\quad\longmapsto\quad
e^{{2\pi i\over8}[(\qu m_1-m_1)-(\qu m_3-m_3)]}
\bigotimes_{j=1}^4\Phi^{l_j}_{m_j,s_j;\qu m_j,\qu s_j}.
\eeq
Now recall our description of Gepner models as orbifolds in
terms of the GSO projection by $\zeta_M$ in \req{gso}, where in our case
$M=4$.
A field $\Phi^{(1)}\otimes\Phi^{(2)}$ of $(2)^2\otimes(2)^2$, with
$\Phi^{(l)}$ belonging to the sector of the
$l^{\mb{\scriptsize{}th}}$ tensor factor $(2)^2$
twisted by $\zeta_4^{b_l}$, is invariant under the
above symmetry iff $b_1=b_2$. Again by our description of Gepner models
this implies directly that the above
orbifold, which was induced by the standard geometric
$\Z_4$-symmetry of $\TTT_{1,0,0,1}$, gives the Gepner model $(2)^4$.
This was already shown in \cite[Theorem 3.5]{nawe00}.
Moreover,
we also obtain directly the result \cite[Theorem 3.3]{nawe00}
that the standard geometric $\Z_2$-orbifold of $\TTT_{1,0,0,1}$,
i.e.\ the orbifold by
$\iota^2$ above, yields the model $(\wh 2)^4$ which
is obtained from $(2)^4$ by means of the $\Z_2$-orbifold by
$[2,2,0,0]$ with notations as in \req{gepnerphase}. In \cite{nawe00},
the proof that $(\wh 2)^4$ agrees with the standard $\Z_2$-orbifold of
$\TTT_{1,0,0,1}$ was given independently of the identification of $(2)^2$.
Note that the above corrected
field identification for $(2)^2$ now directly induces
the precise identification \cite[(3.8)]{nawe00} for $(\wh 2)^4$.
%
%
%
%
\bibliographystyle{jhep}
\bibliography{kw}

\def\polhk#1{\setbox0=\hbox{#1}{\ooalign{\hidewidth
  \lower1.5ex\hbox{`}\hidewidth\crcr\unhbox0}}} \def\cprime{$^\prime$}
\providecommand{\href}[2]{#2}\begingroup\raggedright\begin{thebibliography}{10}

\bibitem{frsh86}
D.~Friedan and S.~Shenker, {\it The integrable analytic geometry of quantum
  string},  {\em Phys. Lett.} {\bf B175} (1986), no.~3, 287--296.

\bibitem{na87}
W.~Nahm, {\it Quantum field theories in one and two dimensions},  {\em Duke
  Math.~J.} {\bf 54} (1987), no.~2, 579--613.

\bibitem{mose88}
G.~Moore and N.~Seiberg, {\it Polynomial equations for rational conformal field
  theories},  {\em Phys. Lett.} {\bf B212} (1988) 451--465.

\bibitem{mose89}
G.~Moore and N.~Seiberg, {\it Classical and quantum conformal field theory},
  {\em Commun. Math. Phys.} {\bf 123} (1989) 177--253.

\bibitem{se88b}
G.~Segal, {\it The definition of conformal field theory},  in {\em Differential
  geometrical methods in theoretical physics (Como, 1987)}, vol.~250 of {\em
  NATO Adv. Sci. Inst. Ser. C Math. Phys. Sci.}, pp.~165--171, Kluwer Acad.
  Publ., Dordrecht, 1988.

\bibitem{gaga00}
M.~Gaberdiel and P.~Goddard, {\it Axiomatic conformal field theory},  {\em
  Commun. Math. Phys.} {\bf 209} (2000) 549--594,
  [\href{http://xxx.lanl.gov/abs/hep-th/9810019}{{\tt hep-th/9810019}}].

\bibitem{se04}
G.~Segal, {\it The definition of conformal field theory},  in {\em Topology,
  geometry and quantum field theory}, vol.~308 of {\em London Math. Soc.
  Lecture Note Ser.}, pp.~421--577, Cambridge Univ. Press, Cambridge, 2004.

\bibitem{bpz84}
A.~Belavin, A.~Polyakov, and A.~Zamolodchikov, {\it Infinite conformal symmetry
  in two--dimensional quantum field theory},  {\em Nucl. Phys.} {\bf B241}
  (1984) 333--380.

\bibitem{gi89}
P.~Ginsparg, {\it Some statistical mechanical models and conformal field
  theories},  in {\em Trieste Superstrings 1989}, (Trieste, Italy),
  pp.~130--196, Apr 3-14, 1989.

\bibitem{fuve69}
S.~Fubini and G.~Veneziano, {\it Level structure of dual resonance models},
  {\em Nuovo Cim.} {\bf A64} (1969) 811.

\bibitem{na69}
Y.~Nambu, {\it Quark model and the factorization of the {V}eneziano model},  in
  {\em Proc. of the Int. Conference on symmetries and quark models} (R.~Chand,
  ed.), (Gordon Breach, NY), p.~269, Wayne State University, 1969.

\bibitem{gsw87}
M.~Green, J.~Schwarz, and E.~Witten, {\em Superstring theory I \& II}.
\newblock Cambridge University Press, 1987.

\bibitem{po98}
J.~Polchinski, {\em String theory. {V}ols. {I \& II}}.
\newblock Cambridge Monographs on Mathematical Physics. Cambridge University
  Press, Cambridge, 1998.

\bibitem{ra71}
P.~Ramond, {\it Dual theory for free fermions},  {\em Phys. Rev.} {\bf D3}
  (1971) 2415--2418.

\bibitem{nesc71}
A.~Neveu and J.~Schwarz, {\it Quark model of dual pions},  {\em Phys. Rev.}
  {\bf D4} (1971) 1109--1111.

\bibitem{bo86}
R.~Borcherds, {\it Vertex algebras, {K}ac--{M}oody algebras and the monster},
  {\em Proc. Nat. Acad. Sci. U.S.A.} {\bf 83} (1986) 3068--3071.

\bibitem{flm88}
I.~Frenkel, J.~Lepowsky, and A.~Meurman, {\em Vertex operator algebras and the
  monster}, vol.~134 of {\em Pure and Appl. Math.}
\newblock Academic Press, Boston, 1988.

\bibitem{lvw89}
W.~Lerche, C.~Vafa, and N.~Warner, {\it Chiral rings in ${N}=2$ superconformal
  theories},  {\em Nucl. Phys.} {\bf B324} (1989) 427--474.

\bibitem{cogp91}
P.~Candelas, X.~D.~L. Ossa, P.~Green, and L.~Parkes, {\it A pair of
  {C}alabi-{Y}au manifolds as an exactly soluble superconformal theory},  {\em
  Nucl. Phys.} {\bf B359} (1991) 21--74.

\bibitem{grpl90}
B.~Greene and M.~Plesser, {\it Duality in {C}alabi-{Y}au moduli space},  {\em
  Nucl. Phys.} {\bf B338} (1990) 15--37.

\bibitem{na86}
K.~Narain, {\it New heterotic string theories in uncompactified dimensions $<
  10$},  {\em Phys. Lett.} {\bf 169B} (1986) 41--46.

\bibitem{se88}
N.~Seiberg, {\it Observations on the moduli space of superconformal field
  theories},  {\em Nucl. Phys.} {\bf B303} (1988) 286--304.

\bibitem{asmo94}
P.~Aspinwall and D.~Morrison, {\it String theory on ${K}3$ surfaces},  in {\em
  Mirror symmetry II} (B.~Greene and S.~Yau, eds.), pp.~703--716, 1994,
  [\href{http://xxx.lanl.gov/abs/hep-th/9404151}{{\tt hep-th/9404151}}].

\bibitem{cent85}
A.~Casher, F.~Englert, H.~Nicolai, and A.~Taormina, {\it Consistent
  superstrings as solutions of the $d=26$ bosonic string theory},  {\em Phys.
  Lett.} {\bf B162} (1985) 121--126.

\bibitem{dhvw85}
L.~Dixon, J.~Harvey, C.~Vafa, and E.~Witten, {\it Strings on orbifolds},  {\em
  Nucl. Phys.} {\bf B261} (1985) 678--686.

\bibitem{dhvw86}
L.~Dixon, J.~Harvey, C.~Vafa, and E.~Witten, {\it Strings on orbifolds {II}},
  {\em Nucl. Phys.} {\bf B274} (1986) 285--314.

\bibitem{eoty89}
T.~Eguchi, H.~Ooguri, A.~Taormina, and S.-K. Yang, {\it Superconformal algebras
  and string compactification on manifolds with ${SU}(n)$ holonomy},  {\em
  Nucl. Phys.} {\bf B315} (1989) 193--221.

\bibitem{ge87}
D.~{G}epner, {\it Exactly solvable string compactifications on manifolds of
  ${SU}(n)$ holonomy},  {\em Phys. Lett.} {\bf 199B} (1987) 380--388.

\bibitem{ge88}
D.~{G}epner, {\it Space-time supersymmetry in compactified string theory and
  superconformal models},  {\em Nucl. Phys.} {\bf B296} (1988) 757--778.

\bibitem{nawe00}
W.~Nahm and K.~Wendland, {\it A hiker's guide to ${K}3$ -- aspects of
  ${N}=(4,4)$ superconformal field theory with central charge $c=6$},  {\em
  Commun. Math. Phys.} {\bf 216} (2001) 85--138,
  [\href{http://xxx.lanl.gov/abs/hep-th/9912067}{{\tt hep-th/9912067}}].

\bibitem{we01}
K.~Wendland, {\it Orbifold constructions of ${K}3$: A link between conformal
  field theory and geometry},  in {\em Orbifolds in Mathematics and Physics},
  pp.~333--358, AMS series Contemporary Mathematics, Providence R.I., 2002,
  [\href{http://xxx.lanl.gov/abs/hep-th/0112006}{{\tt hep-th/0112006}}].

\bibitem{ni75}
V.~Nikulin, {\it On {K}ummer surfaces},  {\em Math. USSR Isv.} {\bf 9} (1975)
  261--275.

\bibitem{wi93}
E.~Witten, {\it Phases of ${N}=2$ theories in two dimensions},  {\em Nucl.
  Phys.} {\bf B403} (1993) 159--222,
  [\href{http://xxx.lanl.gov/abs/hep-th/9301042}{{\tt hep-th/9301042}}].

\bibitem{hu03}
D.~Huybrechts, {\it Generalized {C}alabi-{Y}au structures, {${K}3$} surfaces,
  and {$B$}-fields},  {\em Internat. J. Math.} {\bf 16} (2005), no.~1, 13--36,
  [\href{http://xxx.lanl.gov/abs/math.AG/0306162}{{\tt math.AG/0306162}}].

\bibitem{ber99}
I.~Brunner, R.~Entin, and C.~R{\"o}melsberger, {\it D-branes on
  ${T}^4/\mathbb{Z}_2$ and {T}-duality},  {\em JHEP} {\bf 9906} (1999) 016,
  [\href{http://xxx.lanl.gov/abs/hep-th/9905078}{{\tt hep-th/9905078}}].

\bibitem{wa95}
N.~Warner, {\it Supersymmetry in boundary integrable models},  {\em Nucl.
  Phys.} {\bf B450} (1995) 663--694,
  [\href{http://xxx.lanl.gov/abs/hep-th/9506064}{{\tt hep-th/9506064}}].

\bibitem{bhls03}
I.~Brunner, M.~Herbst, W.~Lerche, and B.~Scheuner, {\it {L}andau-{G}inzburg
  realization of open string {TFT}},
  [\href{http://xxx.lanl.gov/abs/hep-th/0305133}{{\tt hep-th/0305133}}].

\bibitem{kali03a}
A.~Kapustin and Y.~Li, {\it D-branes in topological minimal models: The
  {L}andau-{G}inzburg approach},  {\em JHEP} {\bf 07} (2004) 045,
  [\href{http://xxx.lanl.gov/abs/hep-th/0306001}{{\tt hep-th/0306001}}].

\bibitem{add04}
A.~Ashok, E.~Dell'Aquila, and D.-E. Diaconescu, {\it Fractional branes in
  {L}andau-{G}inzburg orbifolds},  {\em Adv. Theor. Math. Phys.} {\bf 8} (2004)
  461--513, [\href{http://xxx.lanl.gov/abs/hep-th/0401135}{{\tt
  hep-th/0401135}}].

\bibitem{hll04}
M.~Herbst, C.-I. Lazaroiu, and W.~Lerche, {\it D-brane effective action and
  tachyon condensation in topological minimal models},  {\em JHEP} {\bf 03}
  (2005) 078, [\href{http://xxx.lanl.gov/abs/hep-th/0405138}{{\tt
  hep-th/0405138}}].

\bibitem{brga05a}
I.~Brunner and M.~Gaberdiel, {\it Matrix factorisations and permutation
  branes},  {\em JHEP} {\bf 07} (2005) 012,
  [\href{http://xxx.lanl.gov/abs/hep-th/0503207}{{\tt hep-th/0503207}}].

\bibitem{egj05}
B.~Ezhuthachan, S.~Govindarajan, and T.~Jayaraman, {\it A quantum {M}c{K}ay
  correspondence for fractional $2p$-branes on {LG} orbifolds},  {\em JHEP}
  {\bf 08} (2005) 050, [\href{http://xxx.lanl.gov/abs/hep-th/0504164}{{\tt
  hep-th/0504164}}].

\bibitem{brga05b}
I.~Brunner and M.~Gaberdiel, {\it The matrix factorisations of the {D}-model},
  {\em J. Phys.} {\bf A38} (2005) 7901--7920,
  [\href{http://xxx.lanl.gov/abs/hep-th/0506208}{{\tt hep-th/0506208}}].

\bibitem{err05}
H.~Enger, A.~Recknagel, and D.~Roggenkamp, {\it Permutation branes and linear
  matrix factorisations},  [\href{http://xxx.lanl.gov/abs/hep-th/0508053}{{\tt
  hep-th/0508053}}].

\bibitem{msv98}
F.~Malikov, V.~Schechtman, and A.~Vaintrob, {\it Chiral de {R}ham complex},
  {\em Commun. Math. Phys.} {\bf 204} (1999), no.~2, 439--473,
  [\href{http://xxx.lanl.gov/abs/math.AG/9803041}{{\tt math.AG/9803041}}].

\bibitem{boli00}
L.~Borisov and A.~Libgober, {\it Elliptic genera of toric varieties and
  applications to mirror symmetry},  {\em Invent. Math.} {\bf 140} (2000),
  no.~2, 453--485.

\bibitem{goma03}
V.~Gorbounov and F.~Malikov, {\it Vertex algebras and the
  {L}andau-{G}inzburg/{C}alabi-{Y}au correspondence},  {\em Mosc. Math. J.}
  {\bf 4} (2004), no.~3, 729--779, 784,
  [\href{http://xxx.lanl.gov/abs/math.ag/0308114}{{\tt math.ag/0308114}}].

\bibitem{mo98b}
G.~Moore, {\it Arithmetic and attractors},
  [\href{http://xxx.lanl.gov/abs/hep-th/9807087}{{\tt hep-th/9807087}}].

\bibitem{mo98a}
G.~Moore, {\it Attractors and arithmetic},
  [\href{http://xxx.lanl.gov/abs/hep-th/9807056}{{\tt hep-th/9807056}}].

\bibitem{we03}
K.~Wendland, {\it On superconformal field theories associated to very
  attractive quartics},  to appear in the proceedings of the Les Houches
  session {``}Frontiers in Number Theory, Physics and Geometry{''},
  [\href{http://xxx.lanl.gov/abs/hep-th/0307066}{{\tt hep-th/0307066}}].

\bibitem{hi03}
N.~Hitchin, {\it Generalized {C}alabi-{Y}au manifolds},  {\em Q. J. Math.} {\bf
  54} (2003), no.~3, 281--308,
  [\href{http://xxx.lanl.gov/abs/math.DG/0209099}{{\tt math.DG/0209099}}].

\bibitem{in76}
H.~Inose, {\it On certain {K}ummer surfaces which can be realized as
  non-singular quartic surfaces in $\mathbb {P}^3$},  {\em
  J.~Fac.~Sci.~Univ.~Tokyo} {\bf Sec. IA 23} (1976) 545--560.

\bibitem{hewi05}
M.~Headrick and T.~Wiseman, {\it Numerical {R}icci-flat metrics on {K}3},  {\em
  Class. Quant. Grav.} {\bf 22} (2005) 4931--4960,
  [\href{http://xxx.lanl.gov/abs/hep-th/0506129}{{\tt hep-th/0506129}}].

\bibitem{dvv87}
R.~Dijkgraaf, E.~Verlinde, and H.~Verlinde, {\it On moduli spaces of conformal
  field theories with $c\geq1$},  in {\em Perspectives in String Theory},
  (Copenhagen, October, 12 - 16, 1987), pp.~117--137, 1987.

\bibitem{kn92}
A.~Knapp, {\em Elliptic curves}.
\newblock Mathematical Notes. Princeton University Press, 1992.

\bibitem{stp88}
A.~Sevrin, W.~Troost, and A.~V. Proeyen, {\it Superconformal algebras in two
  dimensions with ${N}=4$},  {\em Phys. Lett.} {\bf B208} (1988) 447--450.

\bibitem{ho91}
G.~H{\"o}hn, {\it Komplexe elliptische geschlechter und ${S}^1$ {\"a}quivalente
  {K}obordismustheorie},  Diploma thesis, Rheinische Friedrich-Wilhelms
  Uni\-ver\-si\-t{\"a}t, Bonn, und Valendar, 1991.
\newblock http://baby.mathematik.uni-freiburg.de/papers/diplom.ps.gz.

\bibitem{diss}
K.~Wendland, {\em Moduli spaces of unitary conformal field theories}.
\newblock PhD thesis, University of Bonn, 2000.

\bibitem{ce90}
S.~Cecotti, {\it ${N}=2$ supergravity, type {IIB} superstrings and algebraic
  geometry},  {\em Commun. Math. Phys.} {\bf 131} (1990) 517--536.

\bibitem{ku77}
V.~Kulikov, {\it Surjectivity of the period mapping for {${K}3$} surfaces},
  {\em Uspehi Mat. Nauk} {\bf 32} (1977), no.~4(196), 257--258.

\bibitem{to80}
A.~Todorov, {\it Applications of the {K}\"ahler-{E}instein-{C}alabi-{Y}au
  metric to moduli of {${K}3$} surfaces},  {\em Invent. Math.} {\bf 61} (1980),
  no.~3, 251--265.

\bibitem{lo81}
E.~Looijenga, {\it A {T}orelli theorem for {K}\"ahler-{E}instein {${K}3$}
  surfaces},  vol.~894 of {\em Lecture Notes in Math.}, pp.~107--112, Springer,
  Berlin, 1981.

\bibitem{si81}
Y.~Siu, {\it A simple proof of the surjectivity of the period map of {${K}3$}\
  surfaces},  {\em Manuscripta Math.} {\bf 35} (1981), no.~3, 311--321.

\bibitem{na83}
Y.~Namikawa, {\it Surjectivity of period map for {${K}3$} surfaces},  in {\em
  Classification of algebraic and analytic manifolds (Katata, 1982)}, vol.~39
  of {\em Progr. Math.}, pp.~379--397, Birkh\"auser Boston, Boston, MA, 1983.

\bibitem{ni80}
V.~Nikulin, {\it Integral symmetric bilinear forms and some of their
  applications},  {\em Math. USSR Isv.} {\bf 14} (1980) 103--167.

\bibitem{shmi74}
T.~Shioda and N.~Mitani, {\it Singular abelian surfaces and binary quadratic
  forms},  in {\em Classification of Algebraic Varieties and Compact Complex
  Manifolds} (A.~Dold and B.~Eckmann, eds.), pp.~259--287, Lecture Notes in
  Math. 412, 1974.

\bibitem{we00}
K.~Wendland, {\it Consistency of orbifold conformal field theories on ${K}3$},
  {\em Adv. Theor. Math. Phys.} {\bf 5} (2001), no.~3, 429--456,
  [\href{http://xxx.lanl.gov/abs/hep-th/0010281}{{\tt hep-th/0010281}}].

\bibitem{be88}
J.~Bertin, {\it R\'eseaux de {K}ummer et surfaces ${K}3$},  {\em Invent. Math.}
  {\bf 93} (1988), no.~2, 267--284.

\bibitem{shin77}
T.~Shioda and H.~Inose, {\it On singular ${K}3$ surfaces},  in {\em Complex
  Analysis and Algebraic Geometry} (W.~Bailey and T.~Shioda, eds.),
  pp.~119--136, Cambridge Univ. Press, 1977.

\bibitem{mo84}
D.~Morrison, {\it On ${K}3$ surfaces with large {P}icard number},  {\em Invent.
  Math.} {\bf 75} (1984) 105--121.

\bibitem{pss71}
I.~Pjatecki{\u\i}-{\v{S}}apiro and I.~R. {\v{S}}afarevi{\v{c}}, {\it Torelli's
  theorem for algebraic surfaces of type ${K}3$},  {\em Izv. Akad. Nauk SSSR
  Ser. Mat.} {\bf 35} (1971) 530--572.

\bibitem{mo04}
G.~Moore, {\it {L}es {H}ouches lectures on strings and arithmetic},  to appear
  in the proceedings of the {L}es {H}ouches session {``}Frontiers in Number
  Theory, Physics and Geometry{''},
  [\href{http://xxx.lanl.gov/abs/hep-th/0401049}{{\tt hep-th/0401049}}].

\bibitem{at58}
M.~Atiyah, {\it On analytic surfaces with double points},  {\em Proc. Roy. Soc.
  London Ser. A} {\bf 247} (1958) 237--244.

\bibitem{bhpv84}
W.~Barth, K.~Hulek, C.~Peters, and A.~V. de~Ven, {\em Compact Complex
  Surfaces}.
\newblock Springer-Verlag, Berlin Heidelberg, {S}econd enlarged~ed., 2004.

\bibitem{gi88b}
P.~Ginsparg, {\it Applied conformal field theory},  in {\em Lectures given at
  the Les Houches Summer School in Theoretical Physics 1988}, (Les Houches,
  France), pp.~1--168, June 28 - Aug. 5, 1988.

\bibitem{ya78}
S.~Yau, {\it On the {R}icci curvature of a compact {K}{\"a}hler manifold and
  the complex {M}onge-{A}mp{\`e}re equation, {I}},  {\em Comm. Appl. Math.}
  {\bf 31} (1978) 339--411.

\bibitem{mu88}
S.~Mukai, {\it Fi\-nite groups of auto\-mor\-phisms of ${K}3$ sur\-fa\-ces and
  the {M}athieu group},  {\em Invent. Math.} {\bf 94} (1988) 183--221.

\bibitem{cln92}
E.~Chun, J.~Lauer, and H.~Nilles, {\it Equivalence of ${Z}(n)$ orbifolds and
  {L}andau-{G}inzburg models},  {\em Int. J. Mod. Phys.} {\bf A7} (1992)
  2175--2192.

\bibitem{di87}
L.~Dixon, {\it Some world-sheet properties of superstring compactifications, on
  orbifolds and otherwise},  in {\em Lectures given at the 1987 ICTP Summer
  Workshop in High Energy Physics and Cosmology}, (Trieste, June 29 - August
  7), 1987.

\bibitem{ma89}
E.~Martinec, {\it Algebraic geometry and effective {L}agrangians},  {\em Phys.
  Lett.} {\bf B217} (1989) 431--437.

\bibitem{vawa89}
C.~Vafa and N.~Warner, {\it Catastrophes and the classification of conformal
  field theories},  {\em Phys. Lett.} {\bf B218} (1989) 51--58.

\bibitem{as96}
P.~Aspinwall, {\it ${K}3$ surfaces and string duality},  in {\em Fields,
  strings and duality (Boulder, CO, 1996)}, pp.~421--540, World Sci.
  Publishing, River Edge, NJ, 1997,
  [\href{http://xxx.lanl.gov/abs/hep-th/9611137}{{\tt hep-th/9611137}}].

\bibitem{ei80}
D.~Eisenbud, {\it Homological algebra on a complete intersection, with an
  application to group representations},  {\em Trans. Amer. Math. Soc.} {\bf
  260} (1980), no.~1, 35--64.

\bibitem{re02}
A.~Recknagel, {\it Permutation branes},  {\em JHEP} {\bf 04} (2003) 041,
  [\href{http://xxx.lanl.gov/abs/hep-th/0208119}{{\tt hep-th/0208119}}].

\bibitem{de05}
E.~Dell'Aquila, {\it D-branes in toroidal orbifolds and mirror symmetry},
  [\href{http://xxx.lanl.gov/abs/hep-th/ 0512051}{{\tt hep-th/ 0512051}}].

\bibitem{bfk86}
W.~Boucher, D.~Friedan, and A.~Kent, {\it Determinant formulae and unitarity
  for the ${N}=2$ superconformal algebras in two dimensions or exact results on
  string compactification},  {\em Phys. Lett.} {\bf B172} (1986) 316--322.

\bibitem{dpz86}
P.~D. Vecchia, J.~Petersen, M.~Yu, and H.~Zheng, {\it Explicit construction of
  unitary representations of the ${N}=2$ superconformal algebra},  {\em Phys.
  Lett.} {\bf B174} (1986) 280--295.

\bibitem{zafa86}
A.~Zamolodchikov and V.~Fateev, {\it Disorder fields in two--dimensional
  conformal quantum field theory and ${N}=2$ extended supersymmetry},  {\em
  Sov.~Phys. JETP} {\bf 63} (1986) 913--919.

\bibitem{qi87}
Z.~Qiu, {\it Modular invariant partition functions for ${N}=2$ superconformal
  field theories},  {\em Phys. Lett.} {\bf B198} (1987) 497--502.

\bibitem{kasu89a}
Y.~Kazama and H.~Suzuki, {\it Characterization of ${N}=2$ superconformal models
  generated by the coset method},  {\em Phys. Lett.} {\bf B216} (1989)
  112--123.

\bibitem{kasu89b}
Y.~Kazama and H.~Suzuki, {\it New ${N}=2$ superconformal field theories and
  superstring compactification},  {\em Nucl. Phys.} {\bf B321} (1989) 232--283.

\bibitem{digr88}
J.~Distler and B.~Greene, {\it Some exact results on the superpotential from
  {C}alabi-{Y}au compactifications},  {\em Nucl. Phys.} {\bf B309} (1988)
  295--316.

\bibitem{fks92}
J.~Fuchs, A.~Klemm, and M.~Schmidt, {\it Orbifolds by cyclic permutations in
  {G}epner type superstrings and in the corresponding {C}alabi-{Y}au
  manifolds},  {\em Ann. Physics} {\bf 214} (1992) 221--257.

\bibitem{geqi87}
D.~{G}epner and Z.~Qiu, {\it Modular invariant partition functions for
  pa\-ra\-fer\-mio\-nic field theories},  {\em Nucl. Phys.} {\bf B285} (1987)
  423--453.

\bibitem{te92}
M.~Terhoeven, {\it {G}epner modelle und {W}-algebren},  Diploma thesis,
  Rheinische Friedrich-Wilhelms Universit\"at, Bonn, 1992.

\bibitem{raya87}
F.~Ravanini and S.-K. Yang, {\it Modular invariance in ${N}=2$ superconformal
  field theories},  {\em Phys. Lett.} {\bf 195B} (1987) 202--208.

\bibitem{gvw89}
B.~Greene, C.~Vafa, and N.~Warner, {\it {C}alabi-{Y}au manifolds and
  renormalization group flows},  {\em Nucl. Phys.} {\bf B324} (1989) 371--390.

\end{thebibliography}\endgroup
\end{document}